\documentclass[fleqn,usenatbib]{mnras}

\usepackage{newtxtext,newtxmath}
\usepackage[T1]{fontenc}

\DeclareRobustCommand{\VAN}[3]{#2}
\let\VANthebibliography\thebibliography
\def\thebibliography{\DeclareRobustCommand{\VAN}[3]{##3}\VANthebibliography}

\usepackage{gensymb}    %Inclusion of degree symbol
\usepackage{graphicx}	% Including figure files
\usepackage{amsmath}	% Advanced maths commands
\usepackage{comment}    %comment out a section of the writing
\usepackage{rotating}   %rotate a table
\usepackage{hyperref}   %hyperlink
\usepackage{caption}
\usepackage{subcaption}
\usepackage{soul}
\usepackage[normalem]{ulem}
\usepackage{dirtytalk}
\usepackage[dvipsnames]{xcolor}
\usepackage{float}
\usepackage[dvipsnames]{xcolor}
\usepackage{verbatim}
\title[Consistency of halo perturbation parameter]{The GMRT archive atomic gas survey – IV. Consistency of the dark matter halo perturbation parameter from morphological and kinematic lopsidedness of galaxies }

\author[Biswas et al.]{
Prerana Biswas$^{1}$ \thanks{E-mail: prerana.biswas1994@gmail.com},
Narendra Nath Patra$^{2}$,
Veselina Kalinova$^{3,4}$
\\
$^{1}$Indian Institute of Astrophysics, Koramangala, Bangalore-560034 \\
$^{2}$Department of Astronomy, Astrophysics and Space Engineering, Indian Institute of Technology Indore, Indore 453552, India\\
$^{3}$Max Planck Institute for Radioastronomy, Auf dem Hügel 69, D-53121 Bonn, Germany\\
$^{4}$Institute of Astronomy and National Astronomical Observatory, Bulgarian Academy of Sciences, 72 Tsarigradsko Chaussee Blvd., 1784 Sofia, Bulgaria
}
\date{Accepted XXX. Received YYY; in original form ZZZ}

\pubyear{2015}

\begin{document}
\label{firstpage}
\pagerange{\pageref{firstpage}--\pageref{lastpage}}
\maketitle

% Abstract of the paper
\begin{abstract}
The lopsidedness of galaxies is a commonly observed phenomenon, and through different studies, it has been observed that nearly 30\% of galaxies show this phenomenon. In this work, we study morphological lopsidedness in both stellar and gas disks in the inner and outer regions using Fourier analysis techniques and compare the results for a sample of nearby galaxies with different morphologies and environments. Although lopsidedness can result from diverse factors like tidal interactions, gas accretion, and internal instability, recent studies suggest it is a common feature that is not solely reliant on rare events, and moderate lopsidedness most likely results from the disk's response to a lopsided dark matter halo potential. Assuming lopsidedness originates due to a lopsided halo, we find the morphological and kinematic halo perturbation parameters  in the same radial range. Unlike previous studies, we use 3D kinematic modelled rotation curves for finding kinematic lopsidedness and, hence, kinematic halo perturbation parameter. Although the detected linear correlation between them is not statistically significant for our small sample of eleven galaxies, this approach provides a more uniform and physically consistent framework to test the theoretically expected similarity between morphological and kinematic halo perturbation parameters. Further, within this framework, the discrepancy between them does not appear to depend on the nature of the rotation-curve asymmetry of the two sides of the galaxy, in contrast to trends seen in earlier studies. In future work, we plan to extend this analysis to a substantially larger sample in order to robustly assess these findings.

\end{abstract}

% Select between one and six entries from the list of approved keywords.
% Don't make up new ones.

%%%%%%%%%%%%%%%%%%%%%%%%%%%%%%%%%%%%%%%%%%%%%%%%%%

%%%%%%%%%%%%%%%%% BODY OF PAPER %%%%%%%%%%%%%%%%%%

\section{Introduction}
\label{intro}

It has been frequently observed that the distribution of light and, hence, the matter inside spiral galaxies is not purely symmetrical. This phenomenon named \say{lopsidedness} was first introduced through the pioneering work by \citet{bladwin1980}. From the observation of the atomic hydrogen gas (H~{\sc i}) in two halves of the galaxies, \citet{bladwin1980} reported the asymmetry in the spatial extent of H~{\sc i} in the outer origin. Several studies investigating different samples of galaxies have shown that nearly 30\% of late-type galaxies show this lopsidedness phenomenon \citep[][etc.]{bournaud2005, lavin2022, Dolfi2023}.

Different tracers and different approaches have been followed to quantify this asymmetric distribution of disk galaxies.  The lopsidedness of the galaxies can be identified mainly in two ways: morphologically, i.e., by studying the asymmetric distribution of the light or kinematically, i.e., by checking the asymmetry in the velocity field or the rotation curve in the approaching and receding sides. The Morphological lopsidedness has been seen in both stellar \citep[e.g.,][]{Zaritsky2013} and gas disks \citep{whisp_lop2_morph}. In the earlier studies, lopsidedness in the stellar disk has been examined using both the young stellar population with optical observation (e.g., \citet{sdss_lop} studied the lopsidedness for a sample of $\sim$ 25,000 galaxies using different observation bands of SDSS ) and the old stellar population with infrared observation (e.g., \citet{lop_s4g2013} used 3.6 $\mu$m data from the S4G \citep[ Spitzer Survey of Stellar Structure in Galaxies,][]{s4g1} project and studied lopsidedness of 167 nearby galaxies). Further,   \citet{sdss_lop} noted that lopsidedness found using SDSS $r$ and $i$ band data are, on average, virtually identical and slightly smaller than those obtained using $g$-band.

In the case of morphological lopsidedness, the asymmetry at different radii in stellar and gas disks may differ from each other depending upon the physical origin of the lopsidedness and the response time of the star and the gas disk. From the kinematic point of view also, the response time of the different components may be different, resulting in a difference in morphological lopsidedness of the stellar and the gas disk. Here, in this work, we investigate both the morphological and kinematical lopsidedness of the pilot sample of galaxies from GMRT (Giant Metrewave Radio Telescope) ARChIve Atomic gas survey (GARCIA) \citep{biswas2022}. For the morphological case, we find the lopsidedness in both the stellar and gas disks and compare the results in the inner and outer regions of the stellar and gas disk.

Although the lopsidedness can originate from various phenomena, e.g., tidal interactions, gas accretion \citep[e.g., see][]{Dolfi2025}, internal instability etc., recent studies  \citep{lop_s4g2013} of the stellar lopsidedness of 167 nearby galaxies from the Spitzer survey \citep{s4g1,s4g2,s4g3} suggest that lopsidedness is a generic trait of galaxies and does not solely depend on a rare event, such as the direct interaction of a satellite galaxy onto the disc of the parent galaxy \citep{Zaritsky2013}. Also, in the study of the lopsidedness of the disk galaxies from TNG50  simulation \citep{tng50}, \citet{lavin2022} and \citet{Fontirring2025} suggested that tidal interactions with satellite galaxies or environmental effects are not the primary driving agent of lopsided modes. \citet{Zaritsky2013} also showed that moderate lopsidedness is likely to occur as the response of the disk to a lopsided halo potential. Here in this study, we consider the lopsidedness to be arising due to a lopsided halo and find the halo perturbation parameter \citep{jog2002}. Previous studies by \citet{whisp_lop2_morph} found that the halo perturbation parameters found from the kinematic lopsidedness and morphological lopsidedness can differ significantly depending upon the nature of differences in kinematics between the approaching and receding sides of the galaxies instead of considering the same perturbed halo potential. The kinematic halo perturbation parameter found by them was derived using a rotation curve from 2D kinematic modelling, i.e., by fitting two-dimensional tilted ring model \citep{tiltd_ring1974} to the two-dimensional velocity field. However, \citet{fat} showed there could be significant differences in retrieved kinematics and hence the rotation curve between 2D kinematic modelling and 3D kinematic modelling, i.e., by fitting a three-dimensional titled ring model to the 3D data cube.  Here in this study, we found the kinematic halo perturbation parameter using 3D modelled rotation curve and the morphological halo perturbation parameter through Fourier analysis techniques and compared the results with the previous work and with past theoretical predictions.

Section \ref{data} describes the data used for this study; section \ref{morph_lop} and section \ref{kin_lop} demonstrate the methodology for finding the morphological and kinematic lopsidedness and describe the lopsidedness found in our sample. Using the results of the above sections, in section \ref{halo_parameter}, we find the halo perturbation parameter from morphological and kinematic lopsidedness and compare the results with previous studies and theoretical considerations. Finally, in section \ref{conclusion}, we discuss and conclude our results. In all the figures and tables of the paper, the order of the galaxies are according to their ascending order of R.A. i.e., the order of the galaxies is NGC 0784, NGC 1156, NGC 3027, NGC 3359, NGC 4068, NGC 4861, NGC 7292, NGC 7497, NGC 7610, NGC 7741, NGC 7800.

\section{Data}
\label{data}
As mentioned in the introduction, for this study, we are using the sample of galaxies from the GARCIA survey \citep{biswas2022}. The sample consists of eleven nearby bright galaxies with different morphologies, inclinations and have a broad range of H~{\sc i} mass and radius \citep[see table 5,][for the details of the sample]{biswas2022}. For investigating the morphological lopsidedness of the gas disk, we use H~{\sc i} moment zero maps from this survey.  Further, to find the lopsidedness in the stellar disk, we mainly used Spitzer IRAC 3.6 $\mu m$  data (NGC 0784, NGC 1156,  NGC 3027,  NGC 3359, NGC 4068, NGC 4861,  NGC 7497,  NGC 7741, NGC 7800) from the S4G  project \citep{s4g1,s4g2,s4g3} and Super mosaic data products from Spitzer Heritage Archive \citep{spitzer_heritage_arch}. However, for two galaxies (NGC 7292 and NGC 7610), the IRAC 3.6 $\mu m$ data was not readily available; hence we used the SDSS r-band \citep{sdss_iv_overview} data for them.  For the kinematic lopsidedness, we followed the procedure described in \citet{biswas2023} to extract the extended H~{\sc i} rotation curve using 3D kinematic modelling.

\section{Morphological lopsidedness}
\label{morph_lop}
The deviation of the isophotes or constant surface brightness curves from  circles for face-on type galaxies suggests the lopsidedness of the disk. The earlier studies related to morphological lopsidedness were limited to the nearly face-on type galaxies; e.g., \citet{rix_zuritsky1995} selected their sample of 18 galaxies with inclination $\leq 20 ^{\circ}$ for their work related to morphological lopsidedness. This is because the inclined galaxies, instead of having circular disks, may appear elliptical observationally and indicate false lopsidedness. But, in the later works in this regard \citep[e.g.,][]{whisp_lop2_morph}, the galaxies are deprojected before checking its lopsidedness. This enables the inclusion of galaxies with all ranges of inclination angles for the study of lopsidedness. Here, in our work, too, we deproject the galaxies from our sample before investigating if they appear to be lopsided or not. 

The asymmetric distribution of light or matter is seen in both star disks using optical or infrared observation \citep[e.g.,][]{rix_zuritsky1995, Zaritsky2013} and in gas disks using H~{\sc i} observations \citep[e.g.,][]{whisp_lop2_morph}.  The most natural way to quantify this asymmetry in the circular disks is through Fourier analysis. In the past, the Fourier analysis technique has been used extensively to define the morphological lopsidedness of the disk \citep[][etc.]{rix_zuritsky1995, Zaritsky2013}. Using this technique, the stellar or H~{\sc i} surface brightness at some radius can be expressed in different Fourier modes in the following way \citep{whisp_lop2_morph}:
\begin{equation}
    \Sigma (R, \phi) = I_0(R) + \sum_m I_m(R)\cos(m\phi - \phi_m(R)) , 
    \label{eqn:lop_four}
\end{equation}

where $\Sigma (R, \phi)$ is stellar, or H~{\sc i} surface brightness at radius $R$; $I_0(R)$ is the mean surface brightness; $I_m(R)$ and $\phi_m$ are respectively the amplitude, and the phase of the $m$th Fourier mode; and $\phi$ represents the azimuthal angle in the plane of the galaxy. If the disk is purely  circular, then the surface brightness, $\Sigma (R, \phi)$, will be equal to $I_0(R)$ and all other Fourier modes will become zero. The $m=1$ type of distortion suggests the deformation of a circular disk to a specific direction that may happen due to structures like Magellanic spirals or due to interactions with its surrounding environment or some other galaxies in a specific direction.  The $m=2$ mode may arise due to two-armed spirals or bar-like structures or due to elliptical disks that may appear even after the successful deprojection. Correspondingly, in three-armed spirals the $m=3$ mode will contribute significantly as seen in the galaxies ESO436, NGC 1379 and NGC 7309 from \citet{rix_zuritsky1995}. Thus, these fourier modes give us valuable information about the structure of the galaxies. Here below, we discuss our procedure for finding the morphological lopsidedness in stellar disks and in gas disks through the Fourier decomposition method. 

\subsection{Lopsidedness of the stellar disk}
For the stellar disk, we first mask the brightest stars in the foreground so that they do not contribute to any lopsidedness feature in the Fourier analysis. The resolution in the optical disk is very high   (   for IRAC 3.6 $\mu m$, the resolution is $1.7''$  for S4G \citep{s4g1} data and $2''$ for SPITZER Heritage Archive data \footnote{\url{https://irsa.ipac.caltech.edu/onlinehelp/heritage/}}; and  for SDSS r-band data the resolution is $1.32''$  \footnote{\url{https://www.sdss4.org/dr17/imaging/other_info/}} ) in comparison to H~{\sc i} moment zero maps (with a typical resolution of $\sim$ 35 $^{\prime \prime}$ for an image produced with 5 K$\lambda$ UV cutoff). If we use these higher resolved images in the Fourier analysis to find out the lopsidedness, this will pick the very small fluctuations in the stellar disk across the different radii. These smaller fluctuations are often local and may not necessarily be related to the global lopsidedness phenomenon. That is why we convolve these images to a resolution size of 5$^{\prime \prime}$ $\times$ 5$^{\prime \prime}$ before carrying out the Fourier analysis. We checked for two sources in our sample that the Fourier modes do not change considerably for the change of the resolution (from 2$^{\prime \prime}$ $\times$ 2$^{\prime \prime}$ to 20$^{\prime \prime}$ $\times$ 20$^{\prime \prime}$).  These convolved images are then deprojected.  Our process of deprojection involves several steps. In summary, we first take a grid similar to the data, and then this grid is rotated at an angle equal to the position angle of the galaxy using a rotation matrix. Then the surface brightness in each grid of the original data is interpolated in the new rotated grid, and its value is corrected for the inclination of the galaxy. After that, the rotated grid is deprojected by correcting the length of the y-axis of each grid with the inclination of the galaxy. A new mesh grid is created based on this deprojected grid, and finally, the previously deprojected surface brightness is interpolated in this new grid. For the deprojection, we used the optical centres as found from {\href{https://ned.ipac.caltech.edu/}{NED}} (NASA/IPAC Extragalactic Database). The inclination and position angle used in this process are the kinematic inclination and position angle as found by the best possible 3-dimensional kinematic modelling of these sources \citep[see][for more details]{biswas2023}. It should be noted that our deprojection method does not take into account the effect of the galaxy’s vertical thickness. A better representation of the deprojected galaxy could be achieved by including this effect. However, in the present study, we have limited our deprojection to a simplified approach that ignores thickness. This effect will be incorporated in future analyses of additional sources from the upcoming GARCIA batch-II. Further, we use equation \ref{eqn:lop_four} to find out the different Fourier modes of these galaxies.

 To find the maximum radius up to which the Fourier analysis can be performed, we followed the following procedure. For the stellar disk, we fit an ellipse at the one-sigma contour of the galaxy, and the semi-major axis of the fitted ellipse is considered the maximum traceable radius of the stellar disk. For determining the one-sigma contour label, we first found out the RMS noise in the field of the image. This is done by plotting a histogram of all the pixels in the field. The histogram generally shows the distribution of all the sources in the field. However, the negative values in the histogram come from the noise only. Hence, the full-width half maxima of a Gaussian fitted in the negative part of the histogram gives the RMS noise in the field. After finding the noise, we select a region roughly based on the ellipse found from  {\sc{findgalaxy}}, a routine in the MGE package by \citet{mge_cappellari2002} to determine the region of the galaxy and mask the other regions in the field. The {\sc{findgalaxy}} routine finds the largest region of connected pixels that have flux above a given threshold. This masking helps in improving the fitting quality. 
 
 Initially, we intended to use the radius corresponding to the major axis of the fitted ellipse at the one-sigma contour as the maximum radius for performing the Fourier analysis. The analysis would be carried out in successive annular regions, each separated by the convolved beam size (i.e., 5$^{\prime\prime}$), starting from this radius. However, this radius cannot always be taken as the outer limit for the Fourier analysis. The reason behind it is the following: In some annular regions, which mostly happen at the outer part of the galaxy, there may be some azimuthal angles, $\phi$, where the surface brightness falls below the observational detection threshold. However, emission from one part of the annulus being below the threshold while another is not does indicate morphological lopsidedness, but the emission from regions below the detection limit cannot be treated as statistically significant. Thus, these regions in those annuli will not provide fiducial inputs in the Fourier analysis procedure, and as a result, the corresponding lopsidedness can not be trusted with confidence.  This scenario mostly happens in the last and second last annular regions of the galactic disk. This is why we start from the radius defined from the major axis of the fitted ellipse to the one-sigma contour of the galaxy and check the distribution of the intensity with the azimuthal angle, $\phi$, for each annulus. We choose the annulus as the last annulus for which we have surface brightness above one sigma at all the azimuthal angles. We consider the outer radius of that annulus as the radius up to which Fourier analysis should be performed.
 
 The results of the Fourier analysis for the stellar disk are shown in figure \ref{fig:opt_lop}. The associated galaxies are also shown below each of the plots, showing the radial variation of the ratio of the different Fourier modes. The red concentric ellipses demonstrate every second ring where Fourier analysis has been performed.  For most of the cases, we found that $m=2$ mode and hence the ratio of the amplitudes for $m=2$ mode to $m=0$ mode dominates over the other modes, suggesting the presence of the two spiral arms or a two-fold symmetry. For some galaxies (i.e., NGC 1156, NGC 3027, NGC 4861, NGC 7610),  the odd modes are generally not significant in the inner radius, but towards the higher radius, they become significant.  For some of the galaxies in our sample in the outermost radius, the $m=1$ mode becomes significant, i.e., $I_{m}/I_{0}$ $\geq 0.2$, which is a cutoff ratio for $m=1$ mode for classifying if a galaxy is lopsided or not by \cite{rix_zuritsky1995}. Hence, six of the galaxies from our sample, i.e., NGC 1156, NGC 3027, NGC 4068, NGC 4861, NGC 7610, and NGC 7800, appear to have a lopsided stellar disk according to the ratio of the amplitude of $m=1$ to $m=0$ mode at the maximum traceable radius of the stellar disk.  In addition to evaluating lopsidedness at the outermost traceable radius, we also, adopted a more robust classification based on the average amplitude ratios of the different Fourier modes in the inner and outer regions of the stellar disk as presented in detail in subsection~\ref{subsec:comp_stellar_gas}. Further, it is to be noted that, for NGC 4861, the bright knot in the southeast part of the stellar disk resulting from a star-forming region of this galaxy \citep{lopn4861_2004, lopn48612006} may be responsible for strong asymmetry of the optical disk.

\begin{figure*}
    \centering
    \begin{tabular}{ccc}
        \includegraphics[width=0.3\textwidth]{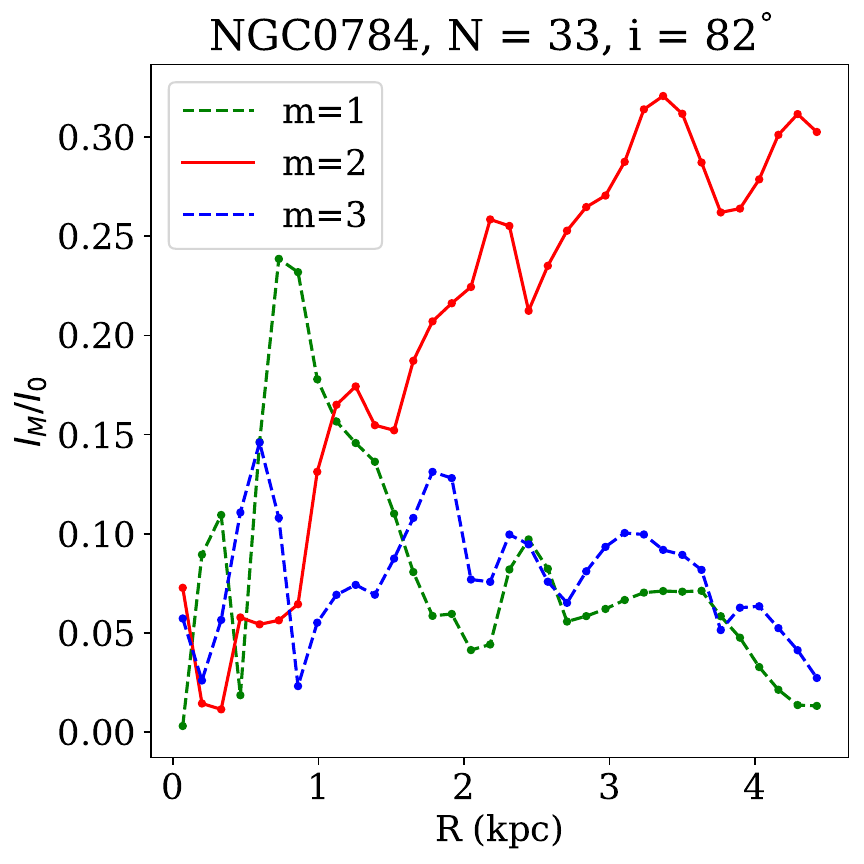} & 
        \includegraphics[width=0.3\textwidth]{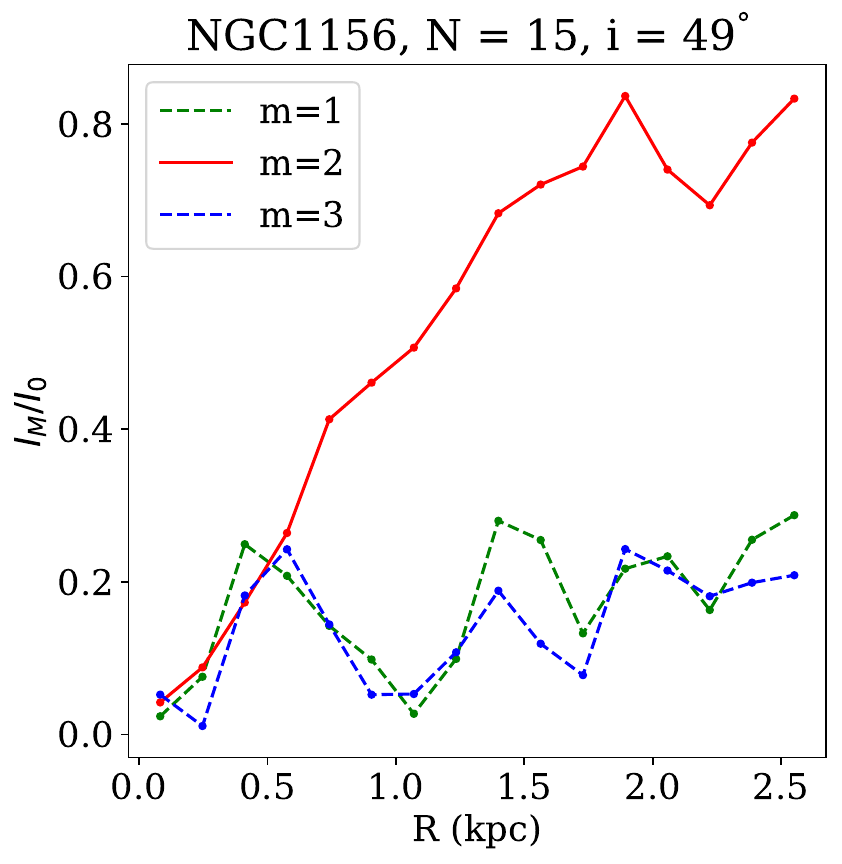} & 
        \includegraphics[width=0.3\textwidth]{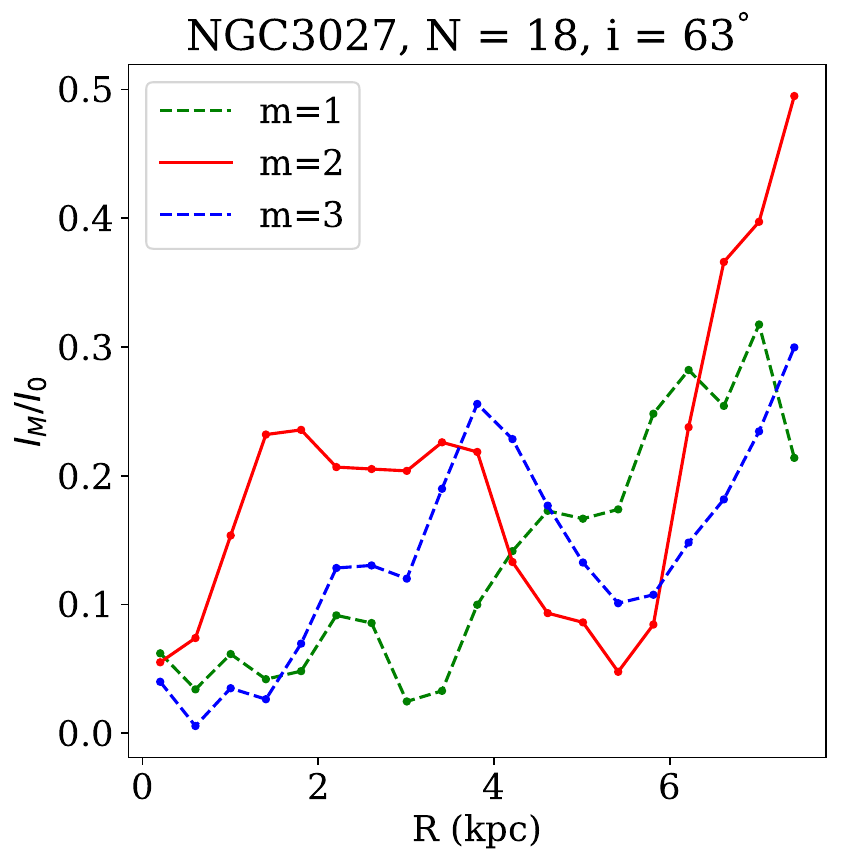} \\
        \includegraphics[width=0.3\textwidth]{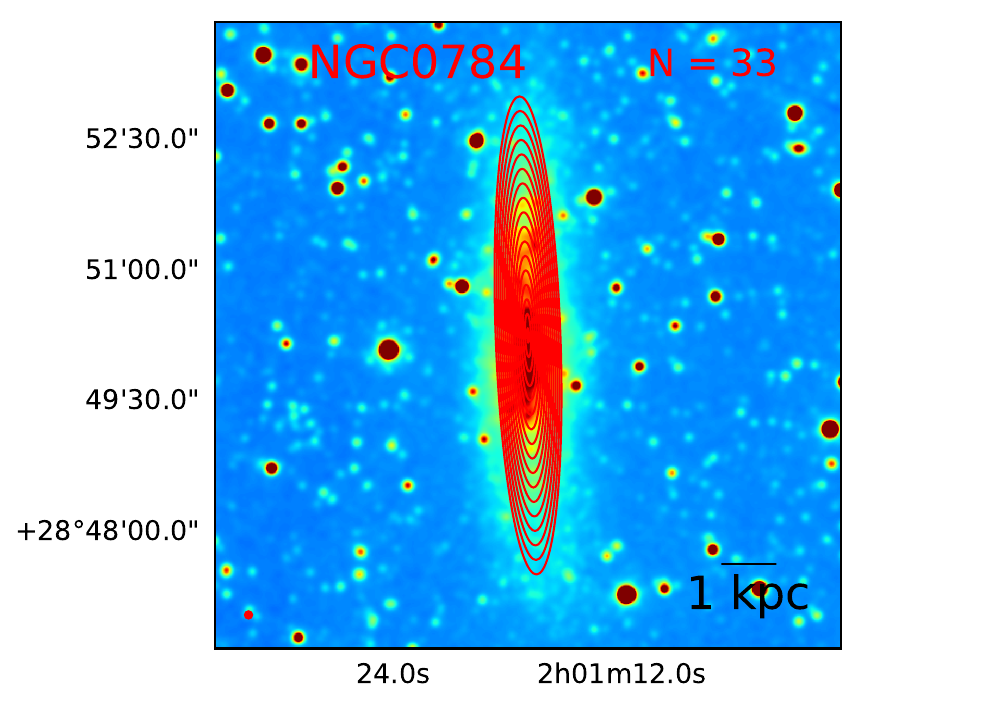} &
        \includegraphics[width=0.3\textwidth]{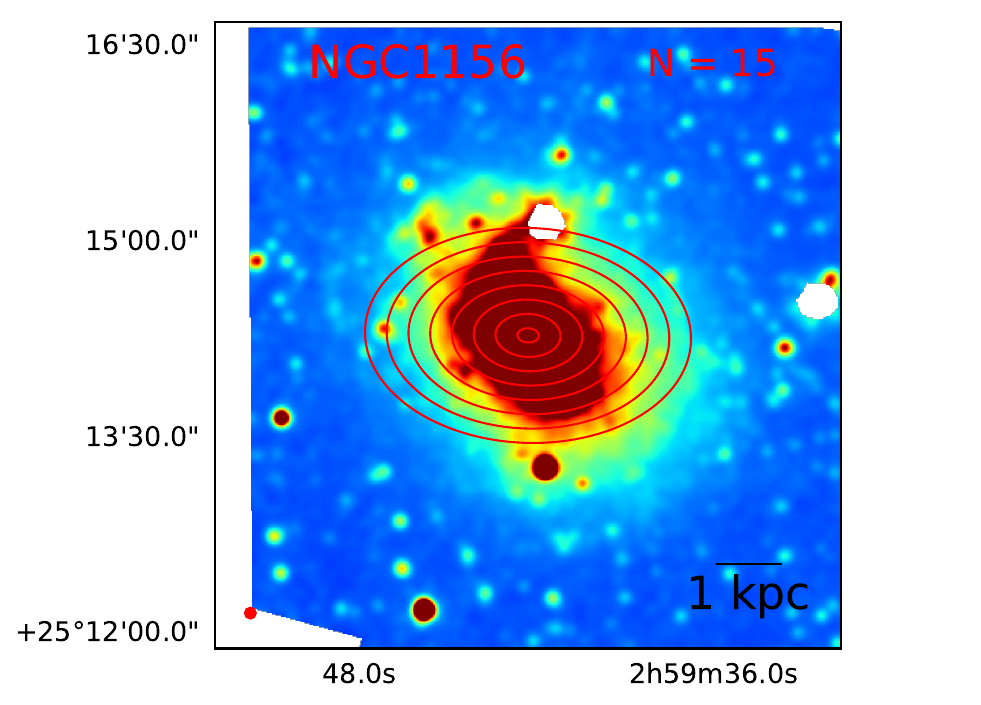} &
        \includegraphics[width=0.3\textwidth]{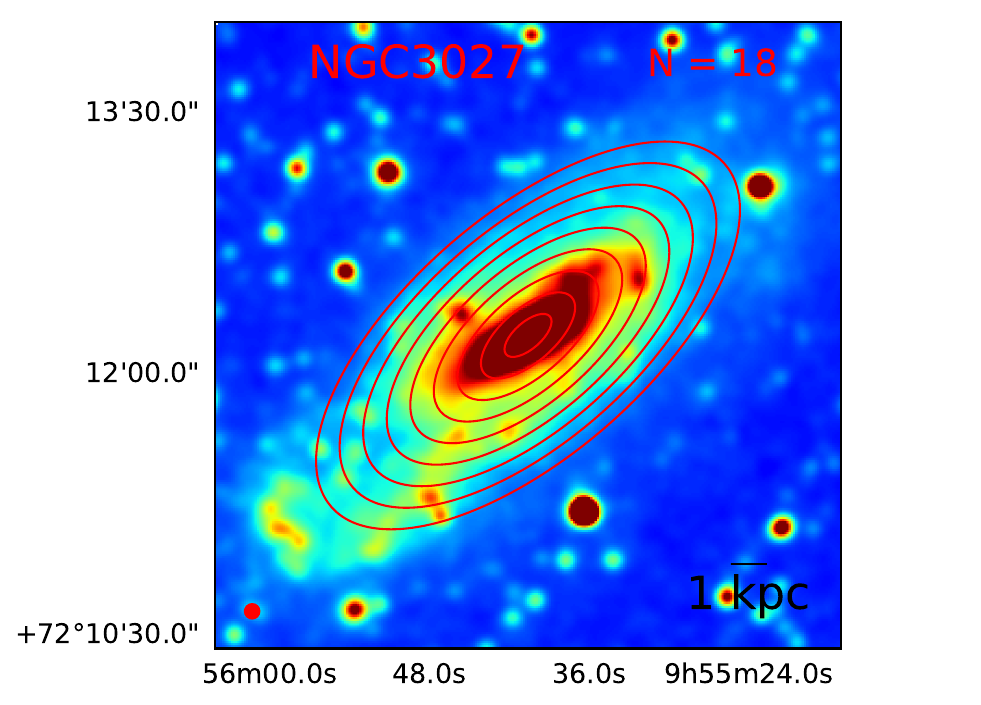} \\

        \includegraphics[width=0.3\textwidth]{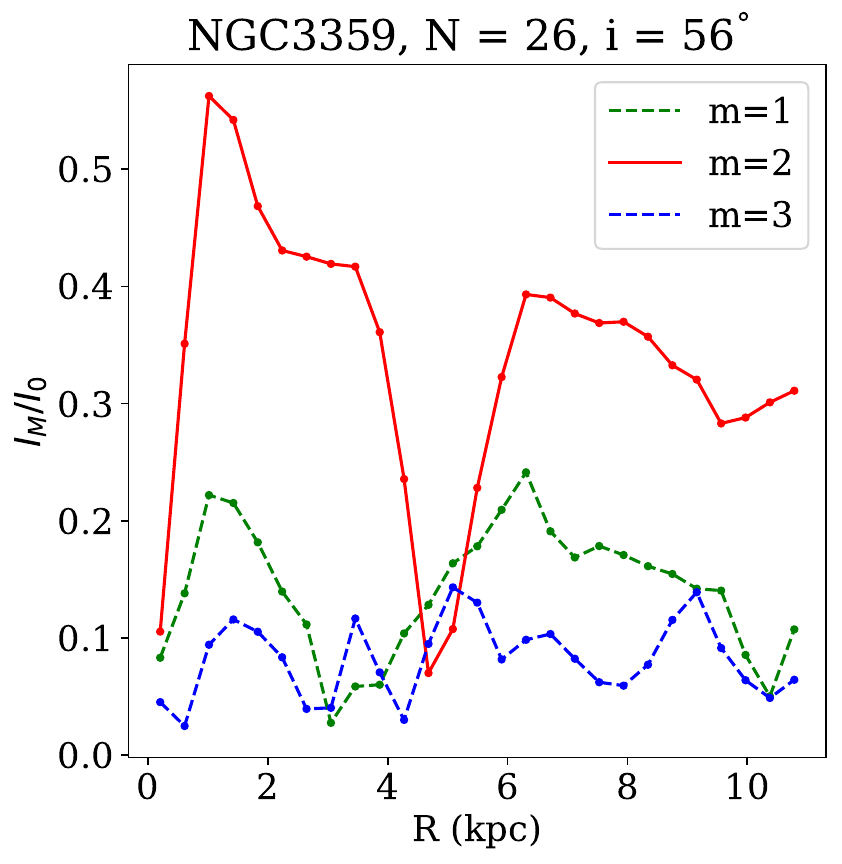} & 
        \includegraphics[width=0.3\textwidth]{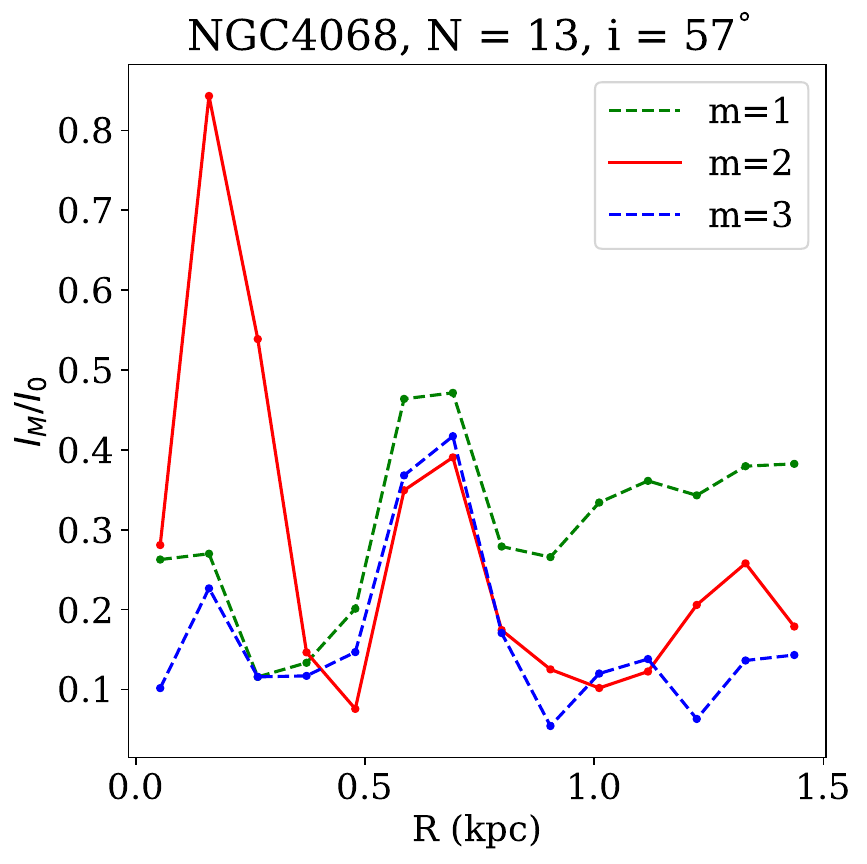} & 
        \includegraphics[width=0.3\textwidth]{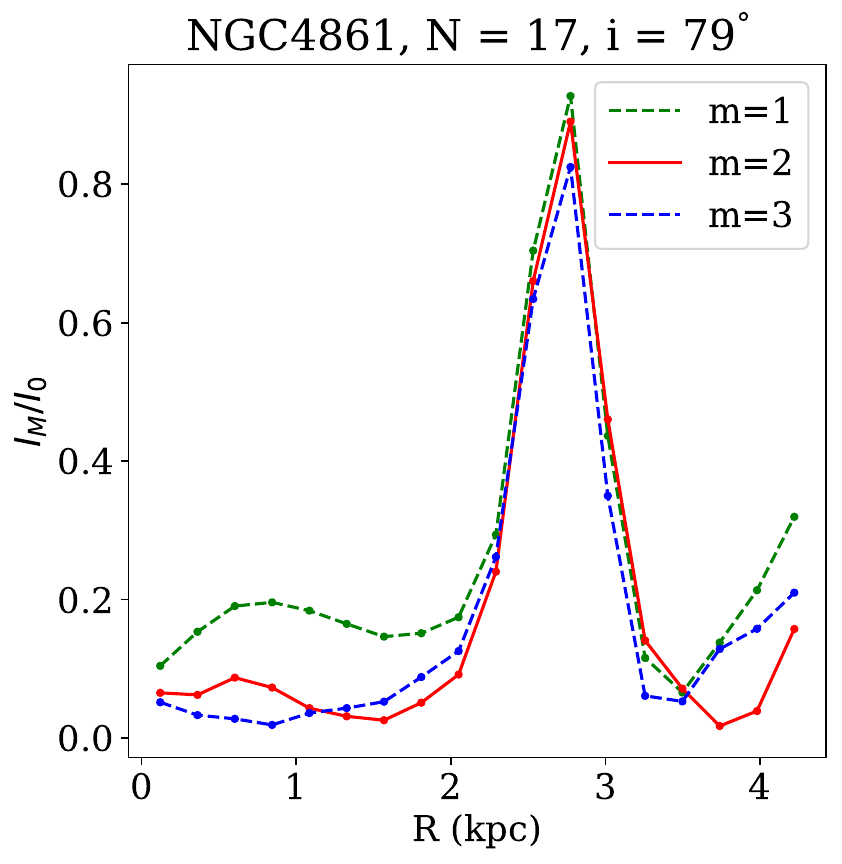} \\
        \includegraphics[width=0.3\textwidth]{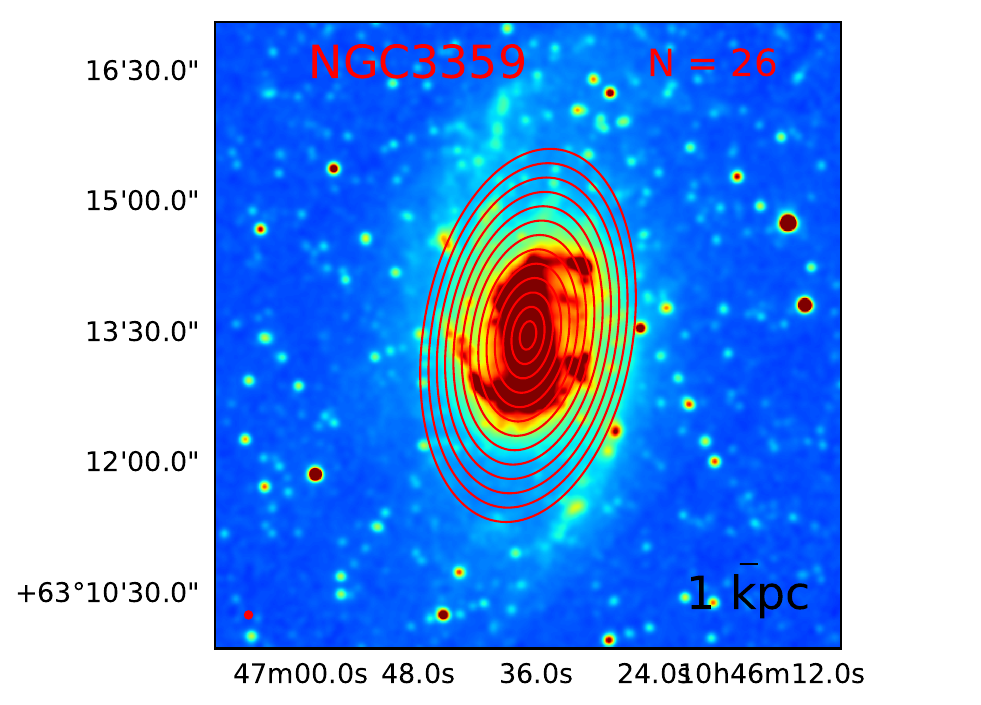} &
        \includegraphics[width=0.3\textwidth]{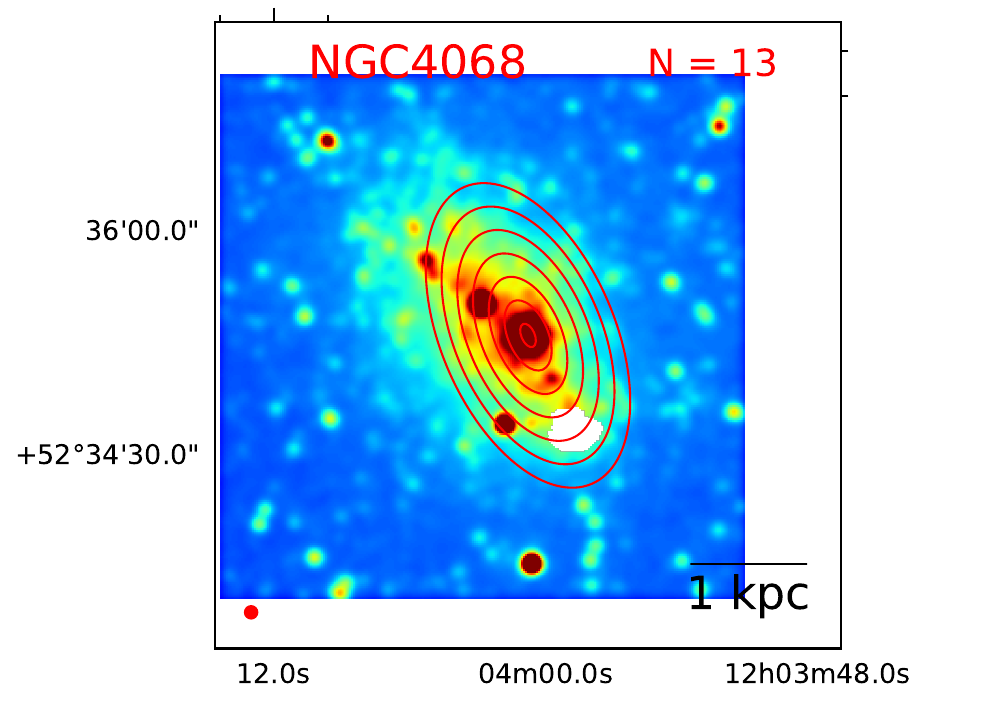} &
        \includegraphics[width=0.3\textwidth]{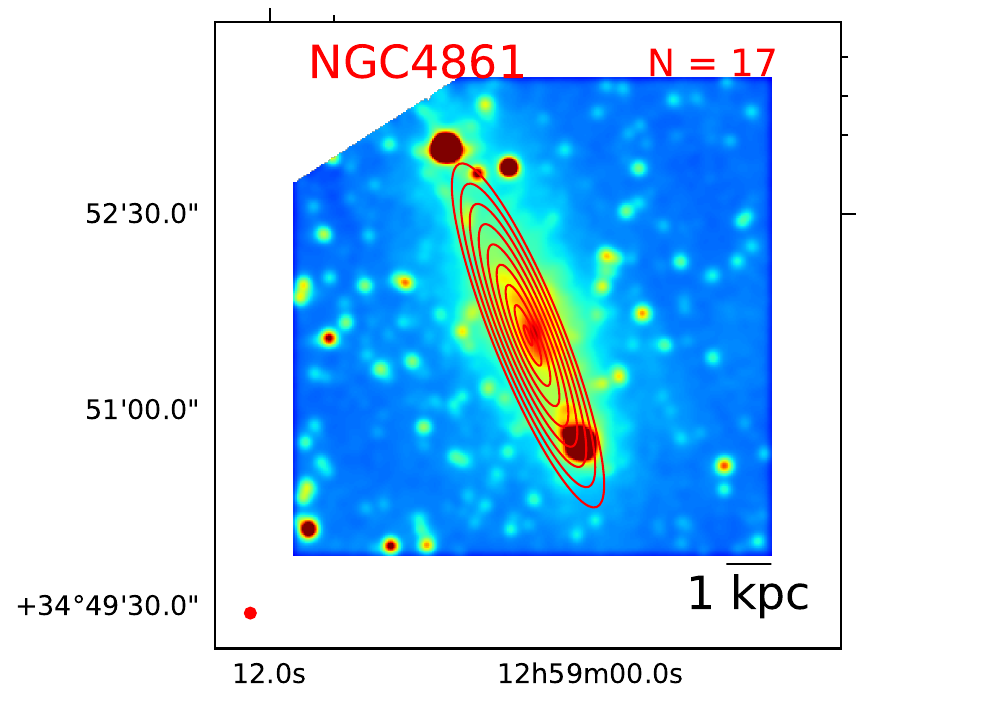} \\

    \end{tabular}

    \caption{The radial distribution of the ratio of amplitudes of different Fourier modes to the $m=0$ mode found in the Fourier analysis of the stellar disk. The images of the respective galaxies for which the Fourier analysis is done are shown just below each plot of the radial distribution of different modes. Optical images for most galaxies, excluding NGC 7292 and NGC 7610, were taken from SPITZER IRAC-3.6$\mu$m data, while SDSS r-band data was used for NGC 7292 and NGC 7610. The concentric red ellipses over-plotted in the images of the stellar disks, are centred at the optical centre of the galaxies, separated by an angular distance of 5$^{\prime \prime}$, projected to an angle equal to the inclination angle of the galaxies and rotated according to the position angle of the galaxies. In these plots, every second such ellipse has been shown, starting from the last ellipse where Fourier analysis should be performed. \emph{(cont.)} }
    \label{fig:opt_lop}
\end{figure*}

\begin{figure*}
\ContinuedFloat
    \centering
    \begin{tabular}{ccc}
        \includegraphics[width=0.3\textwidth]{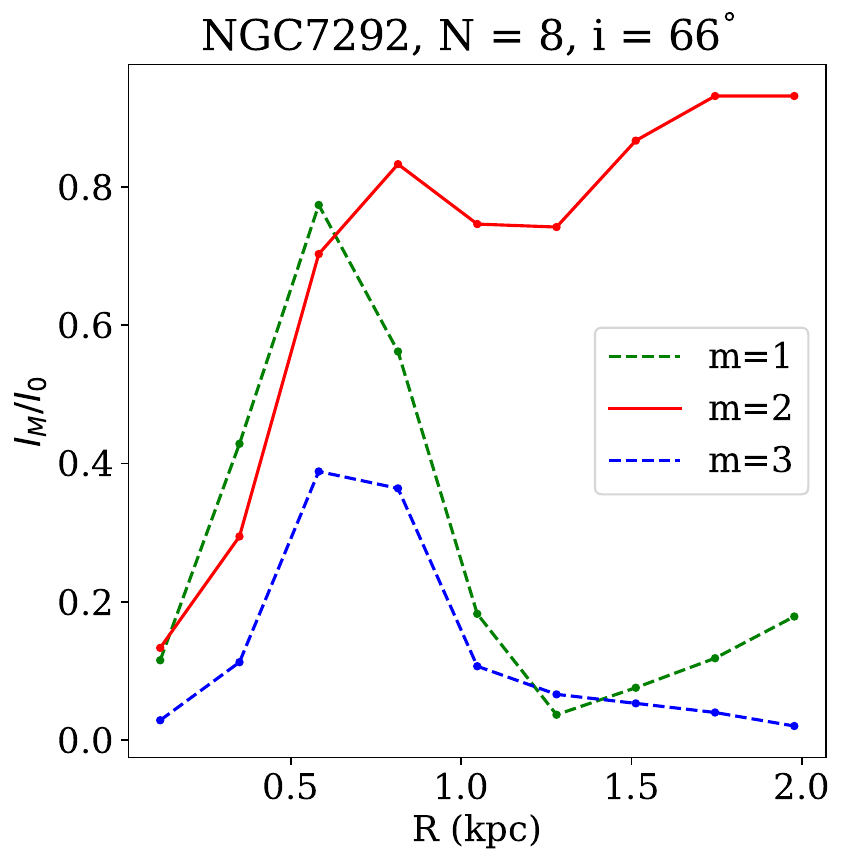} & 
        \includegraphics[width=0.3\textwidth]{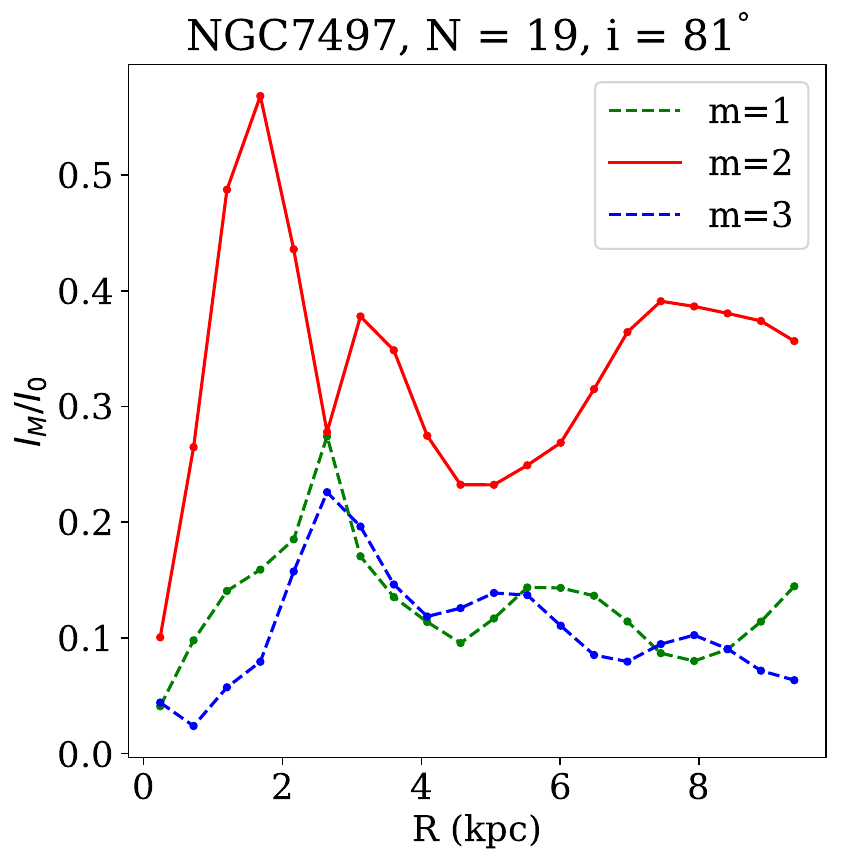} & 
        \includegraphics[width=0.3\textwidth]{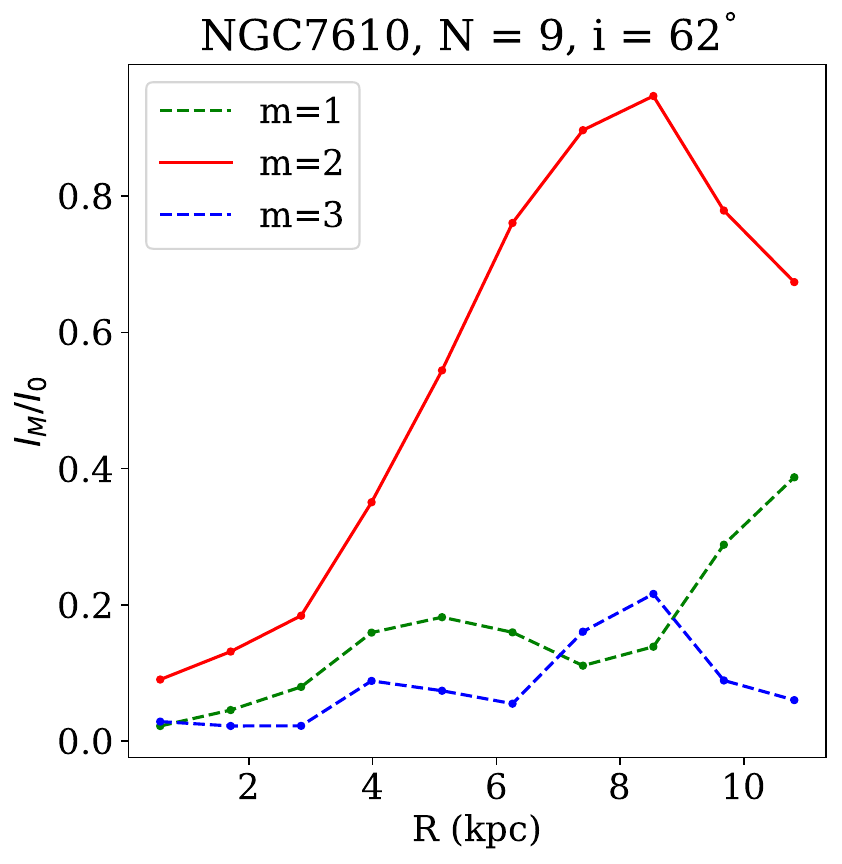} \\
        \includegraphics[width=0.3\textwidth]{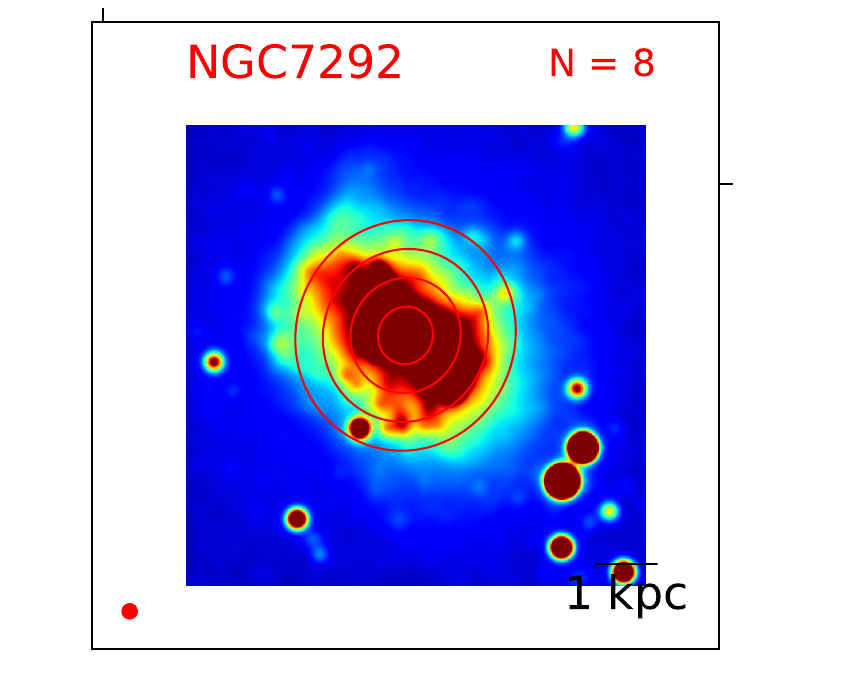} &
        \includegraphics[width=0.3\textwidth]{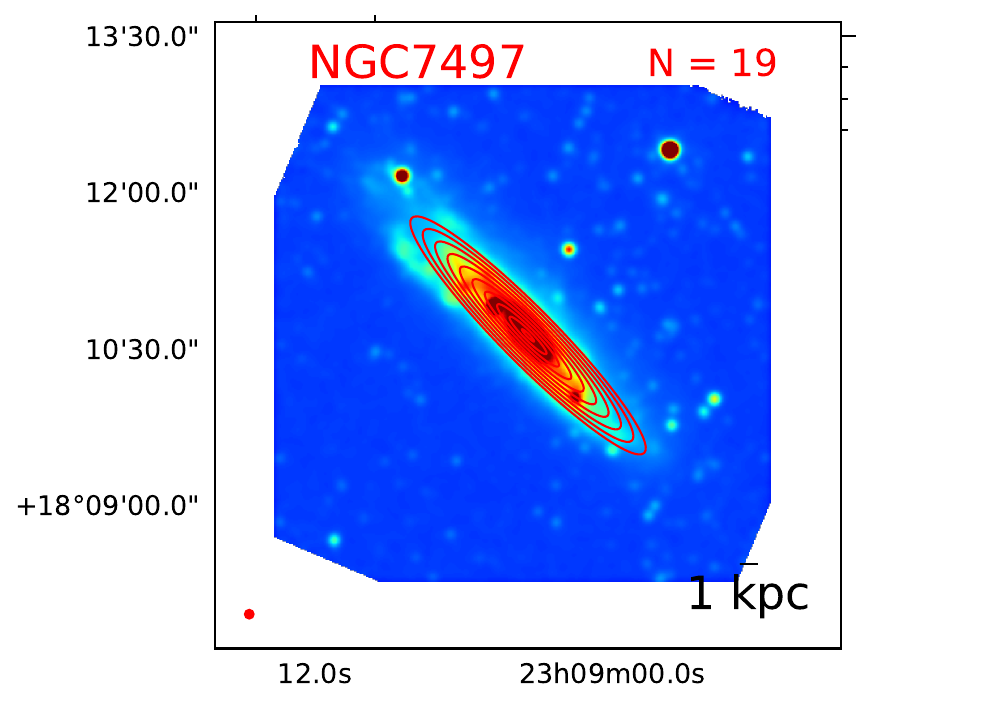} &
        \includegraphics[width=0.3\textwidth]{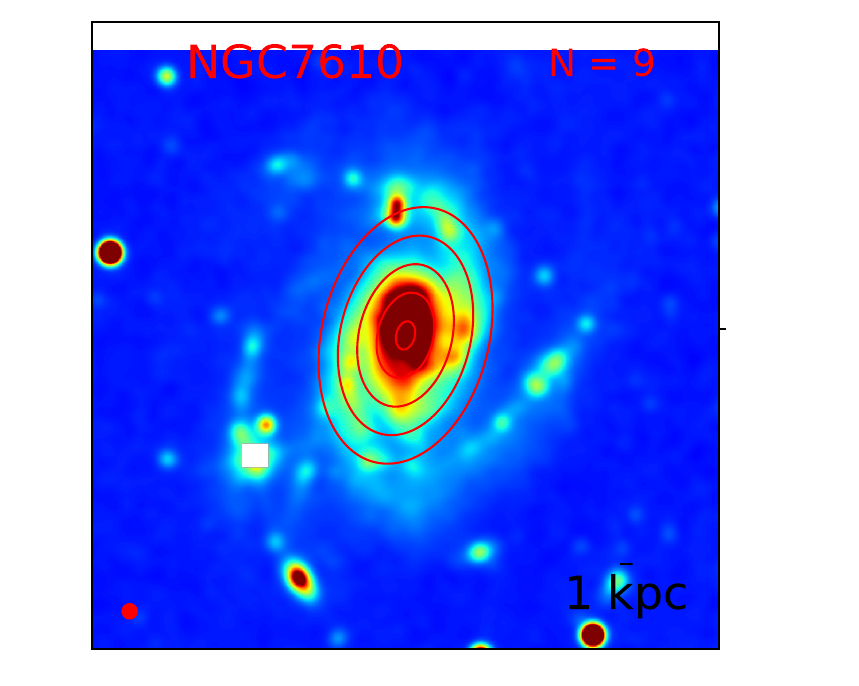} \\
    \end{tabular}
    \begin{tabular}{cc}
        \includegraphics[width=0.3\textwidth]{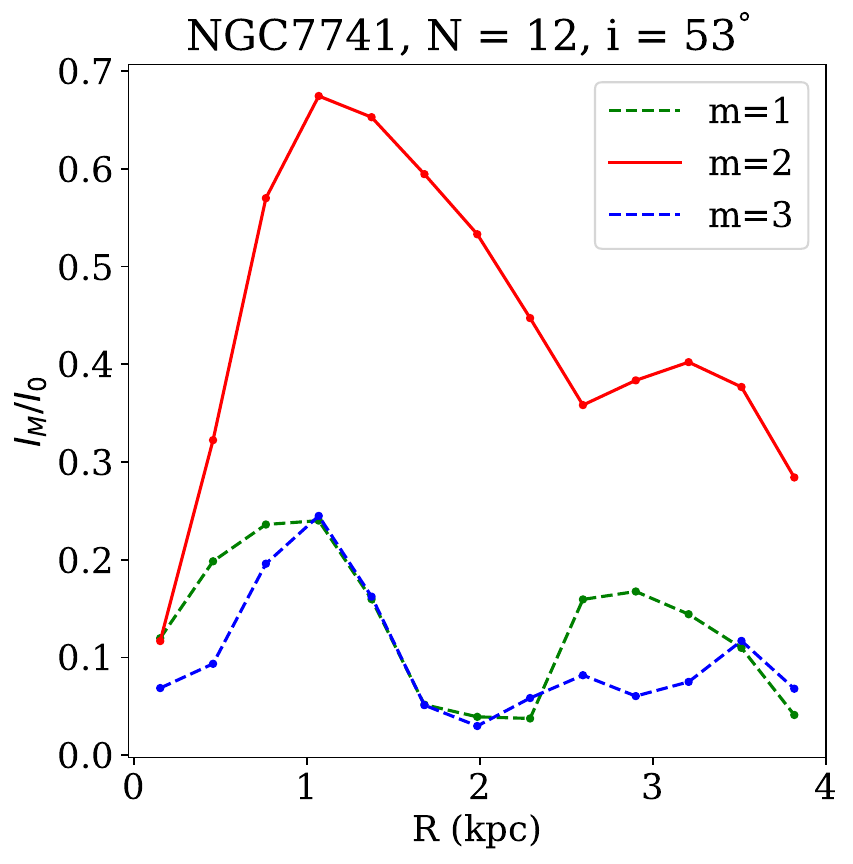} & 
        \includegraphics[width=0.3\textwidth]{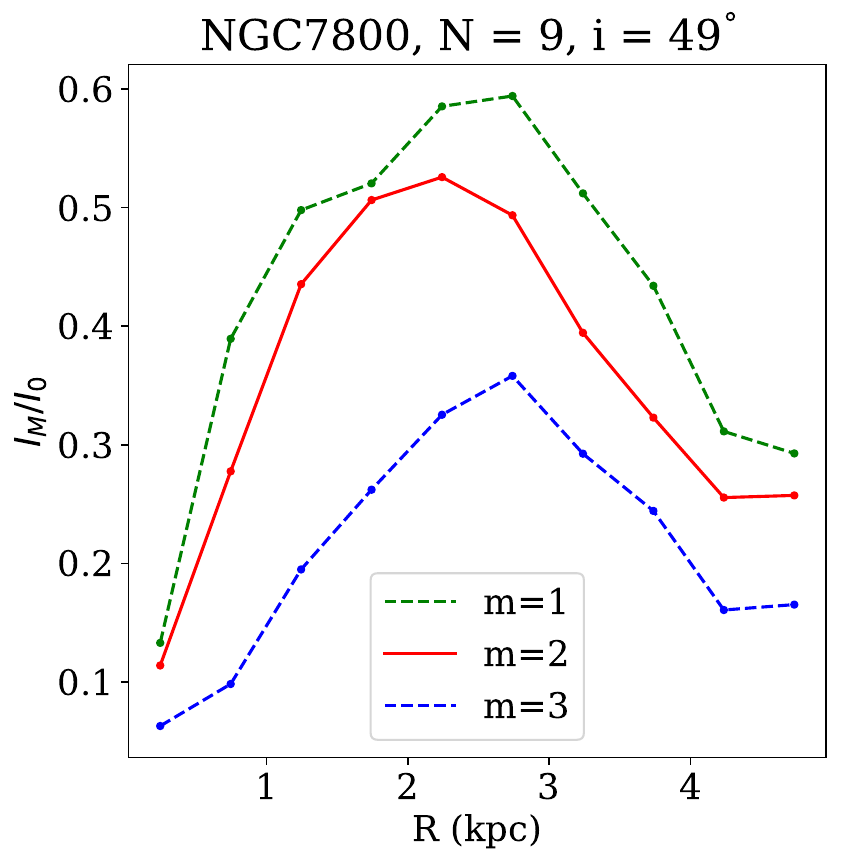}  \\
        \includegraphics[width=0.3\textwidth]{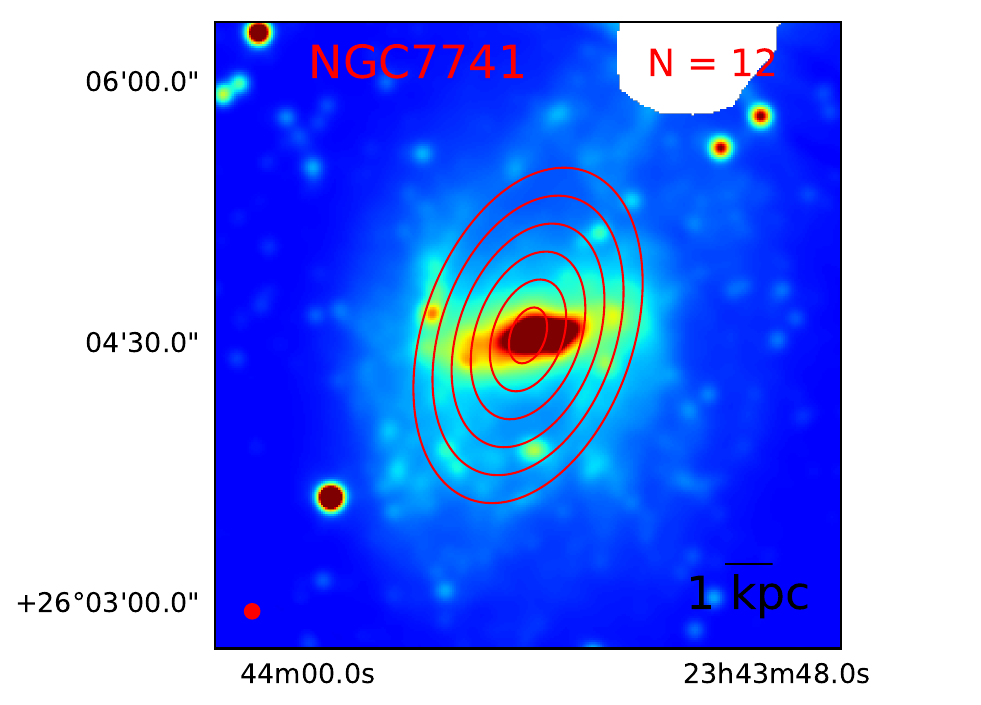} &
        \includegraphics[width=0.3\textwidth]{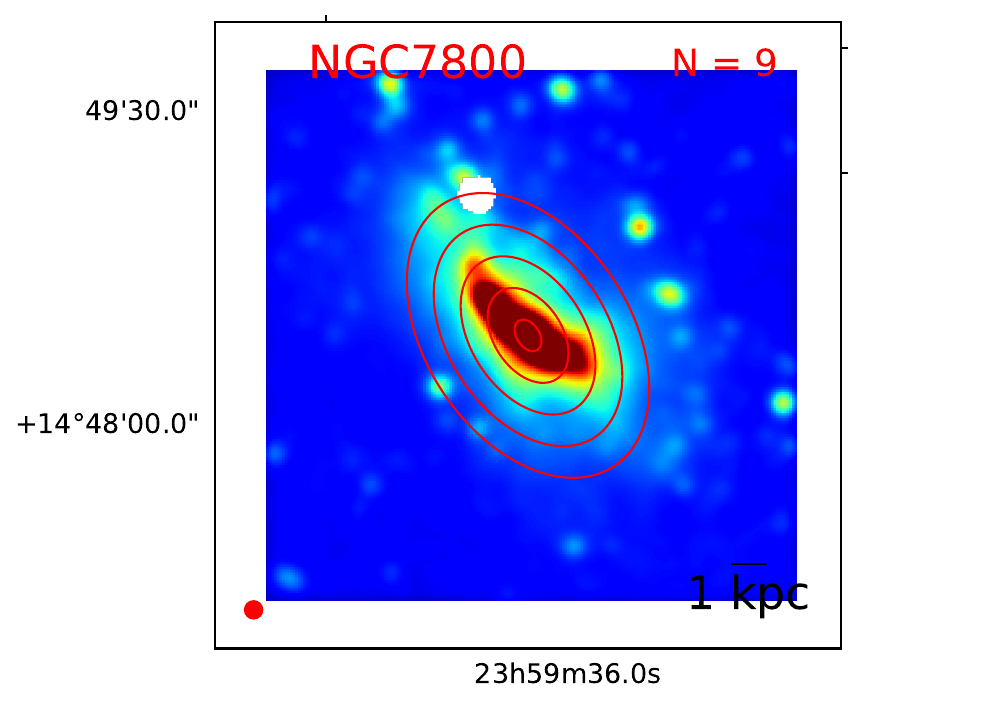}  \\

    \end{tabular}

    \caption{The radial distribution of the ratio of amplitudes of different Fourier modes to the $m=0$ mode found in the Fourier analysis of the stellar disk. The images of the respective galaxies for which the Fourier analysis is done are shown just below each plot of the radial distribution of different modes. Optical images for most galaxies, excluding NGC 7292 and NGC 7610, were taken from SPITZER IRAC-3.6$\mu$m data, while SDSS r-band data was used for NGC 7292 and NGC 7610. }
\end{figure*}

\subsection{Lopsidedness of the gas disk}
As mentioned earlier, to investigate the lopsidedness in the gas disk, we use the H~{\sc i} moment zero maps from the data products of the GARCIA pilot survey \citep{biswas2022}. As lopsidedness is a global phenomenon and should not be confused with the local fluctuations of the galaxies, we use the lowest resolved maps from our analysis, i.e., the images prepared from a cube made with a 5K$\lambda$ UV cutoff and thus have a typical beam size of $\sim$ 35$^{\prime \prime}$. However, the Fourier analysis is performed at radii separated by half of the beam size of respective images. Further, unlike the optical data, the H~{\sc i} moment maps do not include any other sources from the background or in the foreground as they are prepared from the continuum subtracted data cubes (see section 3 of \citet{biswas2022}). As the H~{\sc i} disk extends far beyond the stellar disk, the Fourier analysis was successfully performed up to a much larger radius compared to the stellar disk. For tracing the maximum traceable radius, up to which the Fourier analysis should be performed in the H~{\sc i} disk, we fit an ellipse to the $1\times10^{19}$ cm$^{-2}$ column density contour level of the H~{\sc i} disk. We have checked that in our data, $1\times10^{19}$ cm$^{-2}$ column density matches closely with the $3 \sigma$ sensitivity level of our moment zero maps.

For the HI disk, we follow the same procedure as described for the stellar disk, except that the initial outer radius is defined by the major axis of the fitted ellipse at the $1\times10^{19}$ cm$^{-2}$ column density contour. Starting from this radius, the Fourier analysis is performed in successive annuli separated by half the beam size. As in the stellar case, we exclude some of the outer annuli where the surface brightness is not detected at all azimuthal angles. The outermost annulus that shows H~{\sc i} emission above the detection limit at every azimuthal angle is adopted as the limiting radius for the Fourier analysis.

The results of the Fourier decomposition of the gas disks are shown in figure \ref{fig:hi_lop}. The associated galaxies' H~{\sc i} disks are shown below the radial variation of the ratio of the different Fourier modes. The white concentric ellipses show every second ring where Fourier analysis has been performed.  It is to be noted that the inclination and position angles that were used in the deprojection of the moment zero maps are the same kinematic inclination and position angle as used for the deprojection of the stellar disk. From figure \ref{fig:hi_lop}, we see that the H~{\sc i} disk appears to be lopsided in all cases except NGC 0784 and NGC 7497, considering the outermost radius, with $I_{m}/I_{0}$ $\geq 0.2$ for $m=1$.  For NGC 0784 and NGC 7497, we see that $m=2$ modes dominate over all other modes almost throughout all the radii, suggesting a strong 2-fold symmetry in the gas disk. However, for galaxies NGC 1156, NGC 3027, NGC 3359,  NGC 4068, NGC 7292, NGC 7610, NGC 7741, and NGC 7800, we find that $m=1$ mode either dominates over other modes or becomes comparable to $m=2$ modes in mostly the outer region of the gas disk. This scenario suggests significant lopsidedness in the gas disk in most of the galaxies in our sample. Similar to the stellar disk, for the gas disk also, we adopted a more robust classification based on the average amplitude ratios of the different Fourier modes in the inner and outer regions of the gas disk, which we will describe in the following section.

\begin{figure*}
    \centering
    \begin{tabular}{ccc}
        \includegraphics[width=0.3\textwidth]{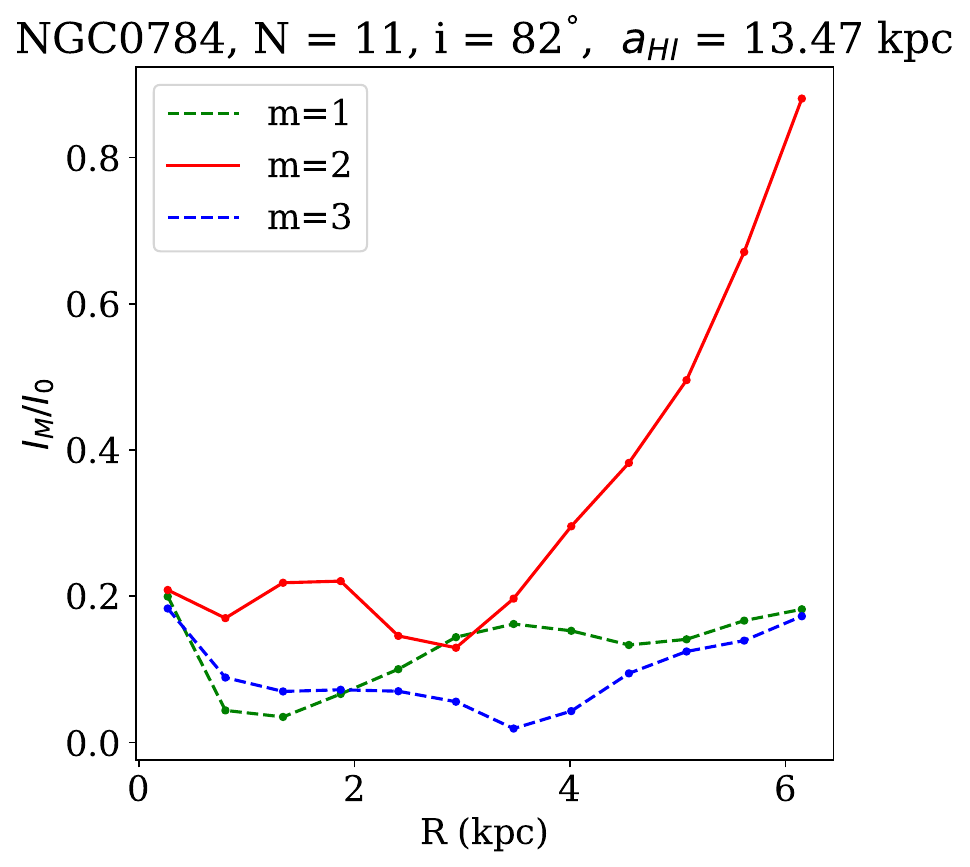} & 
        \includegraphics[width=0.3\textwidth]{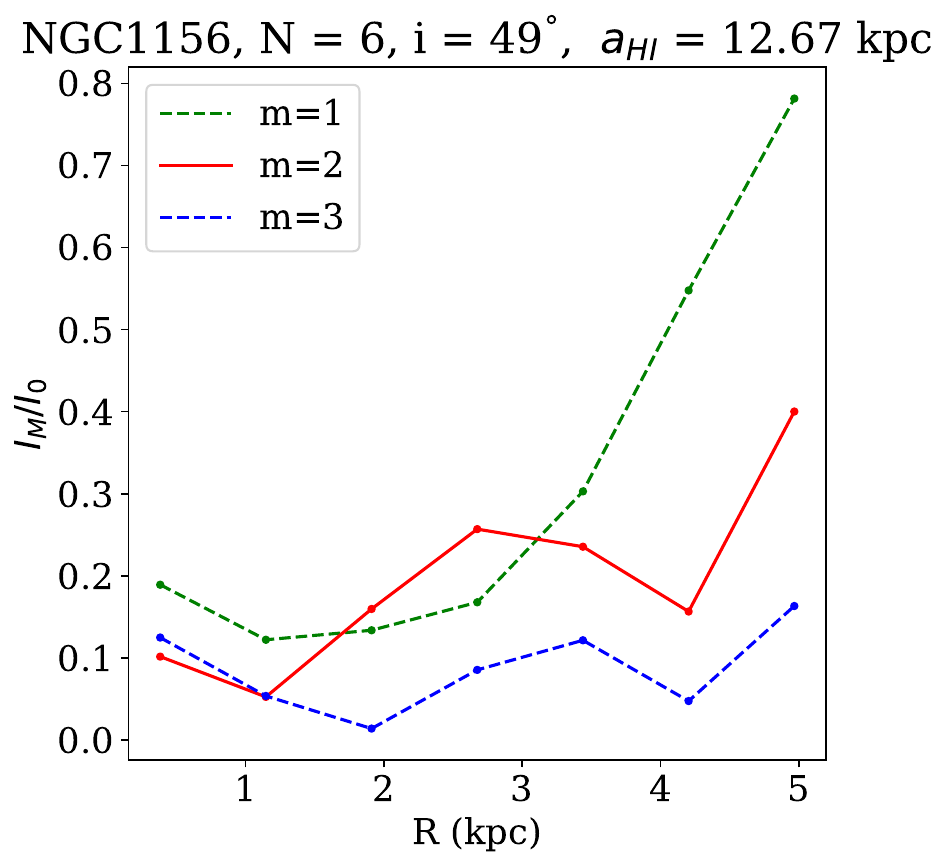} & 
        \includegraphics[width=0.3\textwidth]{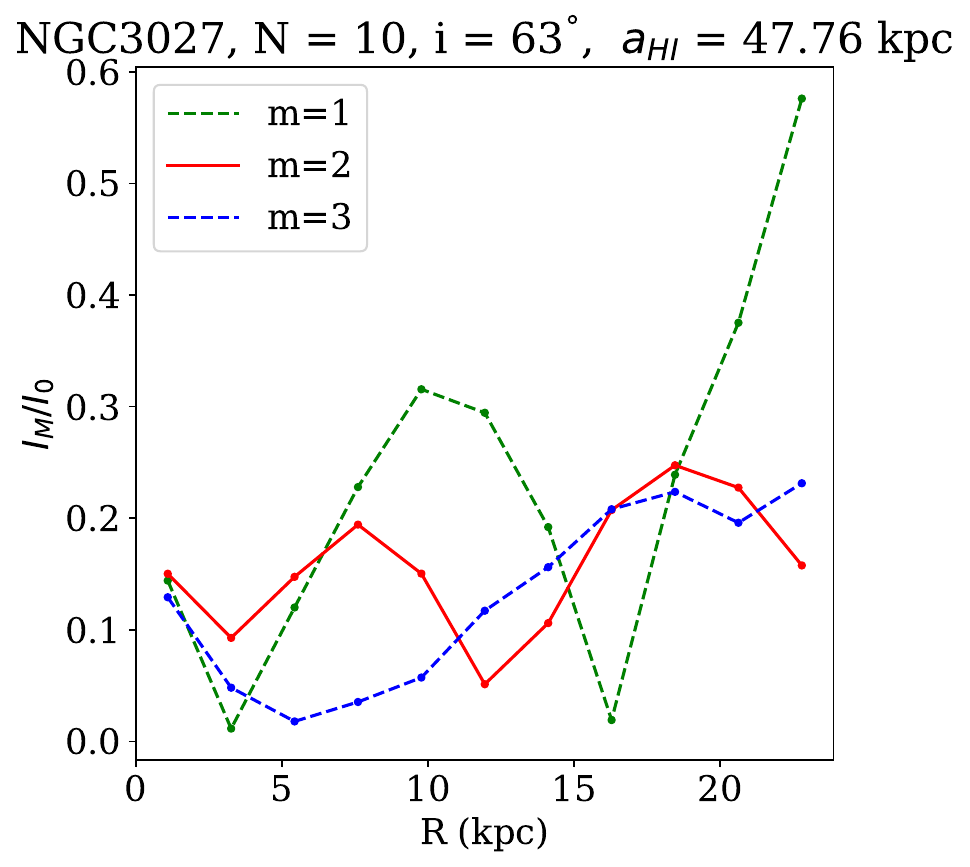} \\
        \includegraphics[width=0.3\textwidth]{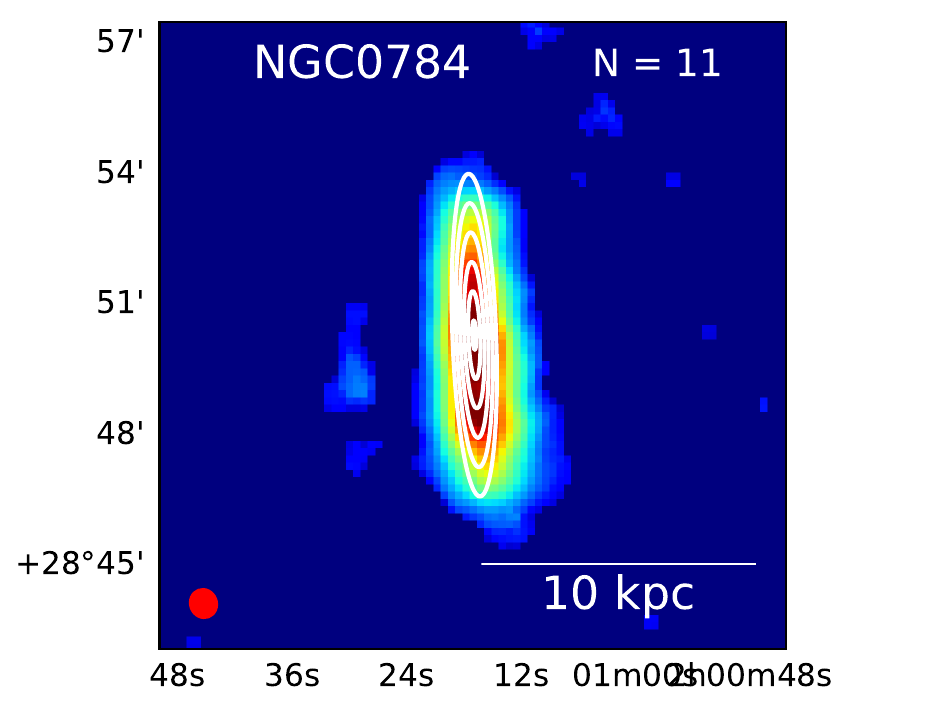} &
        \includegraphics[width=0.3\textwidth]{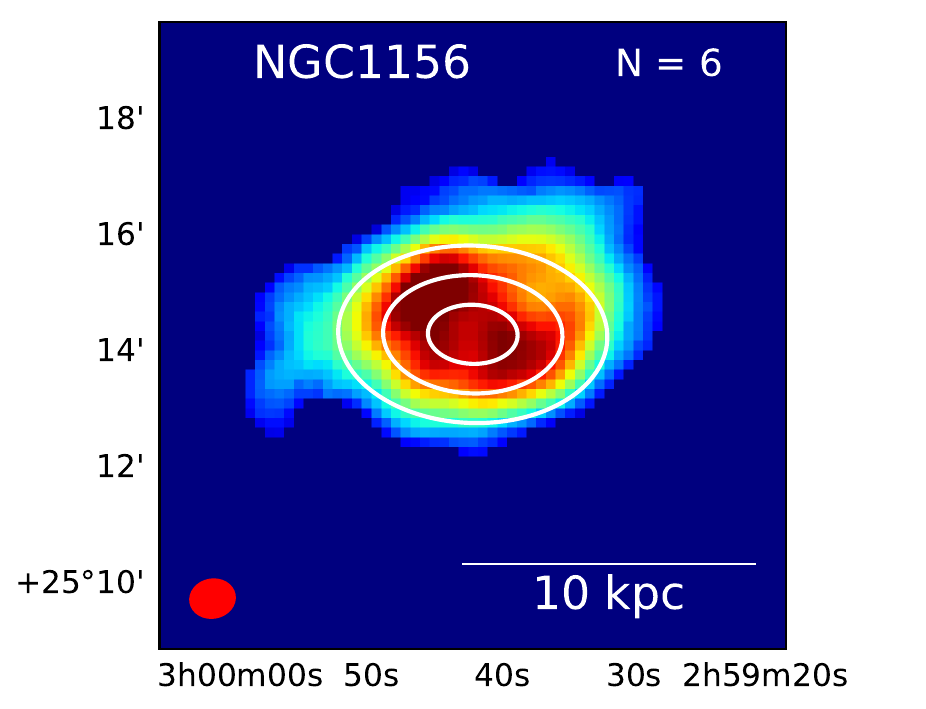} &
        \includegraphics[width=0.3\textwidth]{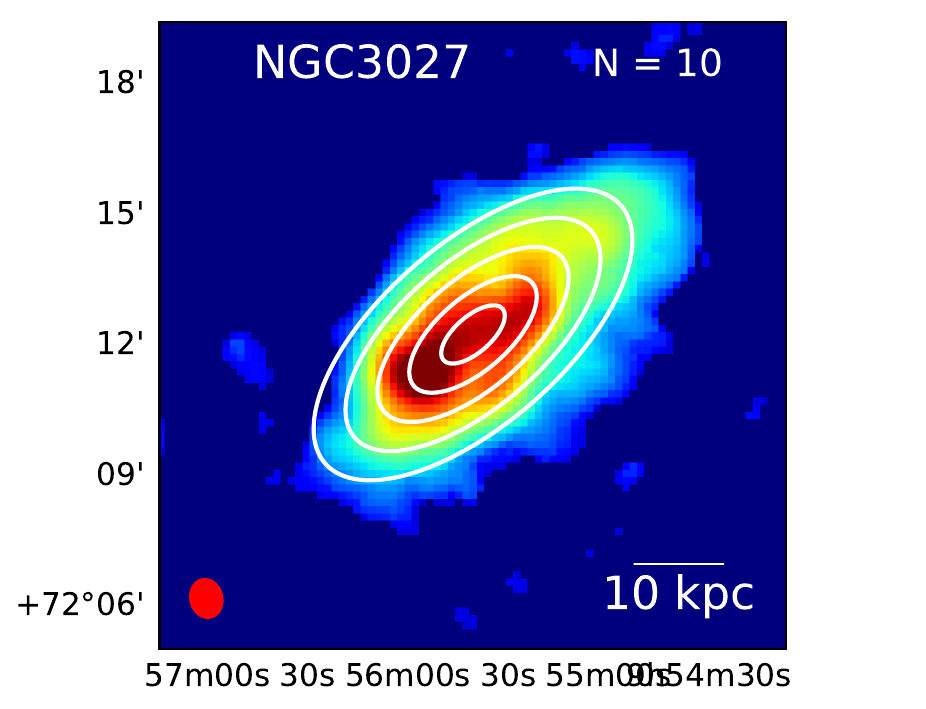} \\

        \includegraphics[width=0.3\textwidth]{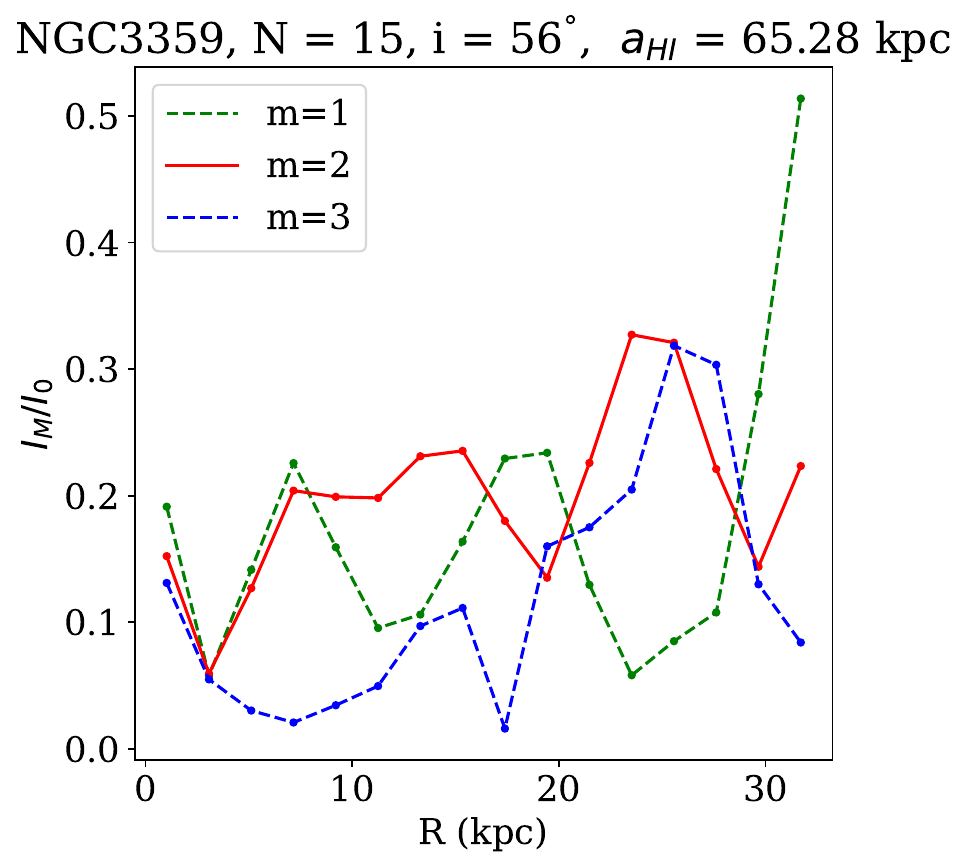} & 
        \includegraphics[width=0.3\textwidth]{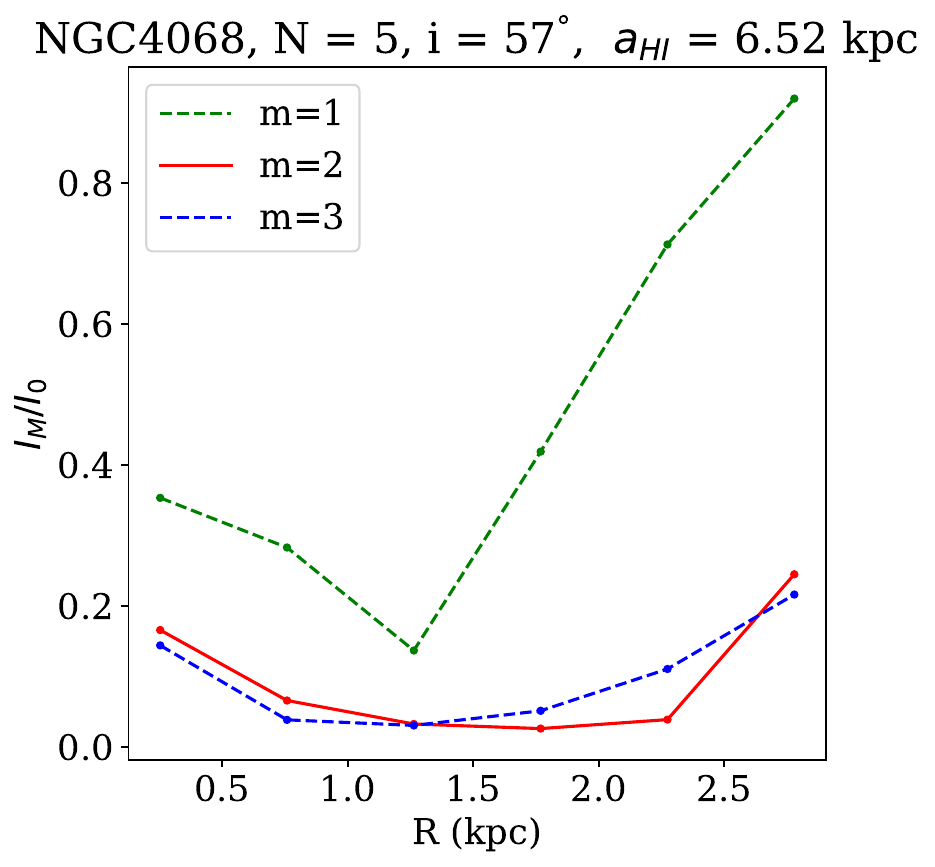} & 
        \includegraphics[width=0.3\textwidth]{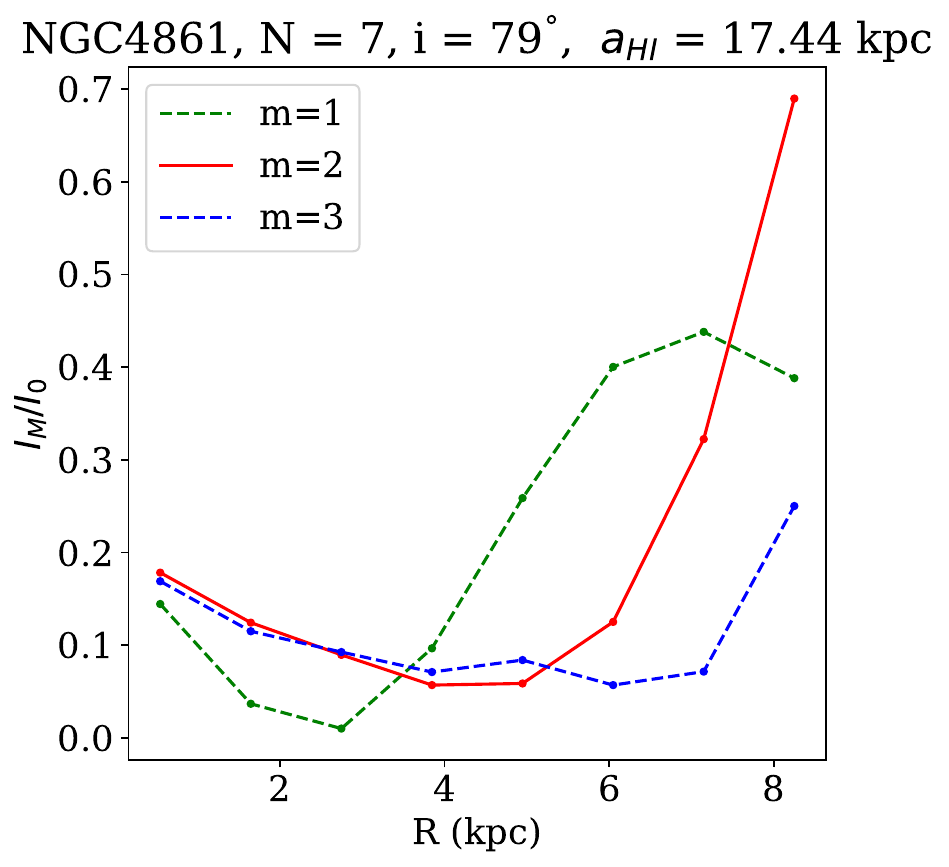} \\
        \includegraphics[width=0.3\textwidth]{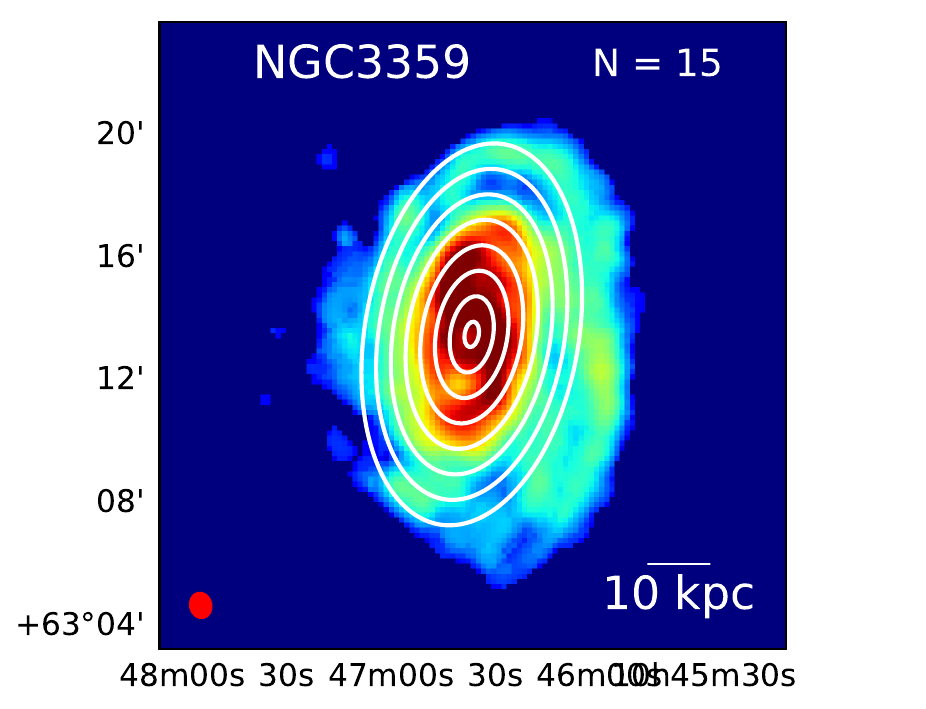} &
        \includegraphics[width=0.3\textwidth]{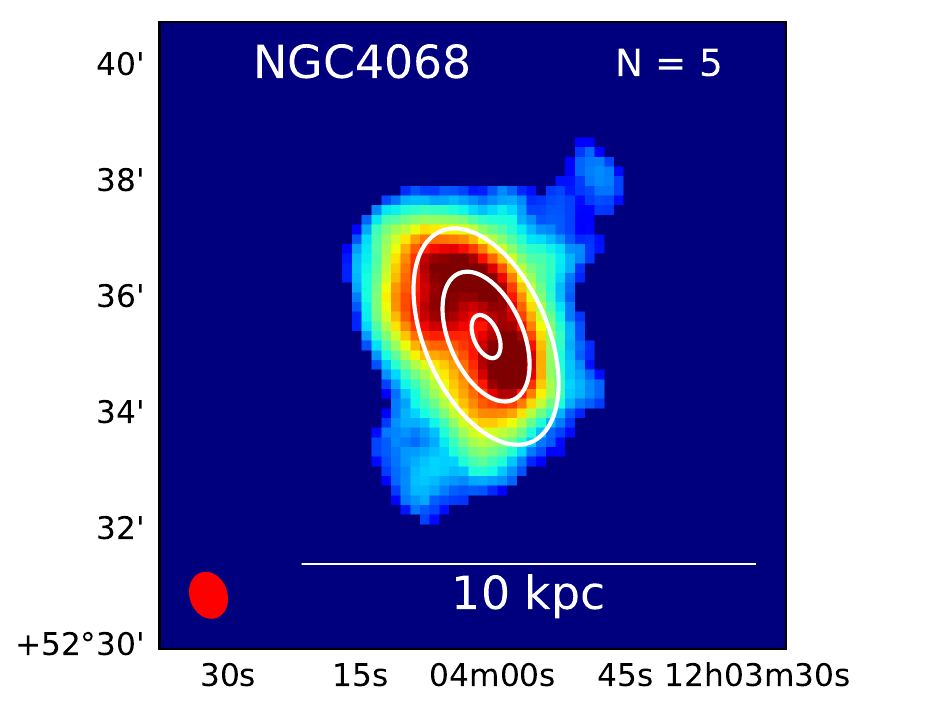} &
        \includegraphics[width=0.3\textwidth]{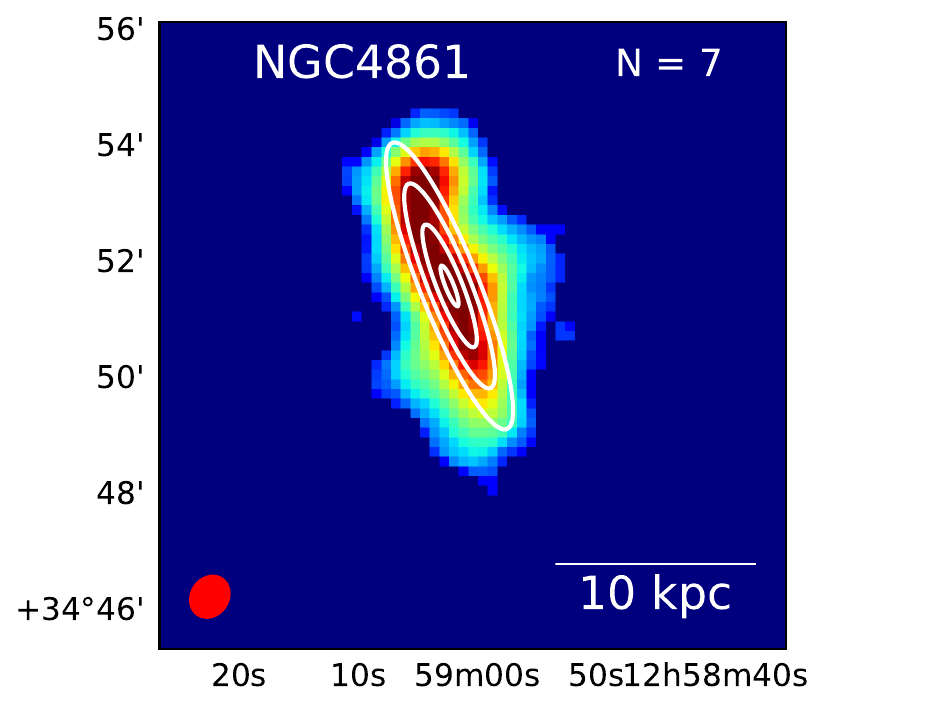} \\

    \end{tabular}

    \caption{The radial distribution of the ratio of amplitudes of different Fourier modes to the $m=0$ mode found in the Fourier analysis of the H~{\sc i} disk. The images of the respective galaxies for which the Fourier analysis is done are shown just below each plot of the radial distribution of different modes. The concentric white ellipses over-plotted in the images of the H~{\sc i} disks, are centred at the optical centre of the galaxies, separated by an angular distance of half of the beam size, projected to an angle equal to the inclination angle of the galaxies and rotated according to the position angle of the galaxies. In these plots, every second such ellipse has been shown, starting from the last ellipse up to which the Fourier analysis should be performed. \emph{(cont.)}}
    \label{fig:hi_lop}
\end{figure*}

\begin{figure*}
\ContinuedFloat
    \centering
    \begin{tabular}{ccc}
        \includegraphics[width=0.3\textwidth]{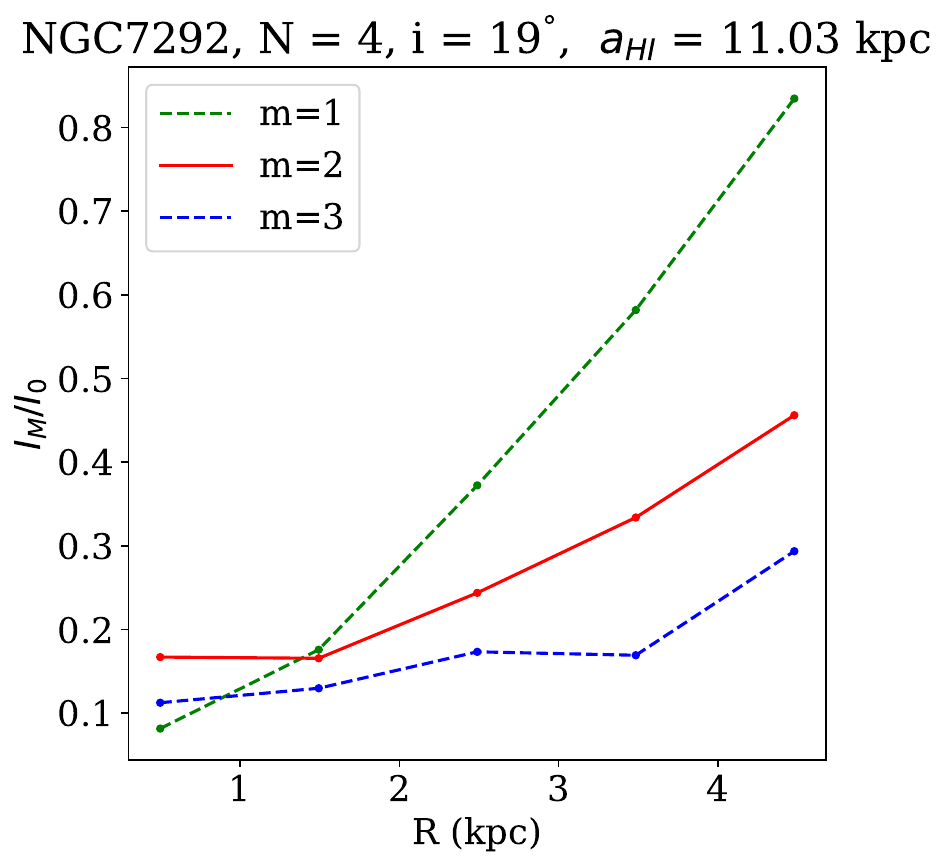} & 
        \includegraphics[width=0.3\textwidth]{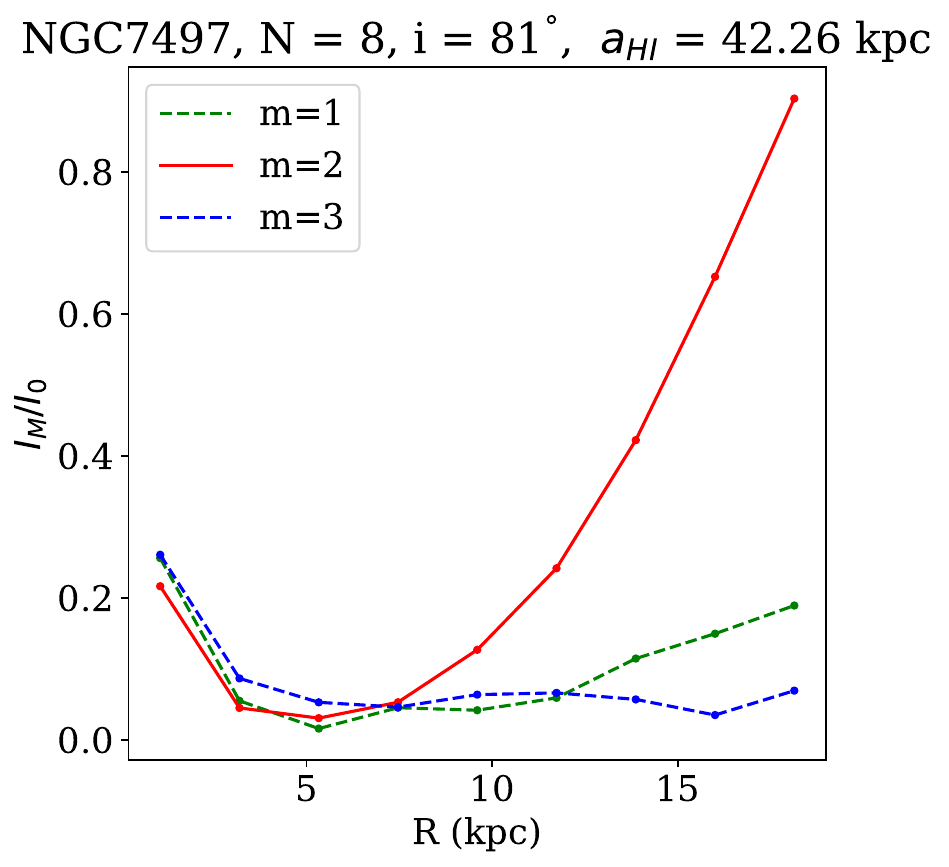} & 
        \includegraphics[width=0.3\textwidth]{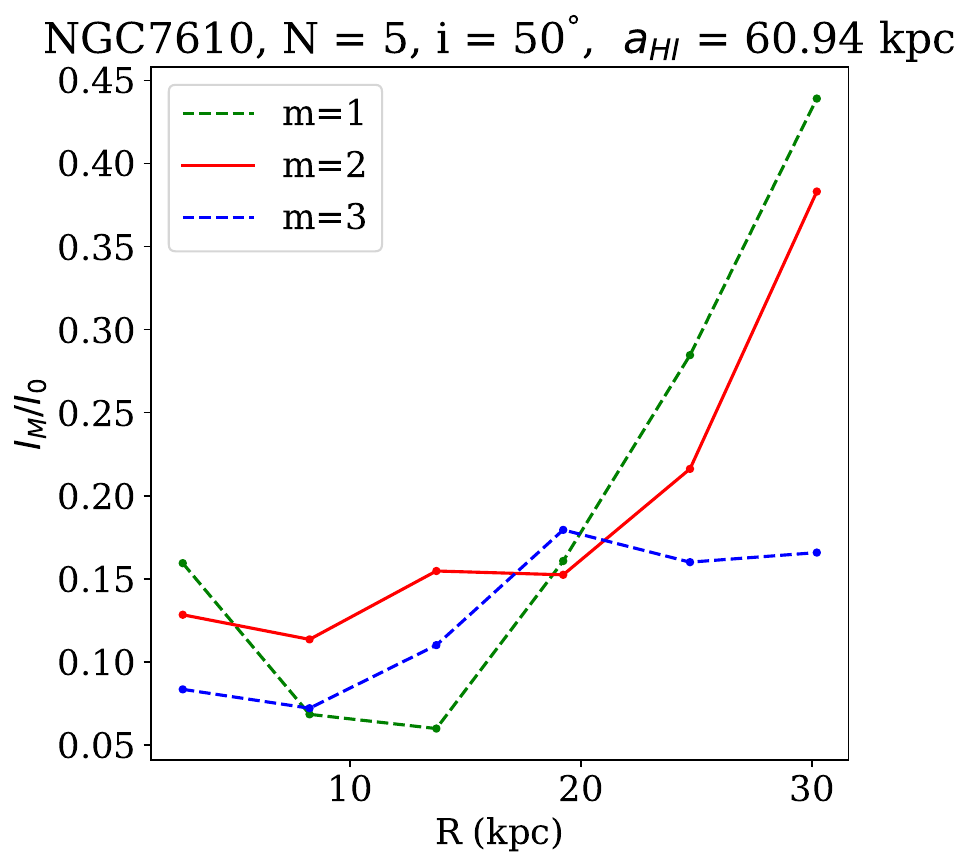} \\
        \includegraphics[width=0.3\textwidth]{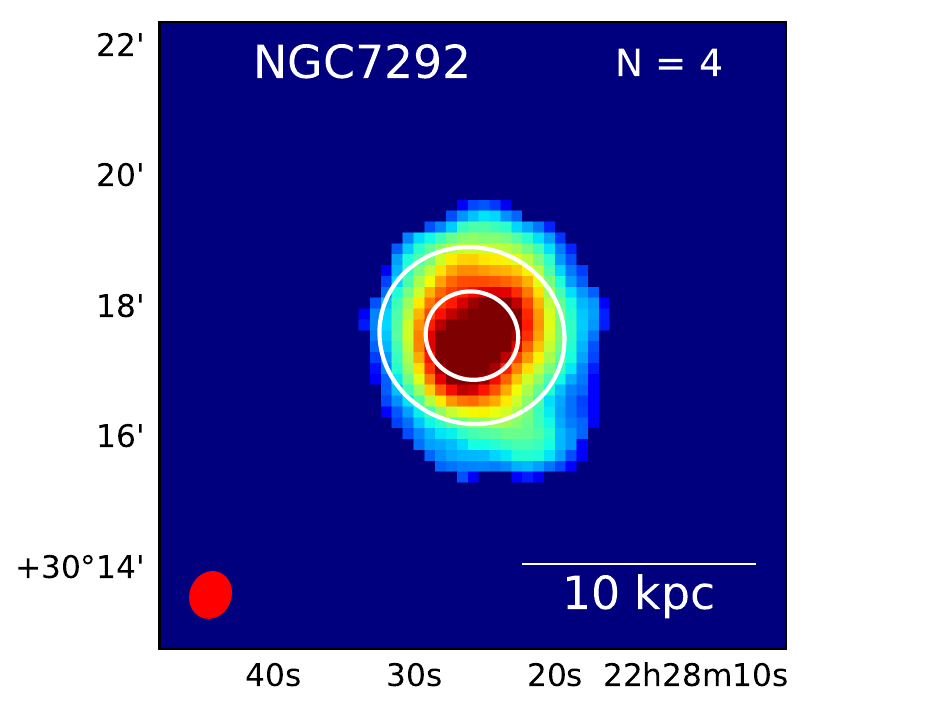} &
        \includegraphics[width=0.3\textwidth]{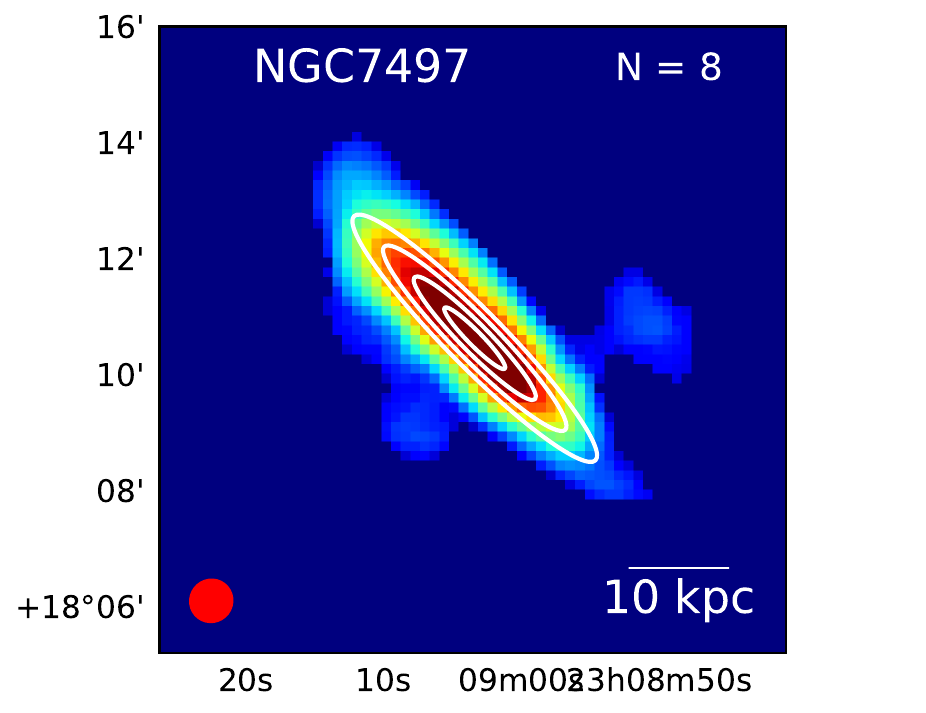} &
        \includegraphics[width=0.3\textwidth]{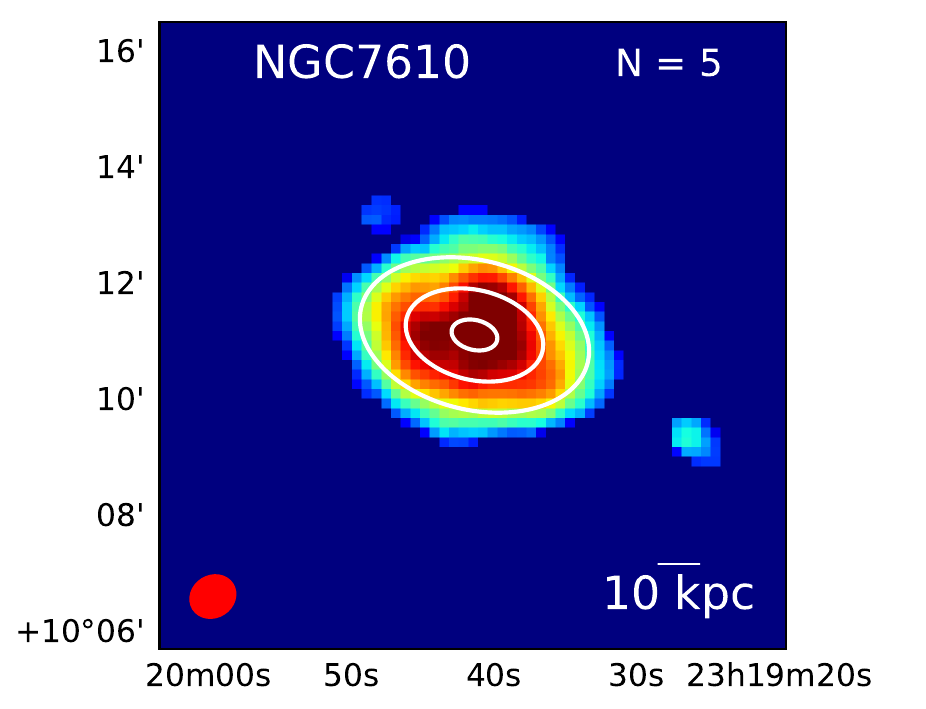} \\
    \end{tabular}
    \begin{tabular}{cc}
        \includegraphics[width=0.3\textwidth]{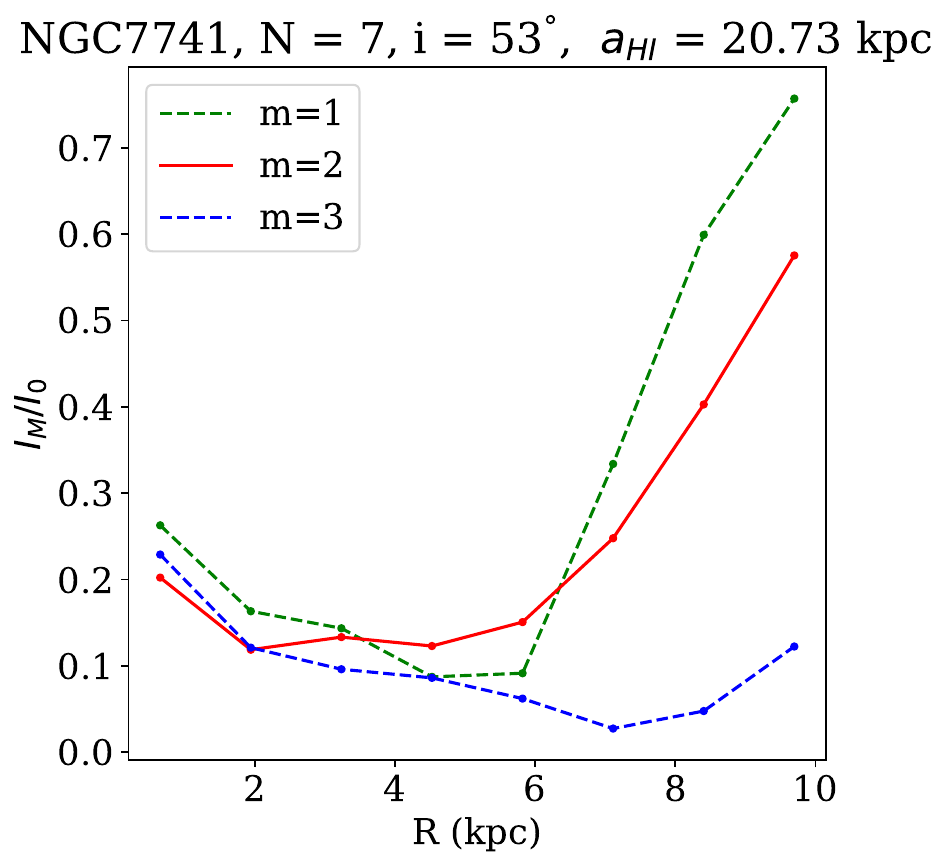} & 
        \includegraphics[width=0.3\textwidth]{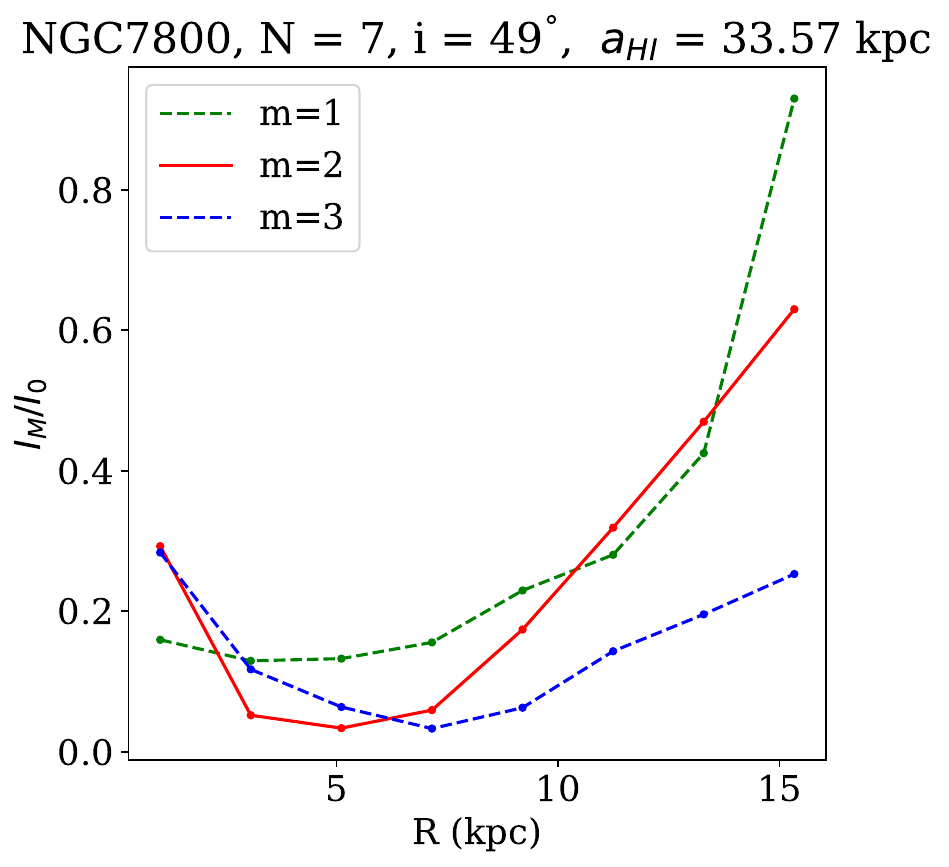}  \\
        \includegraphics[width=0.3\textwidth]{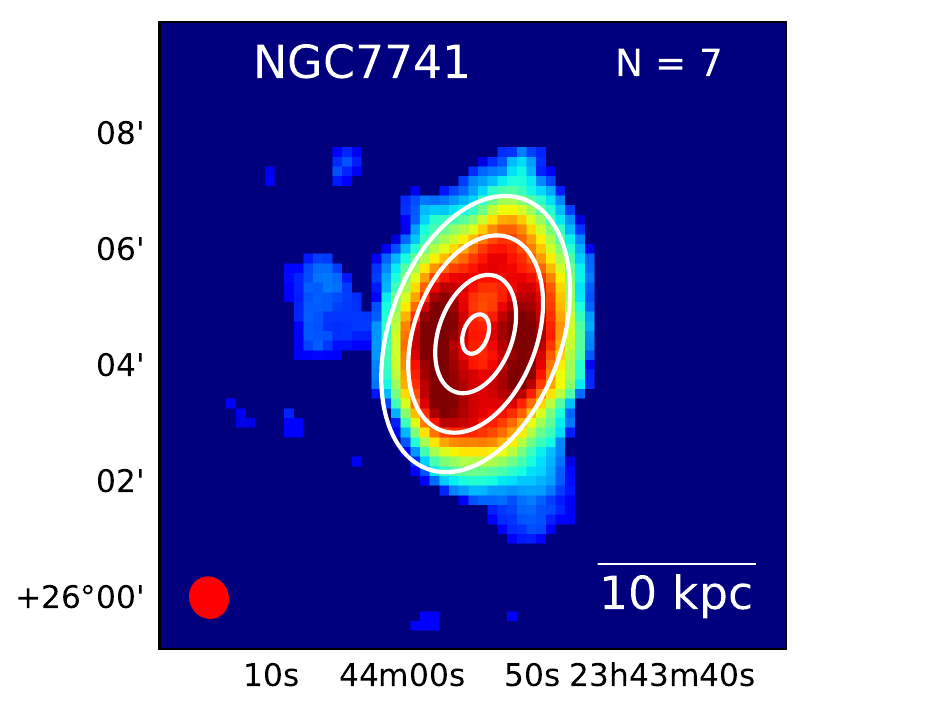} &
        \includegraphics[width=0.3\textwidth]{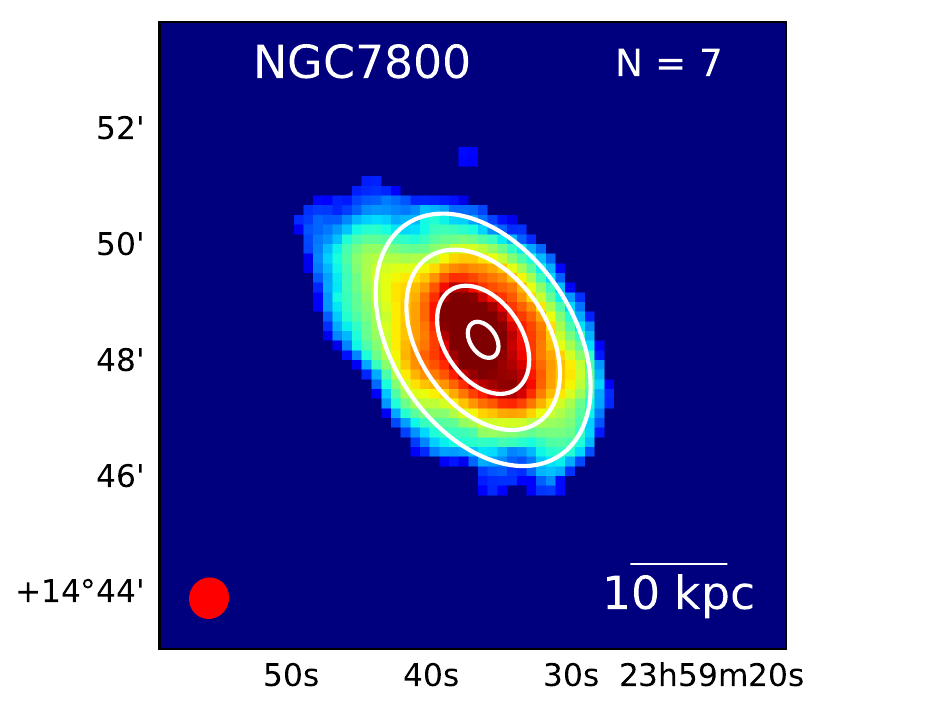}  \\

    \end{tabular}

    \caption{The radial distribution of the ratio of amplitudes of different Fourier modes to the $m=0$ mode found in the Fourier analysis of the H~{\sc i} disk. The images of the respective galaxies for which the Fourier analysis is done are shown just below each plot of the radial distribution of different modes.}
\end{figure*}

\subsection{Comparison of lopsidedness in the stellar disk and the gas disk}
\label{subsec:comp_stellar_gas}

To compare the lopsidedness of the stellar and the gas disk quantitatively, we further found the average of the ratio of the amplitudes of different modes ($m=1$ to $3$) to the amplitude of $m=0$ mode in the inner and outer region of the stellar and the gas disk. 

To define the inner and outer regions of the stellar disk, we first found the disk scale-length of the stellar disk by fitting an exponential profile of the following form to the outer region of the stellar surface brightness profiles:

\begin{equation}
    \Sigma_{star}(R) = \Sigma_{0,star} \exp{ \left(- \frac{R}{R_{h, star}} \right)} , 
\end{equation}
where $ \Sigma_{0,star}$ is the surface brightness at the centre of the galaxy and $R_{h, star}$ is the disk scale-length of the stellar disk.  To find the surface brightness of the stellar disk, we used the MGE method \citep{mge_cappellari2002} as described in sub-section 5.1 of \citet{biswas2023}. These surface brightness profiles can be found on the  {\href{http://www.physics.iisc.ac.in/~nroy/garcia_web/surfbright.html}{ GARCIA website\footnote{\label{garcia-surf2}{\url{http://www.physics.iisc.ac.in/~nroy/garcia_web/surfbright.html}}} }}.  We follow the non-linear least square optimization procedure for fitting the corresponding surface brightness to the exponential profiles and estimate the parameter.  We consider the region inside $R_{h, star}$ to be the inner region of the stellar disk.  We checked that the maximum traceable stellar radius for which the Fourier analysis could be done successfully is $\geq$ $3R_{h, star}$ for all the galaxies. Thus, the region between the radius  $R_{h, star}$  and the maximum radius for which Fourier analysis should be conducted for the stellar disk is defined to be the outer region of the stellar disk.

For the gas disk, one singular exponential function can not fit the H~{\sc i} surface density profiles well. Because of the scarcity of the atomic hydrogen or the presence of H~{\sc i} hole in the central regions of the galaxies, the  H~{\sc i} surface brightness decreases towards the central region of the galactic disk. Considering this, the H~{\sc i} density can be modelled through the following functional form \citep{hiscale-lenght1, hi_scalelenght2} :
\begin{align}
  \begin{split}
    \rho_{HI}(R,z) = \frac{\Sigma_{max, HI}}{4z_{h, HI}} \exp \left( -\frac{R_{m, HI}}{R} - \frac{R}{R_{h, HI}} \right) \\ sech^2(z/2z_{h, HI}) , 
  \end{split}
\end{align}

where $\Sigma_{max, HI}$ is the maximum surface brightness, $R_{h, HI}$ and $z_{h, HI}$ are respectively disk scale-length and vertical scale-height and $R_m$ is such that the maximum surface density is found at a radius equal to $\sqrt{R_{h, HI}R_{m, HI}}$. The profile for the H~{\sc i} surface brightness can be found by integrating this profile in the $z$ direction, as mentioned below:

\begin{equation}
    \Sigma_{HI}(R) = \Sigma_{0, HI} \exp \left( -\frac{R_{m, HI}}{R} - \frac{R}{R_{h, HI}} \right), 
\end{equation}
where $\Sigma_{0, HI}$ is the surface brightness found at the radius $R$ = $\sqrt{R_{h, HI}R_{m, HI}}$. To derive the H~{\sc i} scale length, we take the surface brightness profiles of the gas as derived from the 3D kinematic modelling (see section 3 of \citet{biswas2023}). These, too, can be found in the \href{http://www.physics.iisc.ac.in/~nroy/garcia_web/surfbright.html}{ GARCIA website}.  Looking at the H~{\sc i} surface brightness profiles we can observe that the brightness of the H~{\sc i} disks follows an almost exponential profile in the outer region. Based on this observation, we can enhance the accuracy of our fitting process by focusing on the surface brightness in the outer region only. Since our main goal is to determine $R_{h, HI}$, we use this region only to fit the above-mentioned function. In this case also, we found the best-fitting parameters through a similar way, i.e., by non-linear least square optimization process. Further, for the gas disk, we define the region inside the radius of $R_{h, HI}$ of the gas disk to be the inner region and region in between the radius $R_{h, HI}$ to the maximum radius for which Fourier analysis should be conducted for the gas disk in the Fourier analysis, to be the outer region.

\begin{figure*}
    \centering
    \begin{subfigure}[b]{0.45\textwidth}
         \centering
         \includegraphics[width=\textwidth]{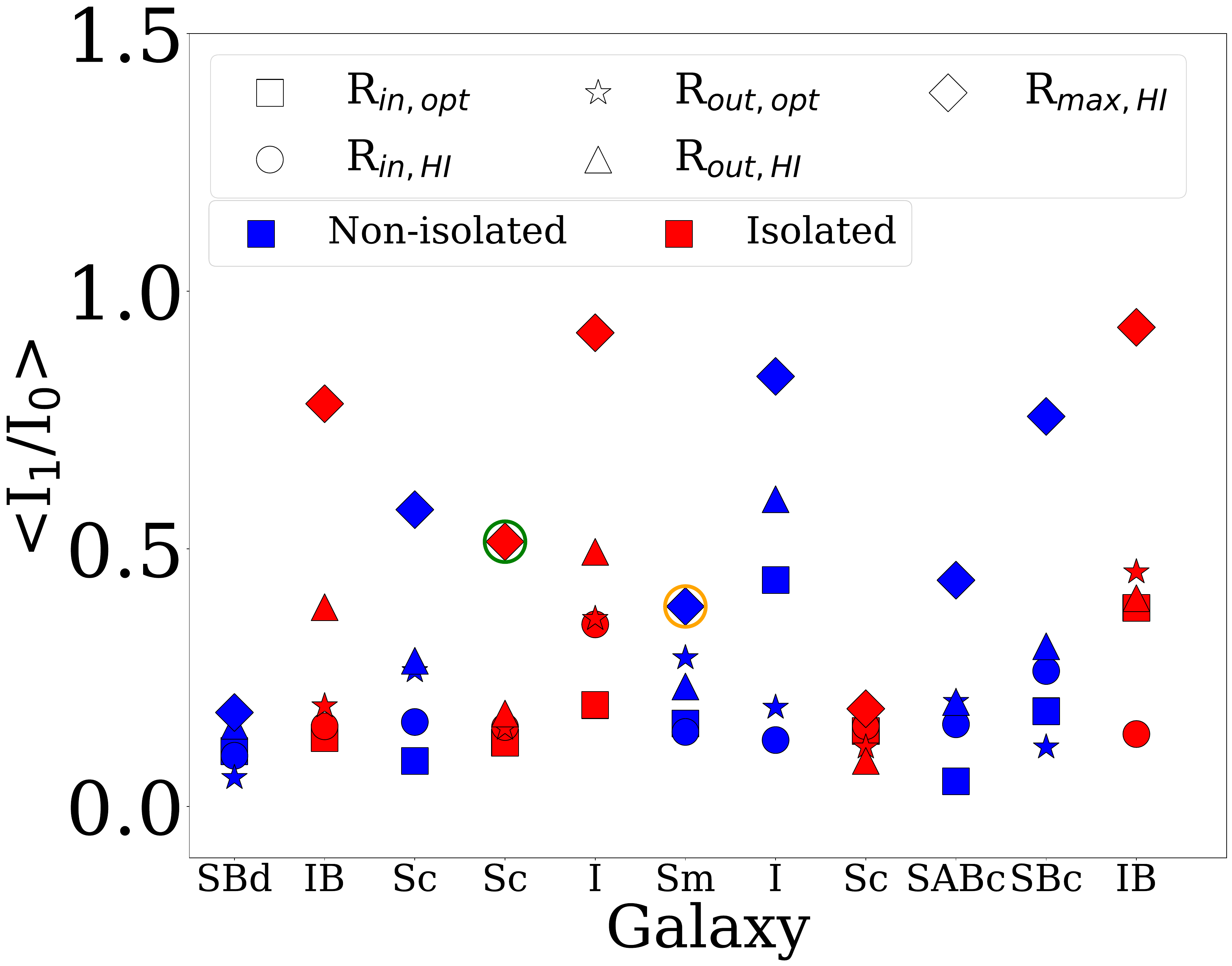}
         %\caption{}
         %\label{}
     \end{subfigure}
     \hfill
      \begin{subfigure}[b]{0.45\textwidth}
         \centering
         \includegraphics[width=\textwidth]{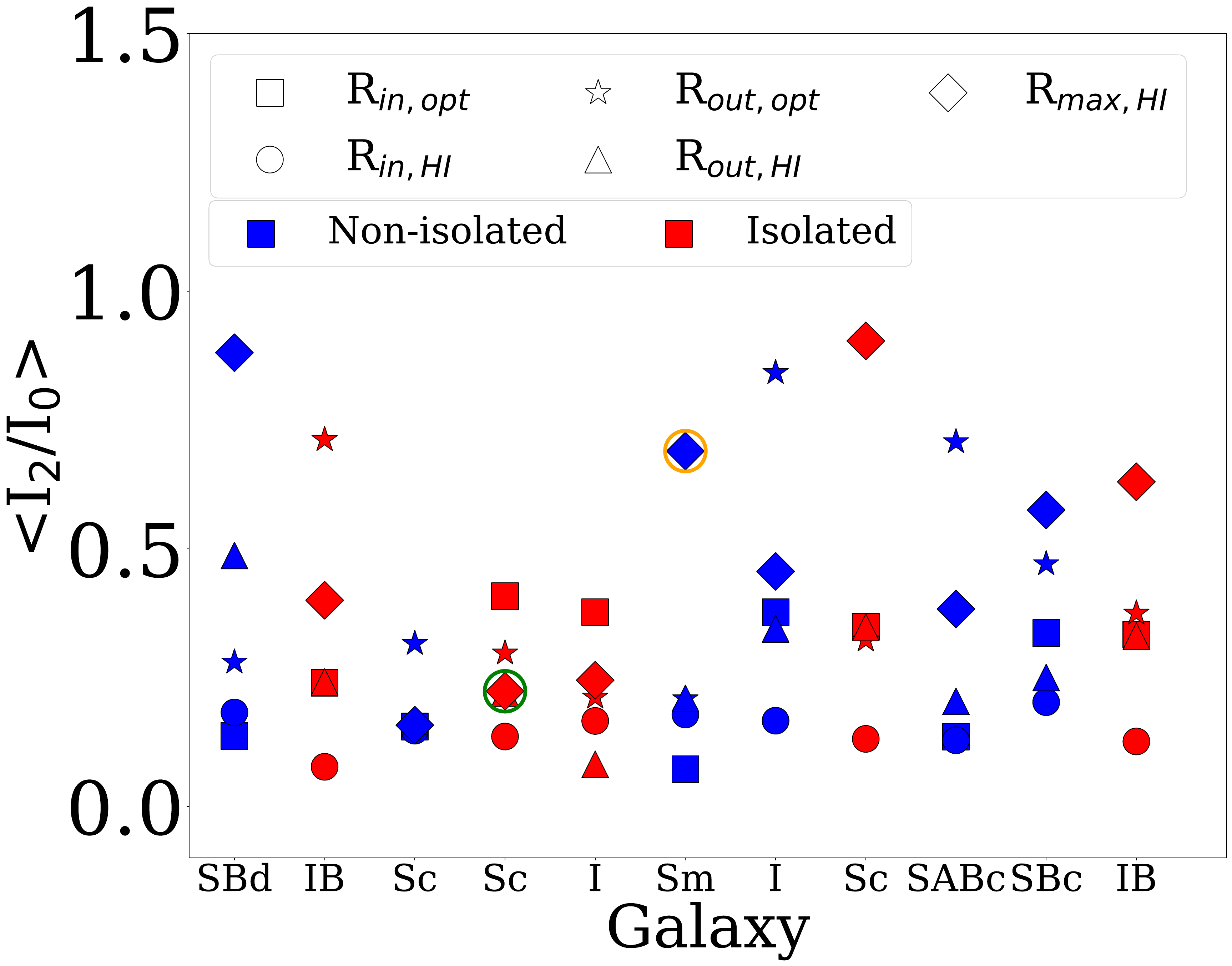}
         %\caption{}
         %\label{}
     \end{subfigure}
     \hfill
      \begin{subfigure}[b]{0.45\textwidth}
         \centering
         \includegraphics[width=\textwidth]{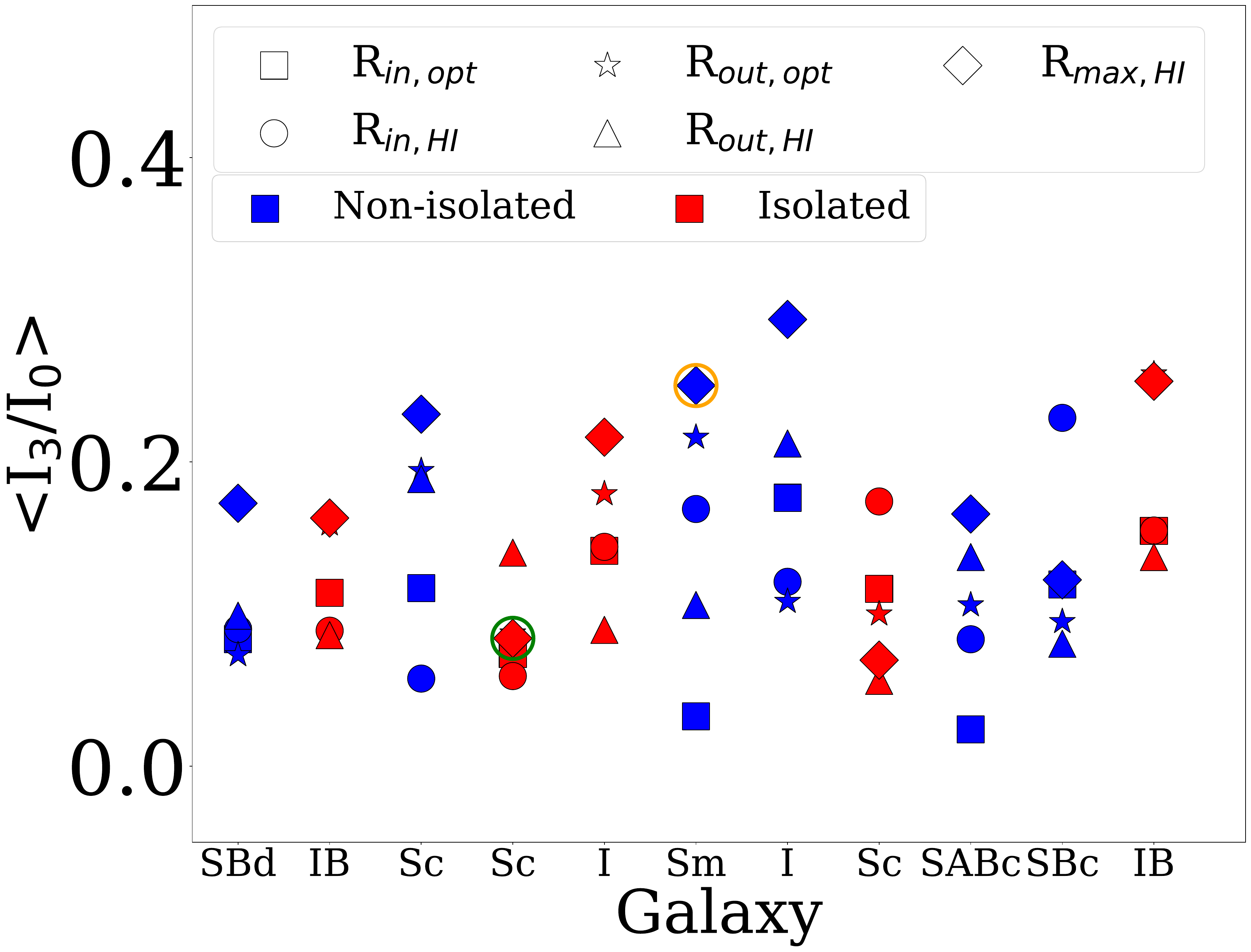}
         %\caption{}
         %\label{}
     \end{subfigure}
     \caption{Distribution of the average of the ratio of the amplitudes  for $m=1,2,3$ modes to $m=0$ mode in the inner and outer regions for both the stellar and gas disk. It is to be noted that the quantities marked as diamond for each galaxy are not an average ratio; these are the values of the ratios at the maximum radius of the H~{\sc i} disk. The morphologies of the galaxies are noted in the x-axis. For each mode, galaxies are colour-coded based on the non-isolated (a part of a group or have nearby companions or interacting) or isolated. The galaxy, marked within the green circle (NGC 3359), is a widely recognized isolated galaxy \citep{n3359_iso}. While the galaxy marked within the orange circle (NGC 4861) has a well-observed nearby companion  (NGC 4861B) \citep{ngc4861_comp}. }
    \label{fig:fmodes}
\end{figure*}

Moreover, we compute the average ratio of the amplitudes for each mode ($m=1$ to $3$) to $m=0$ mode for both the inner and outer regions of both the stellar and gas disks. We also found it in the maximum radius for which Fourier analysis should be conducted for the gas disk. These values are listed in table \ref{tab:lop_inout1}. The distribution of the corresponding quantities  for $m=1,2$ and $3$ modes for all the galaxies is shown in figure \ref{fig:fmodes}. We found that there is no evident correlation between the different modes of the lopsidedness of stellar disk or gas disk defined at different regions and the morphology of the galaxy or presence or absence of the bars or have an effect due to interaction with the companion or presence inside a group for this sample of galaxies.

From table \ref{tab:lop_inout1}, we see that in the inner region of the stellar disk, NGC 4068, NGC 7292, and  NGC 7800 appear to be lopsided, considering the averaged amplitude ratios for $m=1$ mode to be greater than or equal to 0.2 \citep{rix_zuritsky1995}.  However, in the outer part of the stellar disk, some of the galaxies ( NGC 3027, NGC 4068, NGC 4861, NGC 7610, NGC 7800) appear to be lopsided with $m=1$ mode $>$ $0.2$. In the case of the gas disk in the inner region, only NGC 4068 and NGC 7741, show the lopsidedness phenomenon. However, most of the galaxies are lopsided, considering the outer region of the gas disk with $<I_1/I_0>$ $>$ 0.2. For other odd modes, i.e., for $m=3$, we find that stellar lopsidedness in the inner and outer region and the H~{\sc i} lopsidedness in the inner region are comparable for some of the cases, and they, in general, do not depend on the Hubble type or presence of bar or interaction with its surrounding medium. However, H~{\sc i} lopsidedness in the outer region dominates over them in most cases.

In the case of the $m=2$ mode, we find that the average ratio of the amplitudes in the inner region of the stellar disk is higher or comparable with that of the inner region of the H~{\sc i} disk. Similarly, the average ratio of the amplitudes in the outer region of the stellar disk is generally higher or comparable with that of the outer region of the H~{\sc i} disk for most of the cases. However, instead of the average of this ratio, if we consider this ratio at the outermost radius of the H~{\sc i} disk, then we find a significant contribution from $m=2$ mode for the H~{\sc i}  disk for some of the sources. 

\begin{table}
    \centering
    \begin{tabular}{cccccccc}
    \hline \hline
    Galaxy & Region & <$I_{1}/I_{0}$> & <$I_{2}/I_{0}$> & <$I_{3}/I_{0}$> \\
    \hline \hline
NGC 0784 & R$_{opt, in}$ & 0.11 & 0.14 & 0.08 \\
  & R$_{opt, out}$ & 0.06 & 0.28 & 0.07 \\
  & R$_{HI, in}$ & 0.10 & 0.18 & 0.09 \\
  & R$_{HI, out}$ & 0.16 & 0.49 & 0.10 \\
  & R$_{HI, max}$ & 0.18 & 0.88 & 0.17 \\
NGC 1156 & R$_{opt, in}$ & 0.13 & 0.24 & 0.11 \\
  & R$_{opt, out}$ & 0.19 & 0.71 & 0.16 \\
  & R$_{HI, in}$ & 0.16 & 0.08 & 0.09 \\
  & R$_{HI, out}$ & 0.39 & 0.24 & 0.09 \\
  & R$_{HI, max}$ & 0.78 & 0.40 & 0.16 \\
NGC 3027 & R$_{opt, in}$ & 0.09 & 0.15 & 0.12 \\
  & R$_{opt, out}$ & 0.26 & 0.32 & 0.19 \\
  & R$_{HI, in}$ & 0.16 & 0.15 & 0.06 \\
  & R$_{HI, out}$ & 0.28 & 0.17 & 0.19 \\
  & R$_{HI, max}$ & 0.58 & 0.16 & 0.23 \\
NGC 3359 & R$_{opt, in}$ & 0.12 & 0.41 & 0.07 \\
  & R$_{opt, out}$ & 0.15 & 0.30 & 0.09 \\
  & R$_{HI, in}$ & 0.15 & 0.14 & 0.06 \\
  & R$_{HI, out}$ & 0.18 & 0.22 & 0.14 \\
  & R$_{HI, max}$ & 0.51 & 0.22 & 0.08 \\
NGC 4068 & R$_{opt, in}$ & 0.20 & 0.38 & 0.14 \\
  & R$_{opt, out}$ & 0.36 & 0.21 & 0.18 \\
  & R$_{HI, in}$ & 0.35 & 0.17 & 0.14 \\
  & R$_{HI, out}$ & 0.49 & 0.08 & 0.09 \\
  & R$_{HI, max}$ & 0.92 & 0.24 & 0.22 \\
NGC 4861 & R$_{opt, in}$ & 0.16 & 0.07 & 0.03 \\
  & R$_{opt, out}$ & 0.29 & 0.21 & 0.22 \\
  & R$_{HI, in}$ & 0.14 & 0.18 & 0.17 \\
  & R$_{HI, out}$ & 0.23 & 0.21 & 0.11 \\
  & R$_{HI, max}$ & 0.39 & 0.69 & 0.25 \\
NGC 7292 & R$_{opt, in}$ & 0.44 & 0.38 & 0.18 \\
  & R$_{opt, out}$ & 0.19 & 0.84 & 0.11 \\
  & R$_{HI, in}$ & 0.13 & 0.17 & 0.12 \\
  & R$_{HI, out}$ & 0.60 & 0.34 & 0.21 \\
  & R$_{HI, max}$ & 0.83 & 0.46 & 0.29 \\
NGC 7497 & R$_{opt, in}$ & 0.15 & 0.35 & 0.12 \\
  & R$_{opt, out}$ & 0.12 & 0.32 & 0.10 \\
  & R$_{HI, in}$ & 0.16 & 0.13 & 0.17 \\
  & R$_{HI, out}$ & 0.09 & 0.35 & 0.06 \\
  & R$_{HI, max}$ & 0.19 & 0.90 & 0.07 \\
NGC 7610 & R$_{opt, in}$ & 0.05 & 0.14 & 0.02 \\
  & R$_{opt, out}$ & 0.20 & 0.71 & 0.11 \\
  & R$_{HI, in}$ & 0.16 & 0.13 & 0.08 \\
  & R$_{HI, out}$ & 0.20 & 0.20 & 0.14 \\
  & R$_{HI, max}$ & 0.44 & 0.38 & 0.17 \\
NGC 7741 & R$_{opt, in}$ & 0.18 & 0.34 & 0.12 \\
  & R$_{opt, out}$ & 0.12 & 0.47 & 0.09 \\
  & R$_{HI, in}$ & 0.26 & 0.20 & 0.23 \\
  & R$_{HI, out}$ & 0.31 & 0.25 & 0.08 \\
  & R$_{HI, max}$ & 0.76 & 0.58 & 0.12 \\
NGC 7800 & R$_{opt, in}$ & 0.39 & 0.33 & 0.15 \\
  & R$_{opt, out}$ & 0.45 & 0.37 & 0.26 \\
  & R$_{HI, in}$ & 0.14 & 0.13 & 0.15 \\
  & R$_{HI, out}$ & 0.40 & 0.33 & 0.14 \\
  & R$_{HI, max}$ & 0.93 & 0.63 & 0.25 \\
  \hline \hline
    \end{tabular}
    \caption{The average ratio of the amplitudes for each mode ($m=1$ to $3$) to $m=0$ mode. The first two rows for each galaxy show the average ratio of the amplitude for the inner (R$_{opt, in}$) and outer regions (R$_{opt, out}$) of the stellar disk. The third and fourth rows for each galaxy show the average ratio of the amplitude for the inner (R$_{HI, in}$) and outer regions (R$_{HI, out}$) of the gas disk. Consecutively, the fifth row for each galaxy shows the value of this ratio at the maximum traceable radius of the gas disk. It is to be noted that the quantities in the fifth row for each galaxy are not an average ratio; these are the values of the ratios at the maximum radius of the H~{\sc i} disk.}
    \label{tab:lop_inout1}
\end{table}

%%%%%%%%%%%%%%%%%%%%%%%%%%%%%%%%%%%%%%%%%%%%%%%%%%%%%%%%%%%%%%%%%%%%%%%%%
%%%%%%%%%%%%%%%%%%%%%%%%%%%%%%%%%%%%%%%%%%%%%%%%%%%%%%%%%%%%%%%%%%%%%%%%%
\section{Kinematic lopsidedness}
\label{kin_lop}

The lopsidedness phenomenon, as defined morphologically in the earlier sections, can also be anticipated and defined kinematically. In the past, it has been well observed from single-dish observations that a large fraction of galaxies show asymmetry in their H~{\sc i} line profile. From the observation of nearly 1700 galaxies, \cite{hiline_lop1994} showed that around 50 \% of their sample shows asymmetry in their H~{\sc i} line profile. However, the asymmetry in the H~{\sc i} line profile may arise due to both morphological asymmetry or kinematical asymmetry, and it is not possible to infer if kinematic lopsidedness is the sole origin of the asymmetry in the H~{\sc i} line profile. However, in later times, with the advancement of the radio interferometers and techniques of synthesis imaging, the kinematic lopsidedness has been studied through the difference in H~{\sc i} rotation curve between the approaching and receding sides of the galaxies \citep[e.g.,][]{kinlop_whisp1999}. In recent times, with the H~{\sc i} interferometric data from \say{Westerbork H~{\sc i} Survey of Irregular and SPiral galaxies (WHISP) \citep{paper5_whisp1} }, \citet{kinlop_whisp2011} found the kinematic lopsidedness from the difference between rotation velocity derived from the rotation curve of two halves (approaching and receding) of galaxies for about $\sim 70$ sources. However, the rotation curves used in their studies were derived by fitting the 2-dimensional Tilted-ring model to the 2-dimensional velocity field. As discussed in detail in the study by \citet{fat} and in section 3 of \citet{biswas2023}, kinematics and hence the rotation curve derived from the 2D velocity field may often be affected by the beam smearing and projection effects; as such, the 2D modelling may not provide accurate kinematics of the galaxies. The aforementioned works used samples that, in principle, are limited to intermediate ranges of inclination. Because the rotation curves derived from the 2D-velocity field of the galaxies with higher inclinations are highly affected by projection effects \citep{Sancisi1979, Sofue2001}. Also, galaxies with lower inclination ($<\sim$ 40$^\degree$) are not well suited for doing 2D-kinematical modelling from 2D velocity field \citep{Begeman1987PhDT, Bershady2010, fat}.  In this study, we use the rotation curves from the 3D kinematic modelling to find the kinematic lopsidedness and compare our results with \citet{kinlop_whisp2011}. 

\begin{figure*}
    \centering
    \begin{tabular}{ccc}
         \includegraphics[width=0.32\textwidth]{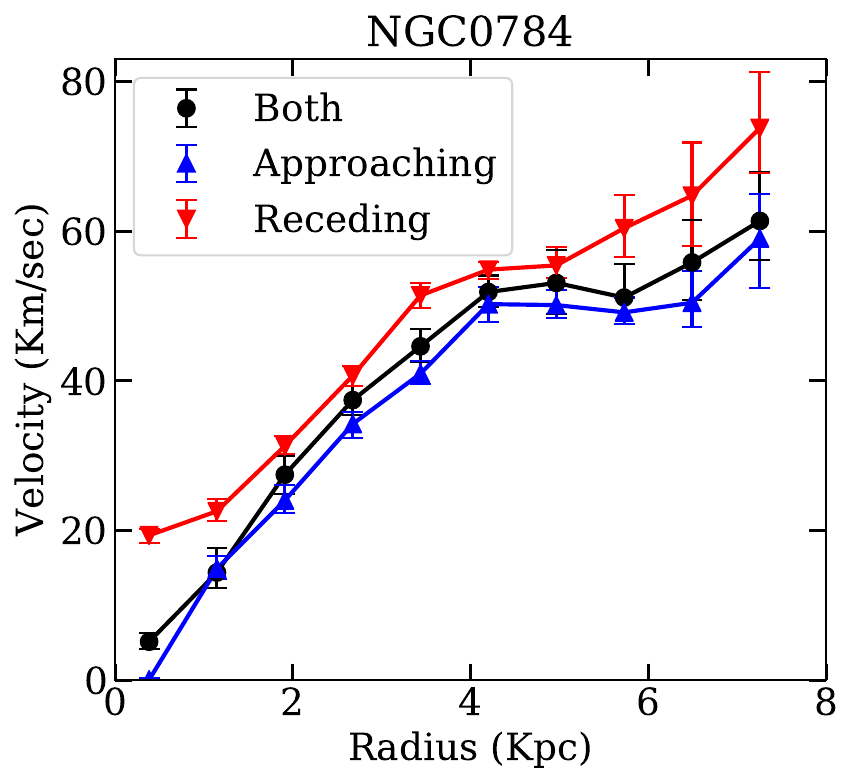} &
         \includegraphics[width=0.32\textwidth]{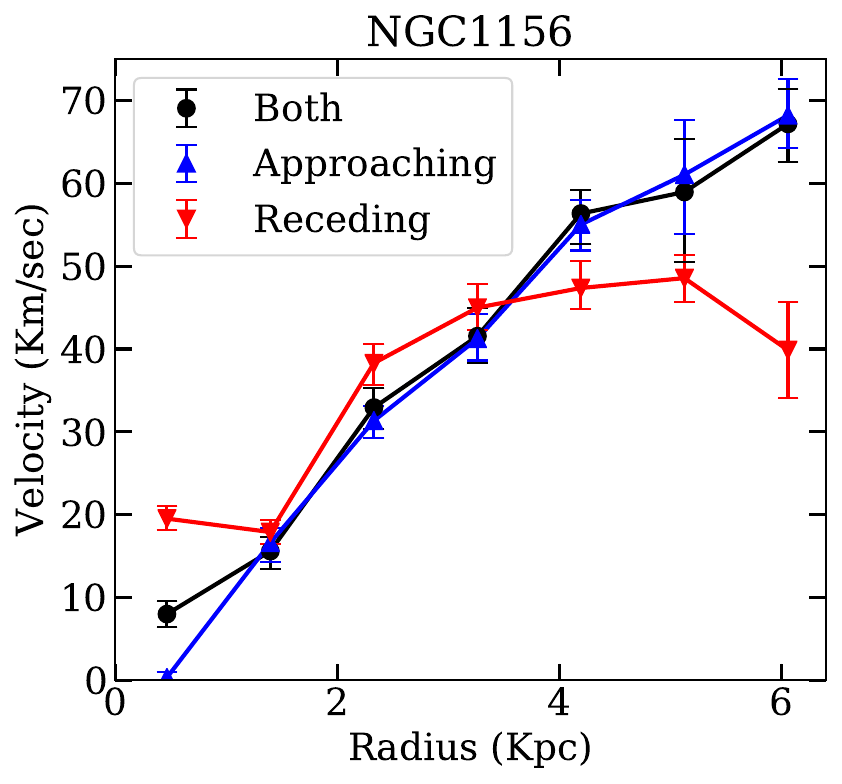} &
         \includegraphics[width=0.32\textwidth]{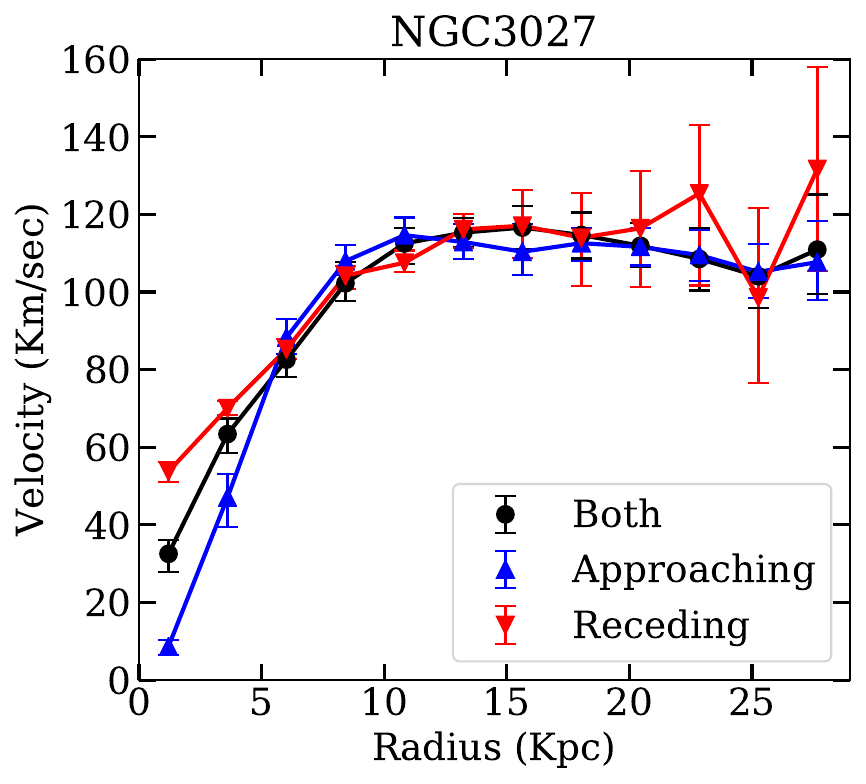} \\

        \includegraphics[width=0.32\textwidth]{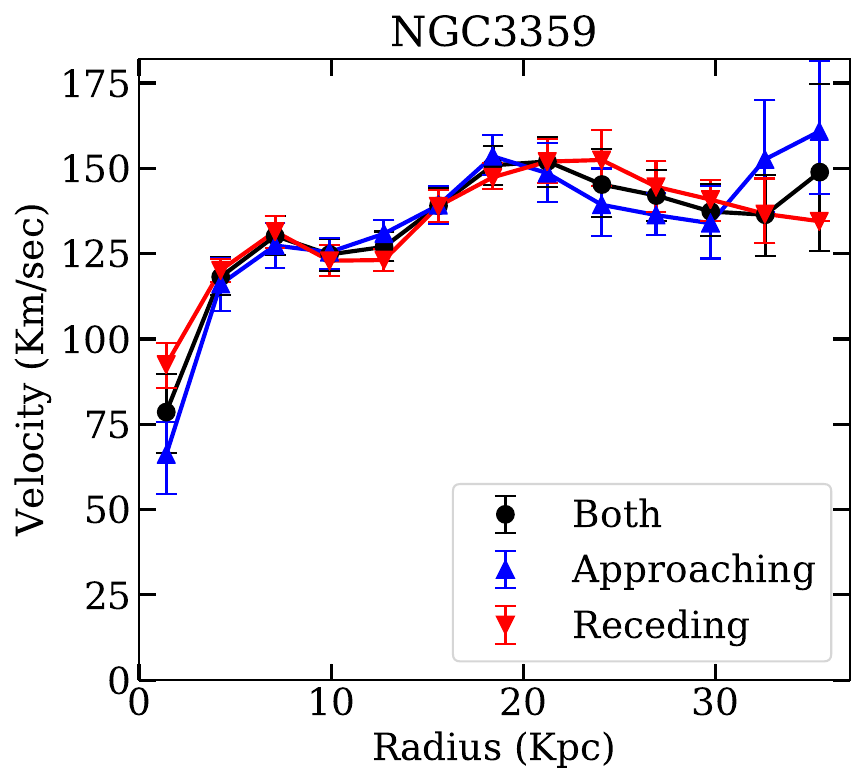} &
         \includegraphics[width=0.32\textwidth]{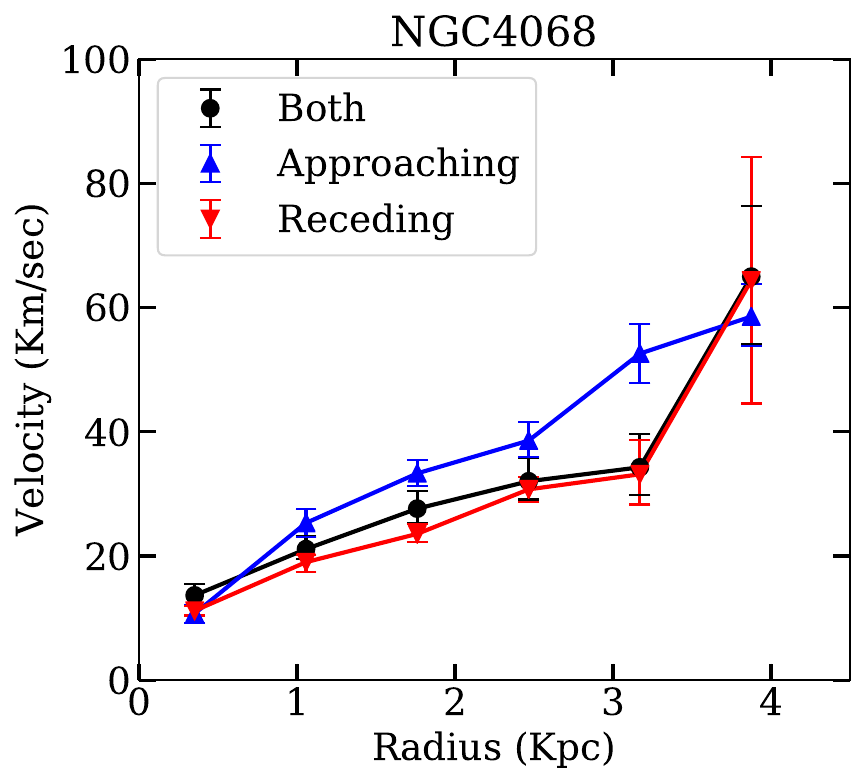} &
         \includegraphics[width=0.32\textwidth]{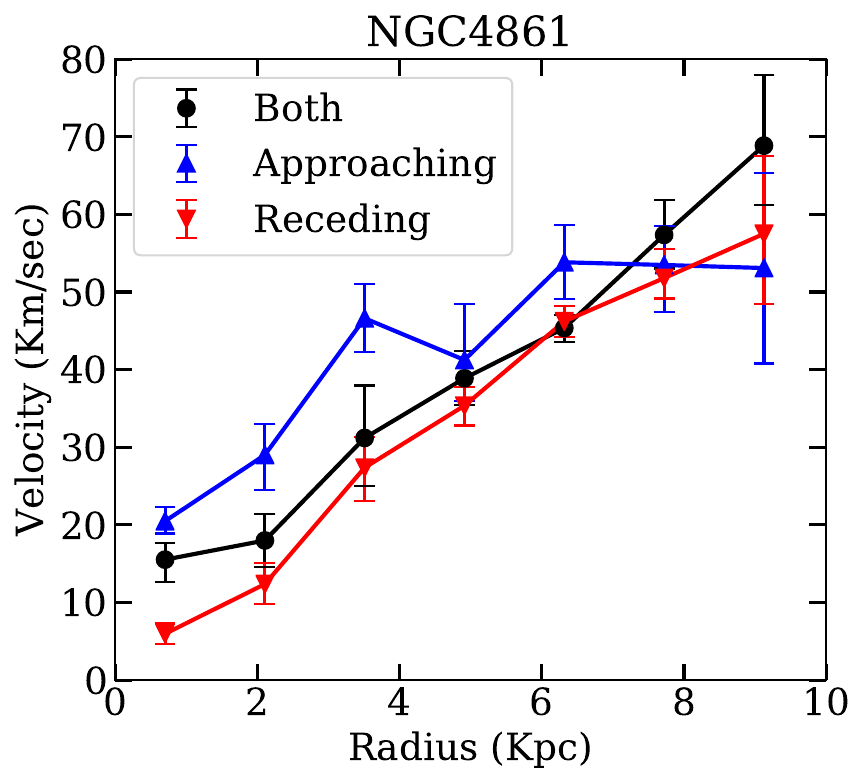} \\

        \includegraphics[width=0.32\textwidth]{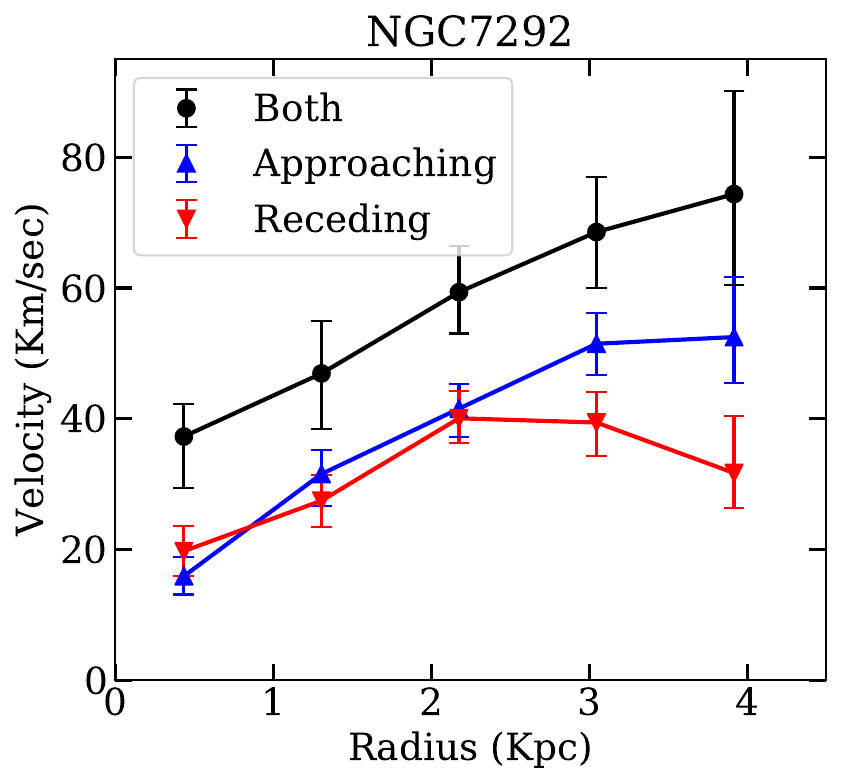} &
         \includegraphics[width=0.32\textwidth]{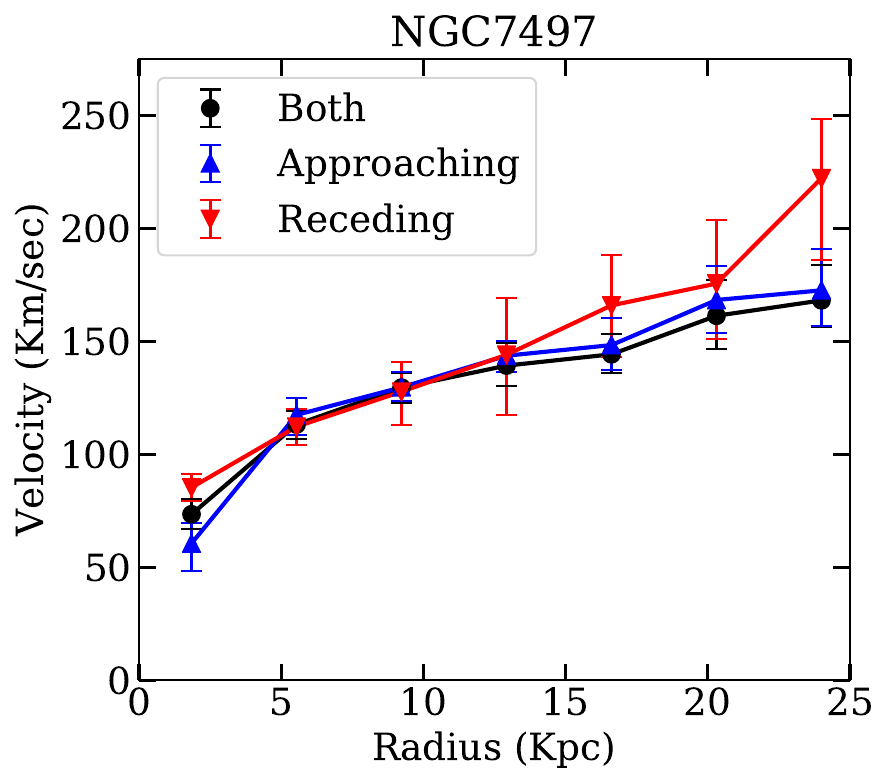} &
         \includegraphics[width=0.32\textwidth]{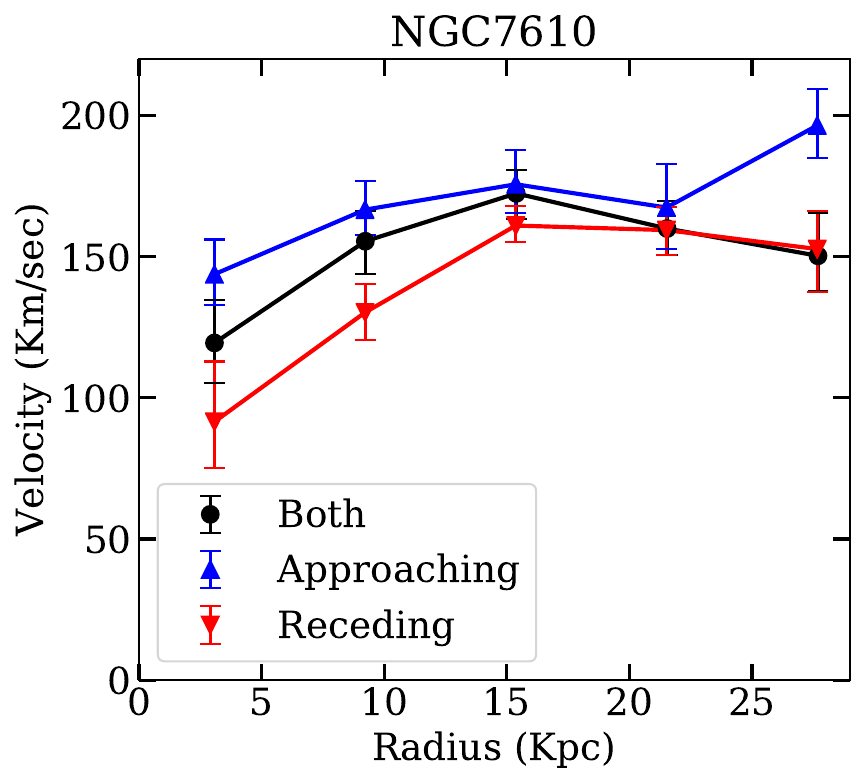} \\
        \end{tabular}
         \begin{tabular}{cc}
         \includegraphics[width=0.32\textwidth]{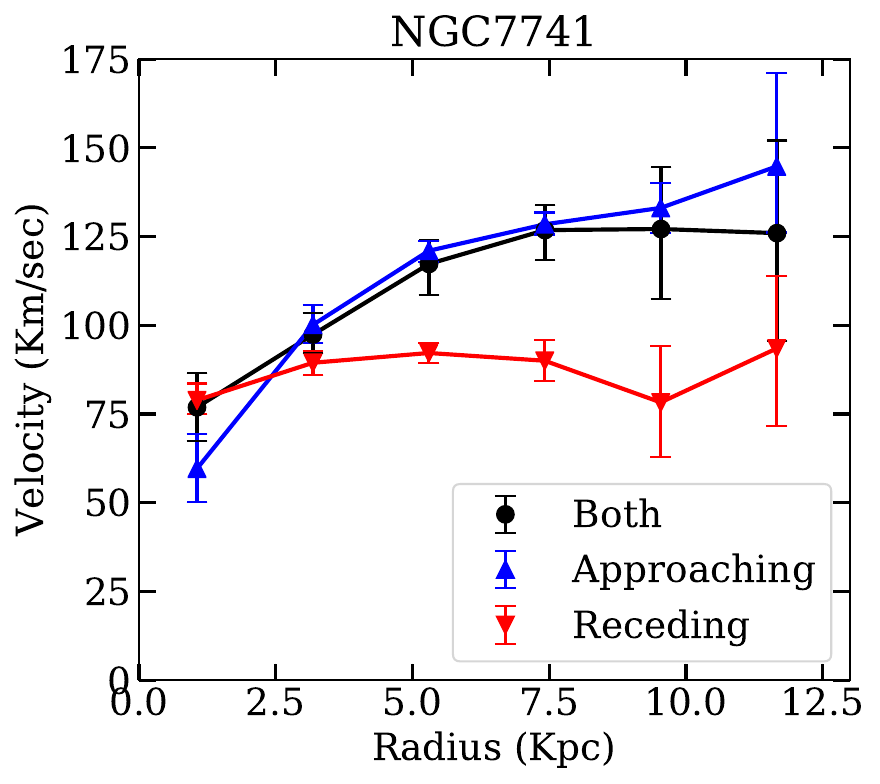} &
         \includegraphics[width=0.32\textwidth]{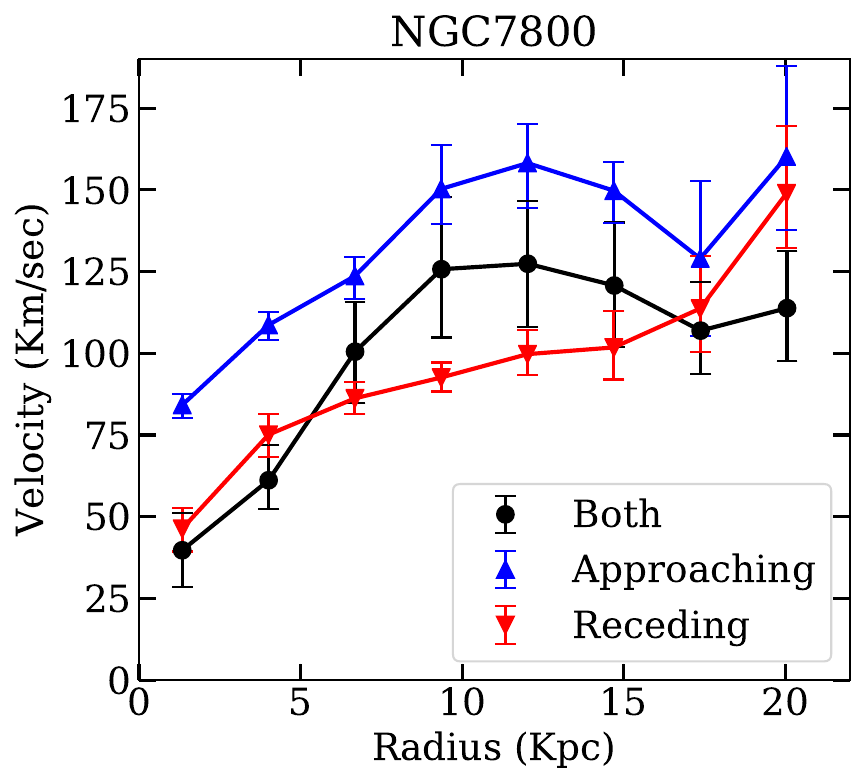} \\
         \end{tabular}
    
    \caption{The rotation curve of the galaxies from the approaching side, the receding side and the whole galactic disk. }
    \label{fig:rotc_lop}
\end{figure*}

In this regard, we use the 3D-Tilted ring model fitting pipeline, BBarolo \citep{BBarolo}, to find the rotation curves from the approaching and receding sides separately. Although the 3D-tilted ring model fitting pipeline, FAT \citep{fat}, in general, gives lower residuals than BBarolo in the fitted models \citep[see section 3, ][]{biswas2023}, this pipeline does not have the option to produce the rotation curve separately in the two halves of the galaxy. Hence, we use BBarolo to fit the rotation curves separately in the two halves of the galaxies. The chosen free parameters and the procedure for guessing the initial values of these parameters and other details for fitting kinematic modelling using BBarolo are discussed in section 3 of \citet{biswas2023}. Figure \ref{fig:rotc_lop} shows the rotation curves derived by fitting the approaching side, the receding side and the whole galactic disk. Further, \citet{kinlop_whisp2011} categorised the differences in the rotation curve arising from both sides in 5 ways:
\begin{itemize}
\itemsep0em
    \item {\it{Type 1}}: Rotation curve from the approaching and receding sides matches in all scales.
    \item {\it{Type 2}}: There is a constant offset between the rotation curves from approaching and receding sides.
    \item {\it{Type 3}}: Rotation curves from approaching and receding sides differ only at large radii.
    \item {\it{Type 4}}: Rotation curves from approaching and receding sides differ only at smaller radii.
    \item {\it{Type 5}}: The rotation curve from one side crosses the rotation curve from the other side.
\end{itemize}

From Fig.~\ref{fig:rotc_lop}, the galaxies in our sample exhibit all five kinds of characteristics. NGC 3359 and NGC 7497 show well-matched rotation curves on both sides ({\it Type 1}). NGC 0784 and NGC 7610 display a constant offset between the two sides ({\it Type 2}). NGC 7292 differs only at large radii ({\it Type 3}), while NGC 3027, NGC 4861, and NGC 7800 differ only at small radii ({\it Type 4}). In NGC 1156, NGC 4861, and NGC 7741, the rotation curves cross each other ({\it Type 5}). These differences in the rotation curves from the approaching and receding sides suggest the lopsided kinematics of the corresponding galaxies. We further derive the halo perturbation parameter from this kinematic lopsidedness as well as from morphological lopsidedness and compare them in the next section.

\section{Halo perturbation parameter}
\label{halo_parameter}

As mentioned in the introduction, lopsidedness can arise due to dynamical instability or interaction with the surroundings; it can also result from a perturbed dark matter halo associated with the galaxy. In this section, we found the perturbation parameter describing a lopsided halo from both morphological and kinematic considerations. 

\subsection{Halo perturbation parameter from morphological consideration}
From a theoretical consideration, \citet{jog2002} showed that both the kinematic and morphological lopsidedness can be explained by the response of the galactic disk to a lopsided halo potential. The perturbed potential ($\psi_{m}$) corresponding to the $m$th mode of the Fourier analysis of the morphological lopsidedness for a non-rotating case can be expressed in the following form \citep{Jog2000}:
\begin{equation}
    \psi_{m} = V_c^2 \epsilon_m \cos(m\phi), 
    \label{eqn:per_potential}
\end{equation}
where $V_c$ is the circular velocity at the flat part of the rotation curve, $\phi$ is the azimuthal angle in the plane of the galaxy, and $\epsilon_m$ is a small dimensionless parameter describing the perturbation parameter in the case of the $m$th mode of the Fourier analysis of the morphological lopsidedness. For an exponential disk, this parameter can be computed considering the Fourier modes arising in the morphological lopsidedness analysis of the disk. For $m=1, 2$ and $3$, we find this parameter following the equations mentioned below \citep{Jog2000}:

\begin{align}
    & \epsilon_1 = \frac{I_1/I_0}{(2R/R_{h}) - 1} , \\
    & \epsilon_2 = \frac{I_2/I_0}{1+R/R_{h}} , \nonumber \\
    &  \epsilon_3 = \frac{I_3/I_0}{1+2R/7R_{h}}, \nonumber
\end{align}

where $R_{h}$ is the scale-length considering an exponential disk. Here in our study we already considered the optical disk to be exponential and the H~{\sc i} disk to be nearly exponential in the outer region of the disk. Thus, using the equations mentioned above,  we find out the halo perturbation parameter for $m=1, 2$ and $3$  modes for both the H~{\sc i} disk and stellar disk. The averages of these perturbation parameters are found only in the outer regions of the star and gas disk, where the ratio of the amplitudes of these modes is significant, as shown in the previous sections. Table \ref{tab:esp_lop_morph} with figure \ref{fig:esp_morph} and figure \ref{fig:esp_morph_apndx} in appendix \ref{apndx2} show the average value of this parameter for different modes for all the galaxies. We do not find any correlation between the halo perturbation parameter of each mode and the morphology of the galaxies or the presence of bars or have any environmental effect for this sample of galaxies.

\begin{table}
    \centering
    \begin{tabular}{ccccc}
        \hline \hline
     Galaxy & Region &  $<\epsilon_1>$ & $<\epsilon_2>$ & $<\epsilon_3>$ \\
     \hline
     \hline
NGC 0784 & R$_{opt,out}$ & 0.03 & 0.11 & 0.05 \\
  & R$_{HI,out}$ & 0.08 & 0.19 & 0.07 \\
NGC 1156 & R$_{opt,out}$ & 0.07 & 0.24 & 0.10 \\
  & R$_{HI,out}$ & 0.08 & 0.06 & 0.05 \\
NGC 3027 & R$_{opt,out}$ & 0.19 & 0.14 & 0.14 \\
  & R$_{HI,out}$ & 0.13 & 0.06 & 0.13 \\
NGC 3359 & R$_{opt,out}$ & 0.07 & 0.11 & 0.06 \\
  & R$_{HI,out}$ & 0.05 & 0.07 & 0.08 \\
NGC 4068 & R$_{opt,out}$ & 0.16 & 0.08 & 0.12 \\
  & R$_{HI,out}$ & 0.08 & 0.02 & 0.04 \\
NGC 4861 & R$_{opt,out}$ & 0.07 & 0.05 & 0.11 \\
  & R$_{HI,out}$ & 0.03 & 0.04 & 0.05 \\
NGC 7292 & R$_{opt,out}$ & 0.11 & 0.29 & 0.08 \\
  & R$_{HI,out}$ & 0.22 & 0.12 & 0.14 \\
NGC 7497 & R$_{opt,out}$ & 0.06 & 0.13 & 0.07 \\
  & R$_{HI,out}$ & 0.01 & 0.07 & 0.03 \\
NGC 7610 & R$_{opt,out}$ & 0.07 & 0.22 & 0.07 \\
  & R$_{HI,out}$ & 0.02 & 0.04 & 0.06 \\
NGC 7741 & R$_{opt,out}$ & 0.05 & 0.16 & 0.06 \\
  & R$_{HI,out}$ & 0.05 & 0.05 & 0.04 \\
NGC 7800 & R$_{opt,out}$ & 0.23 & 0.15 & 0.18 \\
  & R$_{HI,out}$ & 0.11 & 0.10 & 0.08 \\
  \hline \hline
    \end{tabular}
    \caption{Average value of the lopsided halo perturbation parameter found from the morphological distortion of $m=1, 2$ and $3$ modes in the outer region of the stellar and gas disk.  }
    \label{tab:esp_lop_morph}
\end{table}

\begin{figure}
    \centering
   \begin{tabular}{c}
      \includegraphics[width=0.45\textwidth]{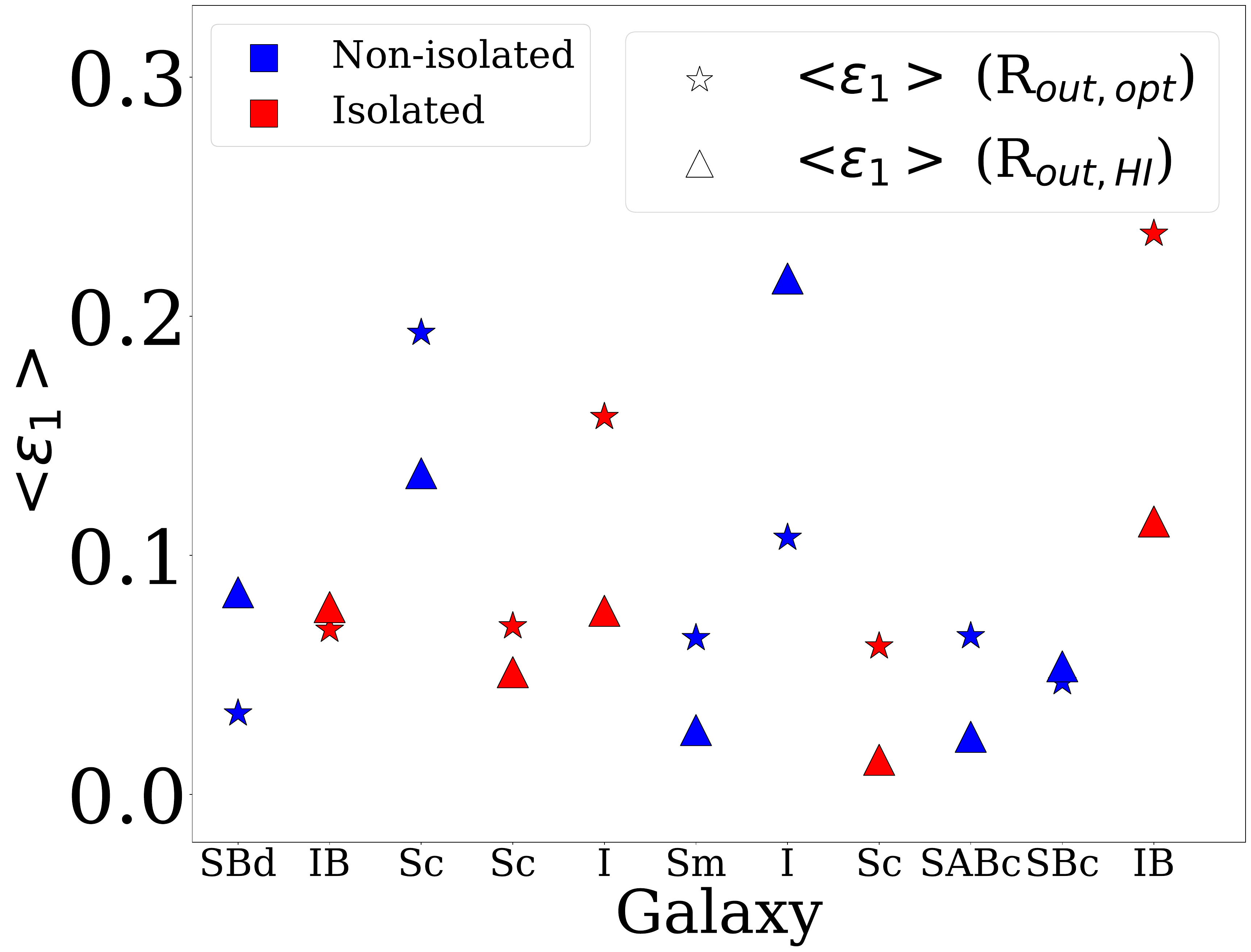}
      \end{tabular}
   
    \caption{Distribution of the average value of the lopsided halo perturbation parameter found from the morphological distortion of $m=1$ mode in the outer region of the stellar and gas disk. Galaxies are colour-coded to indicate if they are non-isolated or isolated. The morphologies of each galaxy are noted on the x-axis.  }
    \label{fig:esp_morph}
\end{figure}

\subsection{Halo perturbation parameter from kinematic consideration}   

Considering the same perturbed potential as mentioned in equation \ref{eqn:per_potential}, the lopsided halo perturbation parameter can be found from kinematical consideration for $m=1$ mode. In this case, the halo perturbation parameter ($\epsilon_{kin}$) can be expressed through the following equation \citep{jog2002}:

\begin{equation}
    \epsilon_{kin} = \frac{|v_{rec} - v_{app}|}{2V_c}, 
\end{equation}
where $v_{rec}$, $v_{app}$ and $V_c$  are the maximum velocity measured from the flat part of the rotation curve rising respectively from the receding side, approaching side and considering both the sides \citep[see section 3.2 of][]{kinlop_whisp2011}. Using the rotation curves derived in section \ref{kin_lop}, we further derive the kinematic halo perturbation parameter for all the galaxies. In cases of those galaxies where the rotation curve does not reach the flat part, the maximum rotation velocity has been considered.

\subsection{Comparison of halo perturbation parameters from morphological and kinematic lopsidedness}
\begin{figure}
    \centering
    \includegraphics[width=0.46\textwidth]{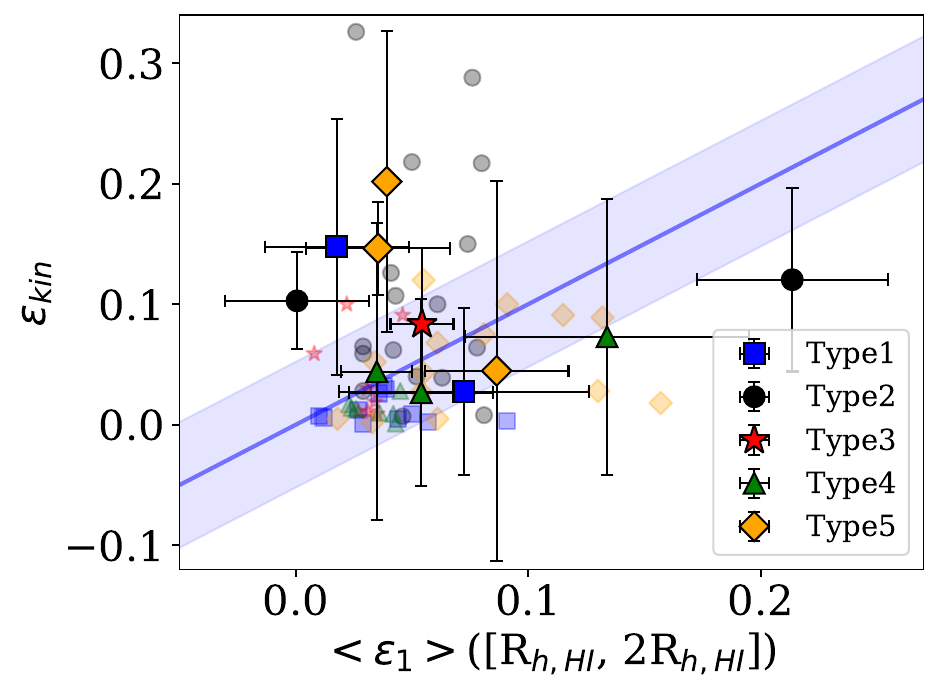}
    \caption{Comparison of $\epsilon_{kin}$ and $<\epsilon_{1}>$ measured between R$_{h,HI}$ and 2R$_{h,HI}$ for various types of differences in the rotation curves. The large data points are from our study, and the small data points are from the study \citet{whisp_lop2_morph}. The solid blue line and the blue shaded region respectively denote where $\epsilon_{kin}$=$<\epsilon_{1}>$ and scatter of represented data points from our study around this line.   }
    \label{fig:whisp_comp}
\end{figure}

We further compare this kinematic halo perturbation parameter arising from the H~{\sc i} rotation curves with the average of the morphological halo perturbation parameter rising from $m=1$ distortion of the H~{\sc i} disk in the outer region. \citet{jog2002} showed that when a halo perturbation is the underlying cause of lopsidedness (equation \ref{eqn:per_potential}), $<\epsilon_1>$ and $\epsilon_{kin}$ should be similar. As mentioned in the introduction, earlier studies by \citet{whisp_lop2_morph} compared the kinematic perturbation parameter and the morphological perturbation parameter and found that they can differ depending upon the nature of the differences of the rotation curves arising from both sides. They found that, galaxies for which $\epsilon_{kin}$ differs most from $<\epsilon_1>$, are either of {\it{Type 2}}, or  {\it{Type 5}}. To compare our results with those from \citet{whisp_lop2_morph}, we first calculate $<\epsilon_1>$ within the region between R$_{h, HI}$ and  2R$_{h, HI}$ and present the distribution of $\epsilon_{kin}$ and $<\epsilon_1>$ through figure \ref{fig:whisp_comp} \citep[similar to figure 5 of ][]{whisp_lop2_morph}. The morphological halo perturbation parameters and kinematic halo perturbation parameters range respectively up to $\sim$ 0.15 and 0.35 in the study by \citet{whisp_lop2_morph}. Both the morphological halo perturbation parameters and kinematic halo perturbation parameters from our study range up to $\sim$ 0.2.  We found that the RMS orthogonal scatter of the data points from our analysis around the line, $\epsilon_{kin} = <\epsilon_1> ([R_{h, HI}, 2R_{h, HI}])$ is 0.062.  Considering the error bars, ten out of eleven sources in our sample lie within the scatter of this line. For consistency, we also calculated the RMS orthogonal scatter of the data points around this line for the galaxies from \citet{whisp_lop2_morph}, which is 0.057. However, there are many sources in their sample that lie way beyond the scatter around this line. Since \citet{whisp_lop2_morph} does not provide individual uncertainties for their data points, a direct error-weighted comparison is not possible; that is why we have not included the uncertainties in our data points either to calculate the scatter around this line.  Given the lack of published uncertainties for the comparison sample \citep{whisp_lop2_morph}, it is not possible at this point to put a firm conclusion whether the data points in our analysis are less scattered than that from \citet{whisp_lop2_morph}.  Additionally, we found that the large discrepancies reported by \citet{whisp_lop2_morph} for {\it{type 2}} and {\it{5}} are not present in our sample when considering the error for these galaxies. 

\begin{figure}
\centering
    \begin{tabular}{c}
      \includegraphics[width=0.45\textwidth]{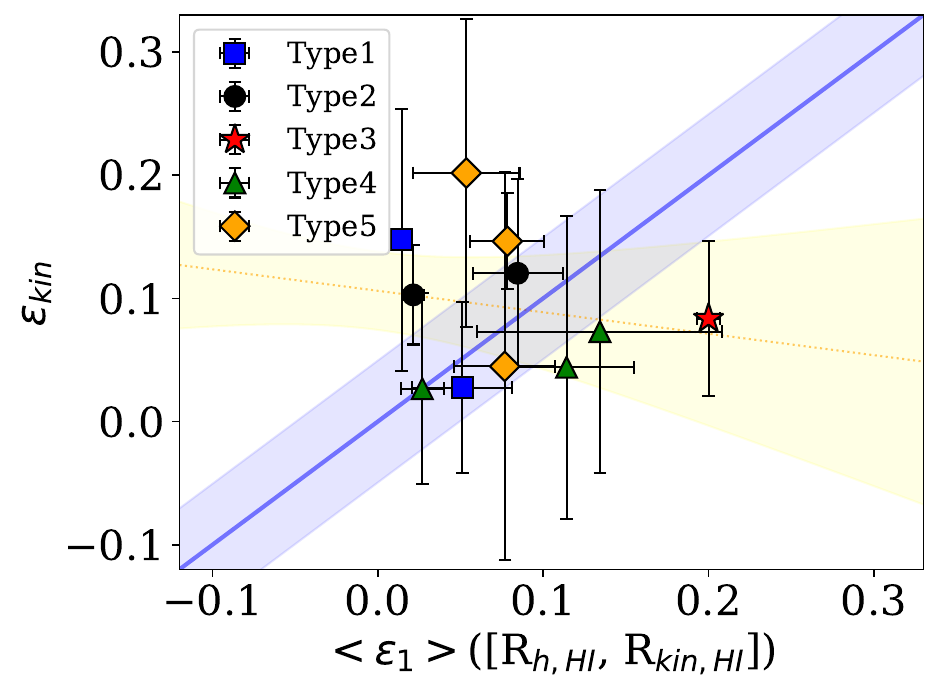}
    \end{tabular}
    \caption{Comparison of $\epsilon_{kin}$ and $<\epsilon_{1}>$ measured between R$_{h,HI}$ and R$_{max,HI}$ for various types of differences in the rotation curves. The solid blue line and the blue shaded region respectively denote where $\epsilon_{kin}$=$<\epsilon_{1}>$ and scatter of represented data points from our study around this line. The dotted yellow line and the yellow-shaded region around it denote the best-fitted line and one-sigma error around it. }
    \label{fig:esp_morph_kin_main}
\end{figure}

It is important to note that the rotation curves used in their studies to find the kinematic lopsidedness and, hence, the kinematic halo perturbation parameter are derived from the 2D-velocity field. For galaxies with very high inclinations or very low inclinations, these rotation curves may be affected due to beam smearing and projection effect as discussed in the study by \citet{fat} and in  section 3 of \citet{biswas2023}. Our study uses the 3D kinematically driven rotation curves to find the kinematic halo perturbation parameter.   Further,  $<\epsilon_1>$ were calculated in the region R$_{h, HI}$ and  2R$_{h, HI}$ in study by \citet{whisp_lop2_morph}. However, we found that the different modes of lopsidedness can vary significantly beyond 2R$_{h, HI}$.  Thus, we compare $<\epsilon_1>$ derived within region R$_{h, HI}$ and R$_{kin, HI}$ with  $\epsilon_{kin}$, where R$_{kin, HI}$ is the radius up to which the 3D kinematic modelling was done successfully. It is to be noted that for some galaxies, R$_{kin, HI}$ is larger than the radius up to which we should perform the Fourier analysis of the gas disk to get physically meaningful results. In those cases, the radius up to which we should conduct the Fourier analysis is taken as the maximum radius to calculate $<\epsilon_1>$. For this case also, it is referred to as R$_{kin, HI}$ to avoid any confusion. However, we checked that this maximum radius up to which the Fourier transform can be done is greater than 2R$_{h, HI}$ for all the cases. Hence, R$_{kin, HI}$ in our study does not necessarily imply radius from the 3D kinematically modelling for all the sources.  The corresponding results are shown in figure \ref{fig:esp_morph_kin_main} and in table \ref{tab:esp_morph_kin}.  From table \ref{tab:esp_morph_kin}, we can see that the halo perturbation parameter found in both ways is in a similar range [$\sim 0.01$ to $\sim$ 0.2]. 

We further fit a straight line to this data to find out the relation between them.  The resulting slope and intercept are $-0.17 \pm 0.34$ and $0.106 \pm 0.032$, respectively, with an RMS orthogonal scatter of $0.0527\pm0.0030$. However, we also find the RMS orthogonal scatter of these points around the line where $<\epsilon_1> = \epsilon_{kin}$,  to be equal to $0.0587$. As our sample size is only eleven, the best-fitted line may not represent the correct relation between them. However, the scatter of the data points around the best-fitted line is a little smaller or nearly similar to the line where $<\epsilon_1>$ is equal to  $\epsilon_{kin}$, and is similar. 

 Further, a Pearson correlation test between $<\epsilon_1>$ measured between [$R_{h, HI}$, $R_{kin, HI}$],  and $\epsilon_{kin}$ yields $r = -0.17$ with $p = 0.62$. Now, if we take $<\epsilon_1>$ measured between [$R_{h, HI}$, $2R_{h, HI}$] and find its correlation with $\epsilon_{kin}$, then its value comes out to be $r = -0.12$ with $p = 0.73$. Finally, for the \citet{whisp_lop2_morph} sample, this correlation coefficient comes out to be $r = 0.098$ with $p = 0.46$. These correlation coefficients indicate that, both for the significantly large sample of \citet{whisp_lop2_morph} and for our smaller sample,  $<\epsilon_1>$ shows little to no linear correlation with $\epsilon_{kin}$. However, comparing the three cases suggests that $<\epsilon_1>$ and $\epsilon_{kin}$ are relatively better linearly correlated when $<\epsilon_1>$ is measured between $[R_{h,\mathrm{HI}}, R_{\mathrm{kin,HI}}]$ and $\epsilon_{kin}$ is derived from 3D kinematically modelled rotation curves, despite the small sample size of eleven galaxies.

Moreover, in \citet{whisp_lop2_morph}, the largest discrepancies between $<\epsilon_1>$ and $\epsilon_{kin}$ were observed for rotation curve differences of \textit{type 2} and \textit{type 5}. In contrast, in our analysis, the discrepancies for types \textit{2} and \textit{5} are not significantly larger than for other types. In fact, these discrepancies are smaller than those found in the previous part, where $<\epsilon_1>$ was computed over the radial range $[R_{h,\mathrm{HI}}, 2R_{h,\mathrm{HI}}]$ instead of $[R_{h,\mathrm{HI}}, R_{\mathrm{kin,HI}}]$. Thus, despite of having small sample size, we can infer that $<\epsilon_1>$ is likely to be consistent with $\epsilon_{kin}$ most when $\epsilon_{kin}$ is derived from the 3D-kinematically modelled rotation curves and $<\epsilon_1>$ is measured in between $[R_{h,\mathrm{HI}}, 2R_{h,\mathrm{HI}}]$.

\begin{table}
    \centering
    \begin{tabular}{ccc}
    \hline
            Galaxy & $<\epsilon_1>$ & $\epsilon_{kin}$ \\
            \hline
    NGC 0784 & $0.08 \pm 0.03$ & 0.12$\pm$0.08 \\
    NGC 1156 & $0.08 \pm 0.02$ &  0.15$\pm$0.04 \\
    NGC 3027 & $0.13 \pm 0.07$ &  0.07$\pm$0.11 \\
    NGC 3359 & $0.05 \pm 0.03$ & 0.03$\pm$0.07 \\
    NGC 4068 & $0.08 \pm 0.03$ & 0.04$\pm$0.16 \\
    NGC 4861 & $0.03 \pm 0.01$ & 0.03$\pm$0.08 \\
    NGC 7292 & $0.200 \pm 0.007$ & 0.08$\pm$0.06 \\
    NGC 7497 & $0.014 \pm 0.005$ & 0.15$\pm$0.11 \\
    NGC 7610 & $0.021 \pm 0.006$ & 0.10$\pm$0.04 \\
    NGC 7741 & $0.05 \pm 0.03$ & 0.20$\pm$0.12 \\
    NGC 7800 & $0.11 \pm 0.04$ & 0.04$\pm$0.12 \\            
  \hline
    \end{tabular}
    \caption{Comparison of the halo perturbation parameter derived from the morphological lopsidedness at the outer region of the H~{\sc i} disk and kinematic lopsidedness derived from the H~{\sc i} observation.}
    \label{tab:esp_morph_kin}
\end{table}

\section{Conclusion}
\label{conclusion}

In conclusion, through this study, we have studied the morphological lopsidedness of the stellar and gas disks in different regions through the Fourier decomposition method. We also found the kinematic lopsidedness from the rotation curves derived by fitting the 3D tilted ring model separately in the approaching and the receding sides. 
We further estimated the halo perturbation parameter independently from both morphological and kinematic lopsidedness. Although no statistically significant linear correlation is detected within our present small sample, the use of 3D kinematic modelling and consistent radial ranges (both evaluated between R$_{h,\mathrm{HI}}$ and R$_{\mathrm{kin,HI}}$) provides a more uniform and physically consistent framework to examine the theoretical expectation that $<\epsilon_{1} >$ and $\epsilon_{kin}$ should be similar. Further , within this framework, the discrepancy between $<\epsilon_{1} >$ and $\epsilon_{kin}$ in independent of the nature of the differences between the rotation curve from two sides of the galaxy, unlike the trends seen in the study by \citet{whisp_lop2_morph}. We emphasize, however, that these conclusions are based on a sample of eleven galaxies. 

We need to include more sources in this study to state whether these results hold only for these selected eleven galaxies or it holds in general for all the galaxies. However, the results of the current paper with these eleven galaxies with different morphology, mass range and inclination angles clearly demonstrate the importance of rechecking the consistency between the halo perturbation parameter found morphologically and kinematically and re-verify the theoretical prediction by \citet{jog2002}. 
Hence, as a future aspect of this work, we would expand our sample of the current star-forming systems to a large number of galaxies across the Hubble sequence and different evolutionary stages \citep[e.g.,][]{kalinova2021, kalinova2022} and in different environments to state whether our results are statistically significant.

\section*{Acknowledgements}
We sincerely thank Dr. Nirupam Roy for his thoughtful insights, scientific guidance and consistent encouragement in shaping the core ideas of this study. His intellectual inputs played a significant role in the development of this manuscript. We also thank Dr. Chanda Jog, Dr. Mousumi Das, and Dr. Soumavo Ghosh for the interesting and helpful discussion regarding this study. We also thank the anonymous reviewers for their suggestions and comments that have helped to improve this paper. Further, this work has utilised the SPITZER and SDSS databases. The Spitzer Space Telescope was operated by the Jet Propulsion Laboratory, California Institute of Technology under a contract with NASA. Support for this work was provided by an award issued by JPL/Caltech. For the Sloan Digital Sky Survey (SDSS), funding has been provided by the Alfred P. Sloan Foundation; the participating institutions are the National Aeronautics and Space Administration, the National Science Foundation, the U.S. Department of Energy, the Japanese Monbukagakusho, and the Max Planck Society. This research has also made use of the NASA/IPAC Extragalactic Database (NED), operated by the Jet Propulsion Laboratory, California Institute of Technology, under contract with the National Aeronautics and Space Administration.

%%%%%%%%%%%%%%%%%%%%%%%%%%%%%%%%%%%%%%%%%%%%%%%%%%
\section*{Data Availability}

All the derived quantities and models produced in this study will be shared at reasonable request to the corresponding author.

%%%%%%%%%%%%%%%%%%%% REFERENCES %%%%%%%%%%%%%%%%%%

% The best way to enter references is to use BibTeX:

\bibliographystyle{mnras}
\bibliography{example} % if your bibtex file is called example.bib

% Alternatively you could enter them by hand, like this:
% This method is tedious and prone to error if you have lots of references
%\begin{thebibliography}{99}
%\bibitem[\protect\citeauthoryear{Author}{2012}]{Author2012}
%Author A.~N., 2013, Journal of Improbable Astronomy, 1, 1
%\bibitem[\protect\citeauthoryear{Others}{2013}]{Others2013}
%Others S., 2012, Journal of Interesting Stuff, 17, 198
%\end{thebibliography}

%%%%%%%%%%%%%%%%%%%%%%%%%%%%%%%%%%%%%%%%%%%%%%%%%%
%%%%%%%%%%%%%%%%% APPENDICES %%%%%%%%%%%%%%%%%%%%%
\appendix

\section{Comparison of the halo perturbation parameter rising from different Fourier modes}
\label{apndx2}
\begin{figure}
    \centering
    \begin{subfigure}[b]{0.45\textwidth}
         \centering
         \includegraphics[width=\textwidth]{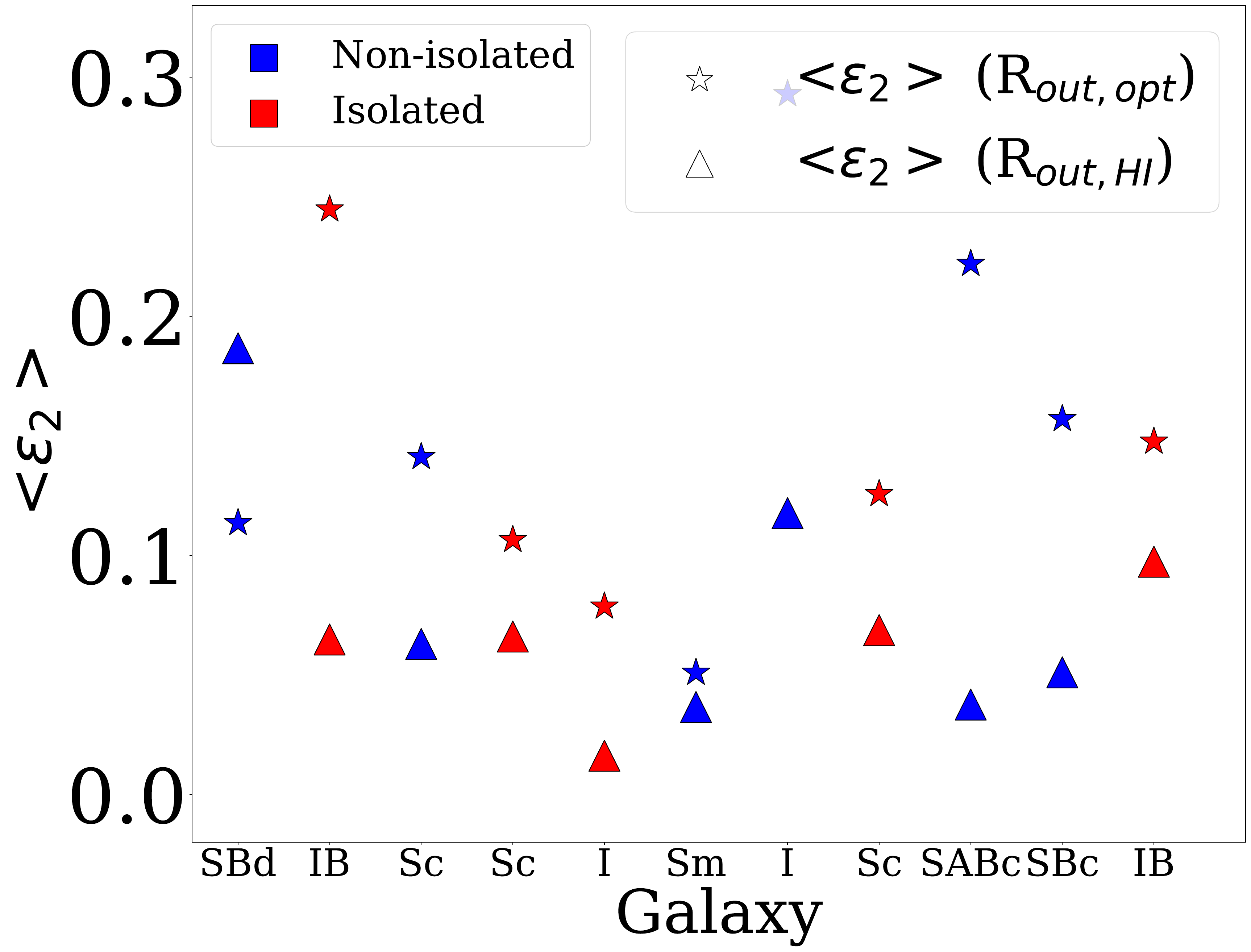}
         %\caption{}
         %\label{fig}
     \end{subfigure}
     \hfill
     \begin{subfigure}[b]{0.45\textwidth}
         \centering
         \includegraphics[width=\textwidth]{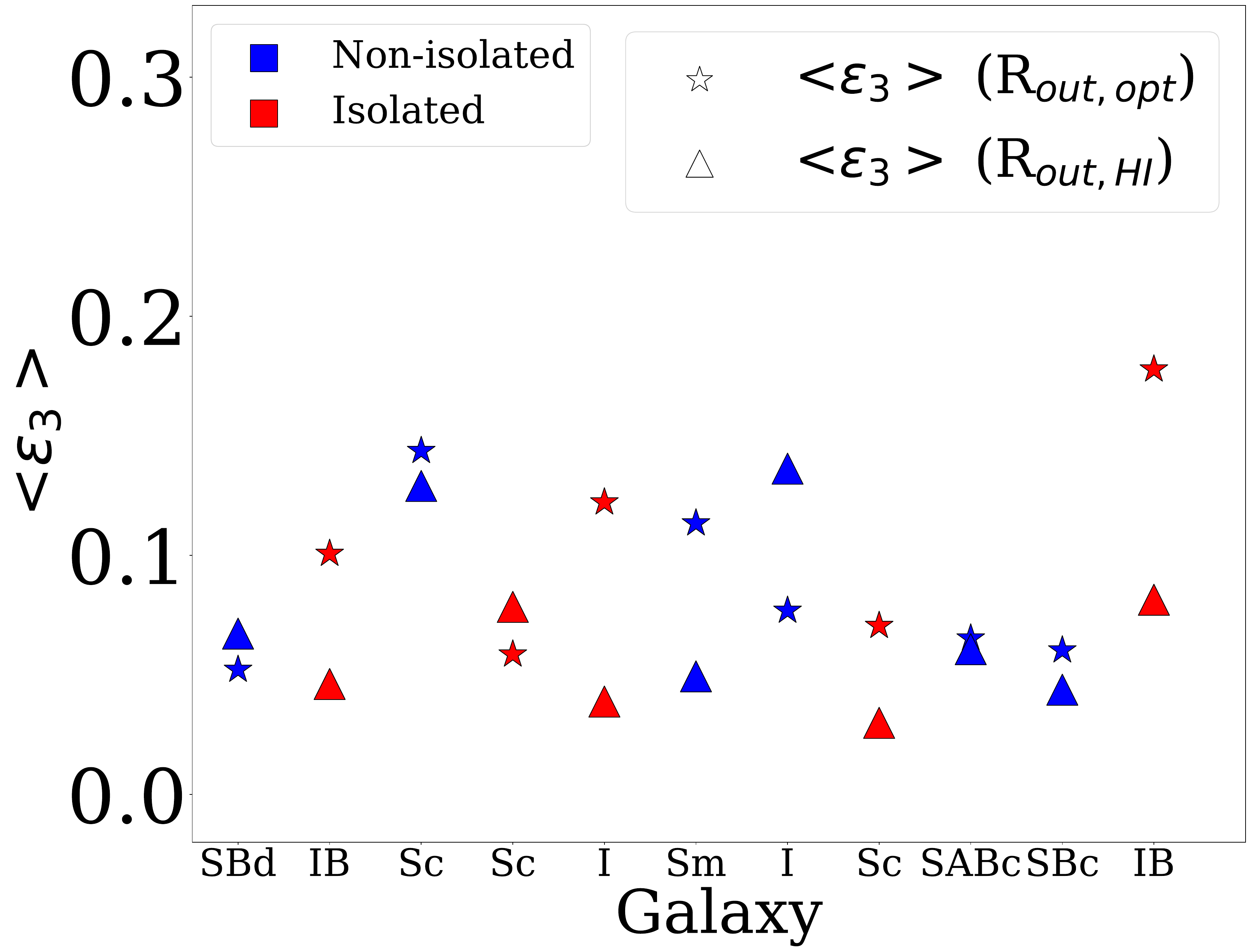}
         %\caption{}
         %\label{fig}
     \end{subfigure}
    \caption{Distribution of the average value of the lopsided halo perturbation parameter found from the morphological distortion of $m= 2$ and $3$ modes in the outer region of the stellar and gas disk. For each figure, galaxies are colour-coded to indicate if they are non-isolated or Isolated. The morphologies of each galaxy are noted on the x-axis.  } 
    \label{fig:esp_morph_apndx}
\end{figure}

Distribution of the average value of the lopsided halo perturbation parameter found from the morphological distortion of $m=2$ and $3$ modes in the outer region of the stellar and gas disk is shown in figure \ref{fig:esp_morph_apndx}. 

%%%%%%%%%%%%%%%%%%%%%%%%%%%%%%%%%%%%%%%%%%%%%%%%%%
%%%%%%%%%%%%%%%%%%%%%%%%%%%%%%%%%%%%%%%%%%%%%%%%%%

% Don't change these lines
\bsp	% typesetting comment
\label{lastpage}

% End of mnras_template.tex

\end{document}